\documentclass[12pt, ChapStyle1, twoside]{./Styles/Dea_Gsm}
\usepackage[latin1]{inputenc}
\usepackage[T1]{fontenc}
\usepackage {longtable}

\usepackage{mathrsfs, amsmath, amssymb, bbm}
\usepackage{pstricks, pst-node, pst-tree, pstcol}
\usepackage[]{graphicx,graphics,epsfig}
\usepackage{floatflt,multirow, array, subfigure, hhline, enumerate, comment, url, pifont}
\usepackage[active]{./Styles/srcltx,rotating}
\usepackage[ruled,vlined,french,titlenumbered,linesnumbered]{algorithm2e}
\usepackage{listings}
\usepackage{lscape}

\usepackage[hidelinks]{hyperref}

\usepackage{epstopdf}

\usepackage[normalem]{ulem}

\makeatletter
\def\hlinewd#1{%
\noalign{\ifnum0=`}\fi\hrule \@height #1 %
\futurelet\reserved@a\@xhline}
\makeatother

\begin{document}
\setcounter{MaxMatrixCols}{10}
\newtheorem{theorem}{Th\'{e}or\'{e}me}
\newtheorem{remark}{Remarque}
\newtheorem{definition}{D\'{e}finition}
\newtheorem{lemma}{Lemme}
\newtheorem{proposition}{Proposition}
\newtheorem{exemple}{Exemple}
\newtheorem{corollaire}{Corollaire}

%%%%%%%%%%%%%%%%%%%%%%%%%%%%%%%%%%%%%%%%%%%%%%%%%%%%%%%%%%%%%%%%%%%%%%%%

\thispagestyle{empty}

%{\setlength{\baselineskip}{0.5\baselineskip} %bill

\begin{center}
\textsc{république tunisienne\\
ministère de l'enseignement supérieur\\
de la recherche scientifique \\
université de tunis el manar
}
%R\'{E}PUBLIQUE TUNISIENNE \\
%MINIST\`{E}RE DE L'ENSEIGNEMENT SUP\'{E}RIEUR,\\
%DE LA RECHERCHE SCIENTIFIQUE ET DE LA TECHNOLOGIE\\
%UNIVERSIT\'{E} DE TUNIS EL MANAR
\end{center}

%\par}
\begin{center}
\begin{figure}[htbp]
    \centering
        \includegraphics[width=4cm,height=2cm]{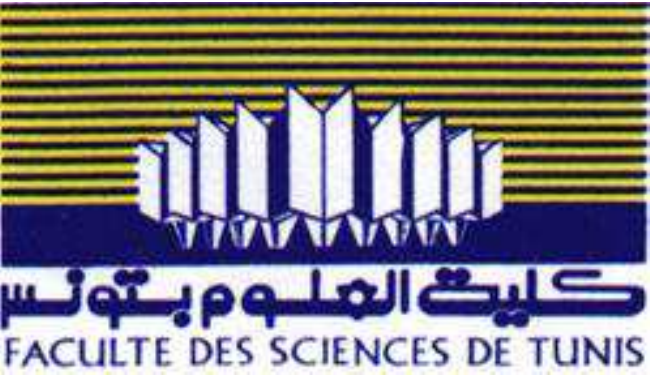}
\end{figure}
%\bigskip
\textsc{faculté des sciences de tunis\\
département des sciences de l'informatique}
%FACULT\'{E} DES SCIENCES DE TUNIS\\
%D\'EPARTEMENT DES SCIENCES DE L'INFORMATIQUE
\end{center}
% \bigskip
\begin{center}
 \LARGE \textbf{\textsc{mémoire de mastère}}%M\'EMOIRE DE MAST\`{E}RE}
\end{center}
% \bigskip
\begin{center}
présenté en vue de l'obtention du\\
\textbf{Diplôme de Mastère en Informatique}\\
%\bigskip
par\\
%\bigskip
\textbf{Imen \textsc{Ben Sassi}}\\
(Maître en Informatique, FST)
\end{center}
\begin{center}
 \LARGE \textbf{La prédiction des intérêts des utilisateurs pour la RI contextuelle et la recommandation d'amis dans un environnement mobile}
\end{center}
 \normalsize
\begin{center}
 Soutenu le 19 Juillet 2012, devant le jury d'examen\\
%\bigskip

%\bigskip

%\begin{longtable} {cll}
% &         &                \\
%%                               &&\\
%%&                &            \\
%%                               &&\\
%%&                &            \\                               &&\\
%%&                &            \\
%% &               &         \\
%\end{longtable}

\begin{longtable} {cll}
\\
Mr. &  Zied \textsc{Bahroun} &  Président \\
Mr. &  Faouzi \textsc{Moussa}   & Rapporteur\\
Mr. & Sadok \textsc{Ben Yahia}  &  Directeur du mémoire\\

\end{longtable}

%\bigskip
Année universitaire: 2011/2012
\end{center}

\newpage
\newpage 

\Remerciements{D\'{e}dicaces} \label{Dedicaces}

\bigskip

\bigskip

\bigskip

\bigskip

\bigskip

\bigskip
\begin{center}
  \emph{Je remercie ma mère et mon père pour leur soutien, leur écoute, et leurs encouragements.
  Je tiens à leur dédier ce mémoire, en guise de reconnaissance éternelle pour m'avoir aidée et soutenue durant cette période.}
\end{center}

\begin{center}
  \emph{Un spécial dédicace à mon cher frère Marwan. Tu es celui qui est là, chaque jour pour me conseiller et pour me protéger. Tu es celui qui essaye de me faire rire dans mes tristesses. Tu es tout simplement celui que j'aimerai pour toujours.
  Je ne peux exprimer à travers ces lignes l'affection et la tendresse que j'éprouve envers toi. Puisse l'amour et la fraternité nous unissent à jamais. Je te souhaite la réussite dans ta vie.
  Que Dieu le tout puissant te garde et te procure santé, bonheur et longue vie.}
\end{center}

\begin{center}
  \emph{Je dédie ce mémoire à toute ma grande famille, en particulier à ma grand-mère, à mes tantes, à mes oncles, à toutes mes cousines, et à mes amis.
  Ils m'ont toujours encouragée, dans les moments de fatigue et de doute.
  Je ne peux les citer tous nommément. Qu'ils sachent que je les aime fort.}
\end{center}

\bigskip

\newpage

\Remerciements{Remerciements} \label{Remerciements}

\bigskip

\emph{C'est avec un grand plaisir que je réserve ces lignes en signe de gratitude et de reconnaissance à tous ceux qui ont contribué de près ou de loin à l'élaboration de ce travail.}
\bigskip

\emph{Je tiens tout d'abord à remercier Monsieur Zied \textsc{Bahroun}, Maître de Conférences à la Faculté des Sciences de Tunis, pour avoir accepté d'être le président de mon jury.}

\bigskip

\emph{J'adresse également mes sincères remerciements à Monsieur Faouzi \textsc{Moussa}, Maître de Conférences à la Faculté des Sciences de Tunis, pour l'intérêt qu'il a porté à mon travail en acceptant d'être le rapporteur de ce mémoire.}

\bigskip

\emph{Toute ma gratitude s'adresse à Monsieur le Maître de Conférences à la Faculté des Sciences de Tunis, Sadok \textsc{Ben Yahia}, mon directeur de mastère qui m'a patiemment formée à la recherche. Il m'a apporté bien plus qu'un encadrement scientifique.
Je le remercie pour sa disponibilité, ses précieux conseils et ses encouragements qui m'ont toujours poussée à progresser et à conserver une grande sérénité pendant toute la durée de ce mastère. Je m'estime très chanceuse de l'avoir pour encadrant.}

\bigskip

\emph{Je suis aussi reconnaissante envers Madame Chiraz \textsc{Trabelsi}, Assistante à l'\'{E}cole Supérieure de Commerce qui a bien voulu être mon codirecteur de mastère. Ses critiques constructives, ses suggestions, ses avis et ses conseils se sont avérés d'une valeur inestimable et m'ont permis d'améliorer considérablement mon travail.}

\bigskip

\emph{Un grand merci à Madame Amel \textsc{Bouzeghoub}, Professeur Agrégé à l'\'{E}cole Télécom SudParis, qui a encadré et suivi de près mes travaux. Ses remarques, son soutien et ses encouragements ont été d'un précieux apport et m'ont beaucoup aidé dans mes investigations. Je la prie de croire à ma sincère reconnaissance pour son acceuil chaleureux et de sa disponibilité tout au long de mon séjour à Paris.}

\bigskip

\bigskip

\emph{Je tiens à remercier tout ceux qui m'ont permis de mener mes travaux dans de bonnes conditions.}

\newpage

%\input{./PagesdeGarde/Resume}
%\Remerciements{} \label{Resume}
%%%%%%%%%%%%%%%%%%%%%%%%%%%%%%%%%%%%%%%%%%%%%%%%%%%%%%%%%%%%%%%%%%%%%%%%
\thispagestyle{empty}
\textbf{Résumé:}
L'émergence des smartphones a donné à l'informatique mobile l'accès à la réalité quotidienne.
Plus précisément, la modélisation du contexte des utilisateurs offre un moyen efficace pour adapter les résultats de recherche et même les éléments recommandés à ces derniers en limitant l'espace des données.
Par ailleurs, ces dernières années, beaucoup de sites sociaux ont intégré la notion du contexte dans leurs recommandations. En effet, avec la disponibilité des appareils mobiles, ces nouveaux sites mobiles ont eu l'avantage de fournir aux utilisateurs des éléments plus pertinents vu leurs situations courantes.
Ainsi, nous introduisons une nouvelle approche de RI contextuelle dans le cadre d'un environnement mobile. Nous proposons d'une part, une approche appelée \textbf{SA-IRI} basée sur la prédiction des intérêts des utilisateurs, à partir de \textsc{Dbpedia}, selon leurs situations courantes. Cette approche applique la technique de la classification associative dans le but de l'enrichissement des requêtes des utilisateurs.
D'autre part, nous introduisons une approche de découverte de communautés appelée \textbf{Foaf-A-Walk} combinant la technique de la marche aléatoire et la modélisation de l'ontologie \textsc{Foaf}, dans le but de la recommandation d'amis.

\textbf{Mots-clés:}
Dbpedia, Situation-courante, classification associative, Foaf, marche aléatoire.

\textbf{Abstract:}
The emergence of smartphones has given mobile computing access to everyday reality.
More specifically, the context modeling offers users an effective way to customize search results and even the recommended elements by limiting the data space. Moreover, in recent years, many social sites have embraced the notion of context in their recommendations. Indeed, with the availability of mobile devices, these new mobile sites have the advantage of providing users with more relevant elements based on their current situations. Thus, we introduce a new approach of contextual IR in a mobile environment. We offer a hand, an approach called \textbf{SA-IRI} based on the prediction of users' interests, from \textsc{DBpedia}, given their current situations. This approach applies the technique of associative classification in order to enrich the users' queries. Secondly, we introduce an approach of communities discovering, called \textbf{Foaf-A-Walk}, combining the random walk technique and the \textsc{Foaf} modeling, for friend recommendation.

\textbf{Keywords:}
Dbpedia, Situation-Aware, associative classification, Foaf, random walk.
\newpage 
\tableofcontents

\listoffigures %
\listoftables

\mainmatter%
%%%%%%%%%%%%%%%%%%%%%%%%%%%%%%%%%%%%%%%%%%%%%%%%%%%%%%%%%%%%%%%%%%%%%%%%

% Intro, chapitres et conclusion
%%%%%%%%%%%%%%%%%%%%%%%%%%%%%%%%%%%%%%%%%%%%%%%%%%%%%%%%%%%%%%%%%%%%%%
%   Fichier :   Introduction.tex
%%%%%%%%%%%%%%%%%%%%%%%%%%%%%%%%%%%%%%%%%%%%%%%%%%%%%%%%%%%%%%%%%%%%%%
\Introduction{Introduction g\'{e}n\'{e}rale} \label{introduction} \vspace{3cm}
\markboth{INTRODUCTION ET MOTIVATIONS}{Introduction et Motivations}

\begin{quotation}
\begin{flushright}
\textit{``Il y a quelque chose de pire dans la vie que de n'avoir pas réussi,\\
c'est de n'avoir pas essayé.''\\
 \textbf{Franklin Roosevelt}}
\end{flushright}
\end{quotation}

\bigskip

\bigskip

%\subsection*{\textsc{Contexte et probl\'{e}matique}}

\PARstart{L}{es} êtres humains ont toujours bien réussi à transmettre les idées entre eux et à réagir de façon appropriée grâce à plusieurs facteurs, \emph{e.g.}, la richesse du langage humain, le sens commun des mots utilisés, la compréhension de l'intention de l'autre via sa situation, etc. Ainsi, dans le cas des discussions entre humains, ces derniers ont tendance à utiliser des informations ayant des relations implicites avec leurs situations. Malheureusement, ce type d'information n'est pas transmis lors de l'intéraction avec les machines. Plus précisément, les ordinateurs n'ont pas de visibilité sur les besoins des utilisateurs, ni sur leurs préférences et leurs intentions de recherche. Par conséquent, les ordinateurs ne sont pas capables de tirer profit du contexte caractérisant l'intéraction de type humain/machine et permettant de produire des services plus utiles.

Une première tentative, visant à améliorer ce type d'intéraction, a exploré deux problématiques: comment enrichir le langage d'intéraction des utilisateurs avec les ordinateurs? et comment augmenter la quantité d'information contextuelle mise à la disposition de l'ordinateur?
Ainsi, cette catégorie d'approches a donné aux utilisateurs la possibilité de communiquer plus naturellement avec les machines dans le but d'améliorer l'intéraction de type utilisateur/ordinateur.
Ce type de communication reste explicite puisque l'ordinateur n'arrive à savoir que les données introduites explicitement par l'utilisateur. Par ailleurs, avec les intéractions de type utilisateur/utilisateur, les informations contextuelles sont indispensables pour comprendre leurs besoins, \emph{e.g.}, les expressions du visage, les émotions, les événements passés et futurs, les personnes proches, les relations sociales, etc. Le cumul de cette compréhension partagée, s'appelle "grounding" \cite{Clark1991}. Ainsi, le besoin d'exprimer explicitement ces informations s'impose obligatoirement vu les limites des machines précédemment mentionnées.
Ces études confirment le fait que la pertinence des résultats de la recherche ou même des ressources recommandées aux utilisateurs ne dépend pas seulement des documents et des requêtes, mais aussi du contexte courant de l'utilisateur.

En outre, l'objectif principal des travaux basés sur le contexte des utilisateurs est d'améliorer la pertinence des résultats en proposant une définition du contexte et en précisant la manière de son exploitation et de son intégration dans le processus de sélection de l'information que ce soit dans le domaine de la recherche d'information (RI) \cite{Daoud2009} \cite{Kraft05} \cite{Liu2004} \cite{White09} ou dans le cadre des systèmes de recommandation \cite{Hariri2011} \cite{Ma2011}.

De nos jours, les appareils mobiles ont évolué afin de pallier aux limites des ordinateurs de bureau et offrir les meilleurs services et fonctionnalités aux utilisateurs (des écrans plus grands, une puissance de traitement accrue et une connexion Internet à large bande). Cette évolution a donné, même aux utilisateurs nomades, la possibilité d'accéder à l'information à partir de n'importe quel emplacement et à tout moment.
De la même manière que les environnements classiques, l'environnement mobile souffre d'un certain nombre de limites. En effet, ce type d'environnement est caractérisé par un ensemble de contraintes à savoir: la zone d'affichage limitée et les difficultés de saisie des requêtes, etc. En effet, vu les spécificités du clavier des appareils mobiles et l'utilisation de la technique "multitap", les utilisateurs ont besoin d'une moyenne de 40.9 touches par requête \cite{Kamvar2006}.
Ces mêmes auteurs ont effectué une étude sur les logs des requêtes des mobinautes \cite{Kamvar2007}, qui a montré que les requêtes des utilisateurs mobiles sont plus courtes et plus ambig\"{u}es avec une moyenne de 1.7 mots et 16.8 caractères par requête. De plus, ces études ont prouvé qu'un grand nombre d'utilisateurs ne consultent que la première page parmi l'ensemble des pages qu'ils ont re\c{c}u comme résultat suite à leurs requêtes.

Dans ce cadre de travail mobile, plusieurs approches ont tenté d'améliorer la précision des systèmes de RI contextuelle. Plus précisement, ils se basent sur la modélisation et l'intégration du contexte des utilisateurs dans le processus de l'extraction de l'information.
La modélisation du contexte comprend le choix des informations contextuelles à utiliser pour définir le contexte de l'utilisateur, ainsi que la technique appliquée pour capter ces informations et les mettre à jours au cours du temps (\emph{i.e.}, changement de la description de l'environnement de l'utilisateur). Ensuite, l'étape de l'intégration du contexte dans le processus de la RI influe directement la précision des systèmes à l'égard des besoins des utilisateurs (le choix des dimensions du contexte et de l'étape au niveau de laquelle ce contexte est incorpéré).

L'émergence de la RI contextuelle a imposé la définition d'un nouveau cadre d'évaluation pour pallier aux limites du cadre ancien. Plus particulièrement, les collections de test des systèmes d'évaluation classiques de la RI, \emph{e.g.}, TREC, CLEF, etc. ne contiennent aucune information con\c{c}ernant le contexte des utilisateurs et se limitent à leurs requêtes et à une collection de documents pour tester la performance des systèmes de RI.

Par ailleurs, les réseux sociaux se sont multipliés en terme de taille et de nombre de fonctionnalités mise à la disposition des utilisateurs.
Comme résultat immédiat, les systèmes de recommandation sociaux se sont imposés dans le paysage du web social, apportant un support d'extension et de renforcement des relations sociales.
Dans ce cadre, un réseau social est une commmunauté d'individus connectés à l'aide d'une infinité de relations, \emph{e.g.}, intérêts, amitiés, localisations, emplois, etc. Ces réseaux sont utilisés dans plusieurs systèmes de recommandation afin de personnaliser leurs ressources proposées aux différents utilisateurs, \emph{e.g.}, articles, produits commer\c{c}iaux, films, extraits musicaux, amis, etc.
En outre, les relations qui existent dans les réseaux sociaux sont incomplètes et seule une partie du cercle social réel de l'utilisateur est modélisée par le biais des connexions virtuelles. Par conséquent, dans le but de rassembler tels utilisateurs, la définition d'un système de recommandation d'amis devient primordiale.
Ainsi, vu les spécificités de l'environnement mobile, plusieurs contraintes s'ajoutent dans la réalisation d'un tel système: la dynamique des relations sociales entre les utilisateurs (les systèmes de recommandation sont fortement concernés par l'instabilité des cercles sociaux et de la description des utilisateurs); et les limites de spécification des relations sociales entre les utilisateurs (dans les cas réels, la plupart des relations des réseaux sociaux sont incomplètes).

La majorité des approches existantes, qui ont proposé des systèmes de recommandation, se sont concentrées sur le but de recommander aux utilisateurs des ressources basées sur leurs contextes mobiles (localisation, temps, etc.) \cite{Zheng2010}, ou bien sur leurs contextes sociaux (intérêts, amis, etc.) \cite{Xin2009}. Et peu d'approches ont trouvé une manière d'en profiter de la description sociale des utilisateurs et de leurs localisations pour pouvoir sélectionner les nouvelles personnes pouvant être des futurs amis avec eux \cite{beach09} \cite{Qiao2011}.

%Dans ce contexte plusieurs questions restent en suspens: sur quelle page web me rendre pour avoir les réponse à mes interrogations? Quels articles pourrais-je lire pour améliorer mes connaissances dans un domaine particulier? Quels films ou musique que je peut découvrir? Ainsi, dans une infinité de domaines, le foisonnement de ressources rend difficiles leur mainipulation et exploitation.
%L'exemple de Google peut illustrer cet explosion de web. En effet, en 2005, Google a indiqué sur sa page d'acceuil de son moteur de recherche que le nombre de page qu'il référen\c{c}ait était de huit milliards. Dans trois ans, et depuis 2008, ce chiffre s'est multiplié pour atteindre plus d'un billion de page.
%De plus, le site Wikipédia possède plus de trois millions d'articles en anglais seulement.

\subsection*{\textsc{Contributions}}

Le travail proposé dans ce mémoire se situe dans le cadre de l'exploitation du contexte de l'utilisateur dans un environnement mobile. Plus précisément, ce contexte est incorporé dans l'enrichissement de ses requêtes et l'extension de son cercle social. Ainsi, nos contributions portent sur les quatre volets suivants:
\begin{enumerate}

  \item \textbf{Définition et modélisation du contexte}:
  dans le cadre de ce mémoire, le contexte de l'utilisateur est défini par sa situation relative à son environnement mobile. Ainsi, pour chaque dimension contribuant à la définition de cette situation, nous donnons nos motivations expliquant le choix d'une telle donnée contextuelle. Par ailleurs, les centres d'intérêt des utilisateurs jouent un rôle indispensable dans la disambig\"{u}isation de leurs requêtes et varient avec le changement de leurs situations. Le défi principal de cette contibution est de choisir les dimensions à concaténer dans la définition de la situation et de trouver la source de connaissance à utiliser dans l'extraction des intérêts des utilisateurs.

  \item \textbf{Exploitation de la situation dans le domaine de la RI}:
  la situation de l'utilisateur est exploitée dans l'enrichissement de ses requêtes. En effet, la technique de \emph{classification associative} est appliquée pour extraire la situation la plus similaire à celle de l'utilisateur dans le but d'identifier son intérêt derrière la requête qu'il a soumise.

  \item \textbf{Exploitation des données contextuelles dans le cadre de la recommandation d'amis}:
  dans un souci de précision de la recommandation, il est important de choisir attentivement les données à intégrer dans ce processus parmi celles disponibles dans les réseaux sociaux.
  Ainsi, l'ensemble des données contextuelles (\emph{i.e.}, localisation de l'utilisateur) et sociales (\emph{i.e.}, amis et intérêts de l'utilisateur) est utilisé dans le but de l'extension de son cercle social. En effet, nous appliquons la technique de la \emph{marche aléatoire} pour la génération des communautés utilisées dans le processus de la recommandation d'amis.

  \item \textbf{Proposition d'un cadre d'évaluation pour la RI contextuelle}:
  Vu l'abssence d'un cadre d'évaluation pour la RI contextuelle, nous avons proposé dans ce mémoire un nouveau cadre d'évaluation visant à évaluer notre travail. Ainsi, nous détaillons une approche d'évaluation basée sur une étude journalière. Durant cette étude, nous avons collecté un ensemble de requêtes utilisateurs associée chacune à une situation mobile. La pertinence des résultats obtenus avec notre approche est jugée par le même ensemble d'utilisateurs.

\end{enumerate}

\subsection*{\textsc{Organisation du Mémoire}}

Les résultats de nos travaux de recherche sont synthétisés dans ce mémoire composé de quatre chapitres:

\subsubsection*{\textit{Chapitre 1: }}
Ce chapitre introduit les principales notions de la RI contextuelle, dans le cadre d'un environnement mobile, indispensables pour la compréhension des travaux présentés dans ce mémoire. Ainsi, nous commen\c{c}ons par présenter la RI en générale. Nous nous focalisons par la suite sur la RI contextuelle dans un environnement mobile. Enfin, nous décrivons brièvement quelques notions liées à la recommandation.

\subsubsection*{\textit{Chapitre 2: }}
Ce chapitre présente une revue critique des différentes approches ayant exploitées le contexte des utilisateurs dans leurs travaux. Nous consacrons une première partie à l'étude des approches qui ont utilisé la notion du contexte dans la personnalisation de la RI. Ensuite, nous passons à la comparaison des travaux qui ont introduit le contexte des utilisateurs dans la personnalisation de leurs systèmes de recommandation. A la fin de chaque partie, nous avons réalisé une étude afin d'identifier les limites de chacune de ces approches et de motiver l'idée que nous avons proposée.

\subsubsection*{\textit{Chapitre 3: }}
Dans ce troisième chapitre, nous proposons notre approche pour l'exploitation du contexte des utilisateurs dans l'enrichissement de leurs requêtes et l'extension de leurs cercles sociaux. En effet, comme il est déjà clair, notre contribution porte sur deux volets. La première partie se base sur la prédiction des intérêts des utilisateurs à partir de leurs situations dans le but de l'enrichissement de leurs requêtes mobiles. Ainsi que la deuxième partie s'articule sur la découverte dynamique des communautés dans la recommandation d'amis aux utilisateurs. Chaque partie de notre approche est suivie d'un exemple illustratif détaillant ses différentes étapes.
Nous représentons, à la fin de ce chapitre, l'intéraction entre ces deux parties.

\subsubsection*{\textit{Chapitre 4: }}
Ce chapitre est consacré à une étude expérimentale menée sur différentes bases de test. Ainsi, nous effectuons une comparaison de notre approche d'enrichissement de requêtes avec le moteur de recherche Google vu la précision des intérêts prédits et des ressources retournées comme résultat. Nous commen\c{c}ons par appliquer cette approche de validation sur les données que nous avons collectées à partir de notre étude journalière. Les mêmes tests sont répétés sur la base obtenue du challenge Quaero pour s'assurer de la précision de notre approche de RI contextuelle. Par ailleurs, afin de valider la qualité de notre approche de recommandation d'amis, nous avons appliqué notre travail sur la base \textsc{FOAF} que nous avons pu collecter du web.
Une comparaison de la portée de notre approche avec les travaux de l'état de l'art est présentée. Cette comparaison est effectuée sur deux volets : la prédiction des intérêts et la découverte des communautés.

\bigskip

Enfin, nous conclurons ce mémoire par un résumé des différents travaux présentés et quelques perspectives de nos futures recherches.

%%%%%%%%%%%%%%%%%%%%%%%%%%%%%%%%%%%%%%%%%%%%%%%%%%%%%%%%%%%%%%%%%%%%%%
%   Fichier : ch1.tex
%%%%%%%%%%%%%%%%%%%%%%%%%%%%%%%%%%%%%%%%%%%%%%%%%%%%%%%%%%%%%%%%%%%%%%
\chapter{Recherche d'information \& Systèmes de recommandation: Notions de base}
\markboth{Recherche d'information \& Systèmes de recommandation}{Notions de base}
\label{ch1} \vspace*{3cm}
\section{Introduction}
L'émergence des smartphones; les téléphones mobiles capables de fournir des fonctionnalités associées à des assistants numériques personnels (PDA) ou même à des ordinateurs personnels (PC); a donné à l'informatique mobile l'accès à la réalité quotidienne. Ainsi, l'informatique mobile a émergé comme un nouveau paradigme de l'informatique personnelle et des systèmes de communication.
Par ailleurs, pour répondre aux questions liées à l'informatique mobile, la modélisation du contexte de l'utilisateur, \emph{i.e.}, localisation, environnement, objectif, etc., offre un moyen efficace pour adapter les résultats de recherche ou les éléments recommandés à ce dernier. Plus précisement, l'intégration du contexte de l'utilisateur permet de limiter l'espace de données et par la suite de fournir des résultats plus conviviaux aux utilisateurs.

D'ailleurs, ces dernières années, beaucoup de sites sociaux ont intégré la notion du contexte dans leurs recommandations. En effet, avec la disponibilité des appareils mobiles, la majorité des utilisateurs arrivent à déterminer l'emplacement de leurs amis ou leurs centres d'intérêts. Par conséquent, ces nouveaux sites mobiles ont l'avantage de fournir aux utilisateurs des éléments plus pertinents vu leurs contextes courants.

Dans ce chapitre, nous introduisons les principales notions de base liées aux domaines de la RI et des systèmes de recommandation.
Ainsi, nous commençons par définir les différents concepts permettant de donner les spécificités d'un système de recherche d'information (SRI) et donnant naissance aux travaux de personnalisation de tels systèmes. Plus particulièrement, nous gardons notre concentration sur la personnalisation de la RI basée sur le contexte des utilisateurs dans un environnement mobile.
Dans le reste de ce chapitre, nous donnons une idée sur les systèmes de recommandation ainsi que leur définition en mettant l'accent sur l'avantage de leur personnalisation face à leur utilisation dans la vie quotidienne.

\section{Recherche d'information}
La RI est un domaine lié à la science de l'information et à la bibliothéconomie qui ont toujours eu des problèmes dans la présentation des documents. Des outils ont été développés, avec l'apparence de l'informatique, permettant le traitement des informations et donnant une représentation des documents au moment de leurs indéxations.

\subsection{Définitions}
Parmi les définitions visant à expliquer et spécifier le sens de la RI, nous citons:
\begin{definition}
La recherche de l'information est une branche de l'informatique qui s'intéresse à l'acquisition, l'organisation, le stockage, la recherche et la sélection d'information \cite{Boubekeur2008}.
\end{definition}

\begin{definition}
La recherche de l'information est une discipline de recherche qui intègre des modèles et des techniques dont le but est de faciliter l'accès à l'information pertinente pour un utilisateur ayant un besoin en information \cite{Daoud2009}.
\end{definition}

Ces définitions partagent l'idée que la RI a comme principal rôle l'extraction des informations pertinentes reflétant un besoin en information, parmi un ensemble de documents.

\subsection{Concepts et principes de base de la RI}
La RI est considérée comme l'ensemble des outils permettant d'extraire, à partir d'une collection de documents, ceux considérés pertinents et qui répondent aux besoins des utilisateurs. Dans ce contexte, la définition de quelques concepts de base devient primordiale.
\begin{itemize}
  \item Collection de documents: la collection de documents constitue l'ensemble des informations accessibles par une entité, \emph{i.e.}, machine, utilisateur, etc. Cette représentation est choisie de façon à rendre l'interrogation et la modification de la base des documents faciles et rapides.

  \item Besoin en information: la notion de besoin en information dans le contexte de la RI constitue le besoin des utilisateurs. Cette notion a été catégorisée par \cite{Ingwersen1994} en trois types:
      \begin{enumerate}
        \item \textit{Besoin vérificatif}: dans ce cas, l'utilisateur cherche à vérifier le texte ou la description des données qu'il possède. Ce type de besoin est stable car il ne risque pas de changer au cours du temps. Un exemple connu de ce besoin est la recherche d'une entrée bibtex d'un article sur internet, dont l'utilisateur connaît déjà le titre ou le nom de son auteur.
        \item \textit{Besoin thématique connu}: dans une telle situation, l'utilisateur cherche à avoir plus d'information dans un domaine donné. Ce type de besoin peut varier au cours du temps. En particulier, les besoins de l'utilisateur peuvent se raffiner ou s'enrichir au cours de la recherche.
        \item \textit{Besoin thématique inconnu}: dans ce cadre de besoin, l'utilisateur cherche de nouvelles informations liées à un domaine non familier. Ce besoin est essentiellement variable car les besoins des utilisateurs sont souvent incomplets vu leur ignorance du domaine de recherche.
      \end{enumerate}

  \item Requête: la requête de l'utilisateur est un ensemble de mots clés permettant d'exprimer son besoin en information en jouant le rôle d'une interface entre ce dernier et le SRI.

  \item Modèle de représentation: un modèle de représentation est une modélisation possible d'un document ou d'une requête, conçu d'une façon à couvrir au mieux le contenu sémantique de ces derniers. Ce processus est appelé \textit{indéxation}, ayant comme résultat des groupes de termes (ou de concepts) de poids différents et rangés dans des structures appelées \textit{dictionnaires}, qui constituent les langages d'indéxation.

  \item Modèle de recherche: ce modèle représente le noyau des SRI. Il permet de faire correspondre un ensemble de documents pertinents à chaque requête utilisateur.
\end{itemize}

\subsection{Modèles de recherche d'information}
Un modèle de la RI est un formalisme permettant de donner une représentation théorique du processus de la RI. Plusieurs modèles ont été cités dans la littérature. Selon \cite{Baeza1999}, ce modèle est décrit par le quadruplet (D,Q,F,R(q,d)), où:
\begin{itemize}
  \item D est l'ensemble des documents
  \item Q est l'ensemble des requêtes
  \item F est le modèle théorique de représentation des requêtes et des documents
  \item R(q,d) est la fonction de pertinence associant le document d à la requête q.
\end{itemize}
Les modèles de la RI peuvent être classés en:
\begin{itemize}
  \item Modèle booléen: Le modèle booléen \cite{Salton1986} est basé sur la théorie des ensembles. Dans ce type de modèles, les documents sont représentés chacun par une conjonction de termes de la forme: $d=t_{1}\wedge t_{2}\wedge ... \wedge t_{n}$. Ainsi, les requêtes sont représentées chacune par des expressions booléennes reliées par des opérateurs logiques (AND, OR ou NOT). La fonction de pertinence R(q,d) est définie pour indiquer la présence ou non des termes de la requête q dans le document d. De ce fait, le résultat de cette fonction est défini par: $$RSV(q,d)=\{0,1\}$$
  \item Modèle vectoriel: Dans ce modèle, la pertinence d'un document d par rapport à une requête q est définie par une distance vectorielle \cite{Salton1986}. De même, les requêtes et les documents sont représentés par des vecteurs à n dimensions. La fonction de pertinence RSV(q,d) est définie par: $$RSV(q,d)=\cos(\overrightarrow{q},\overrightarrow{d})$$
  \item Modèle probabiliste: Dans ce modèle, la pertinence d'un document par rapport à une requête est donnée par un calcul de probabilité \cite{Roberston1976, Salton1983}. Le principe de base de ce modèle est de trouver les documents qui ont une forte probabilité à être pertinents, et en même temps une faible probabilité à être non pertinents. La fonction de pertinence RSV(q,d) est donnée par la formule suivante: $$RSV(q,d)=\frac{P(d/q)}{P(\overline{d}/q)}=\sum_{i=1}^n\log \frac{P(1-q)}{P(1-p)}$$ où: p=P(terme $t_{i}$ present/ d pertinent), q=P(terme $t_{i}$ present/ d non pertinent) et n: le nombre de termes dans la requête.
\end{itemize}

\subsection{Systèmes de recherche d'information}
\subsubsection{Définitions}
Plusieurs définitions ont été données pour préciser le sens et le rôle des SRI. Parmi ces définitions nous pouvons citer:
\begin{definition}
Un système de recherche d'information est défini selon, \cite{Smeaton1990}, par: Le but d'un système de recherche d'information est de retrouver des documents en réponse à une requête des usagers, de manière à ce que les contenus des documents soient pertinents au besoin initial d'information de l'usager.
\end{definition}

\begin{definition}
Selon \cite{Tambellini2007}, un système de recherche d'information est défini par un langage de représentation des documents (qui peut s'appliquer à différents corpus de documents) et des requêtes qui expriment un besoin de l'utilisateur (sous forme de mots clés par exemple), et une fonction de mise en correspendance du besoin de l'utilisateur et du corpus de documents en vue de fournir comme résultats des documents pertinents pour l'utilisateur, c'est-à-dire répondant à son besoin d'information.
\end{definition}

\subsubsection{Personnalisation des systèmes de recherche d'information}
Avec l'augmentation de la quantité d'information mise à la disposition de l'utilisateur, l'adaptation des données à ses besoins devient primordiale.
La personnalisation est une réponse efficace au problème de la surcharge d'informations, d'une part, en injectant dans les requêtes des critères de filtrage plus  restrictifs, et, d'autre part, en reformulant éventuellement les requêtes pour mieux tenir compte des centres d'intérêt et des préférences de  l'utilisateur.
Les SRI les plus utilisés sont basés principalement  sur : le ré-ordonnancement des éléments du résultat, le filtrage du résultat, la recommandation des documents composants le résultat et, finalement, l'enrichissement de la requête avec le profil utilisateur.
\begin{itemize}
  \item Le ré-ordonnancement des éléments du résultat: Le principe du ré-ordonnancement est de modifier l'ordre de l'affichage des résultats retournés aux utilisateurs. Il s'agit d'un post traitement qui, étant donné les éléments retournés par une requête, essaie de trouver une manière d'échanger leurs emplacements en fonction des préférences des utilisateurs. L'échange de l'ordre d'apparition des éléments du résultat est fait généralement en appliquant une fonction qui permet de calculer le nouveau rang de ces éléments.
      Un exemple de SRI basés sur le ré-ordonnancement des résultats est présenté dans \cite{Pretschner1999}. Dans ce cas, le rang du document $d_{i}$ est calculé par la formule \ref{reordonnancement}
      \begin{center}
      \begin{equation}
        $$$\rho(d_{i})$=$\omega(d_{i})$*[0.5+1/4 $\sum_{i=1}^4 $($\pi(c_{i})$*$\gamma(c_{i},d_{i})$)]$$
        \label{reordonnancement}
      \end{equation}
 \end{center}

      où : $\omega(d_{i})$ est l'ancienne position du document $d_{i}$, $\gamma(c_{i},d_{i})$ est la fréquence d'apparition du terme $c_{i}$ dans le document $d_{i}$.
Cette technique de personnalisation est peu utilisée dans la pratique, puisque le calcul du rang doit être appliqué sur chaque élément du résultat, ce qui demande beaucoup de temps de calcul.

  \item Le filtrage du résultat: Cette personnalisation consiste à exécuter la requête sans prendre en considération la personnalisation et, d'appliquer ensuite un post traitement sur le résultat, en éliminant les éléments non pertinents aux utilisateurs. Le filtrage se fait soit en appliquant des requêtes sur le résultat, soit en traitant chaque élément séparément pour étudier sa pertinence. Une représentation des systèmes de filtrage est donnée par la figure \ref{FiltringResults}.

\begin{figure*}[htbp]
\centering
\epsfig{file=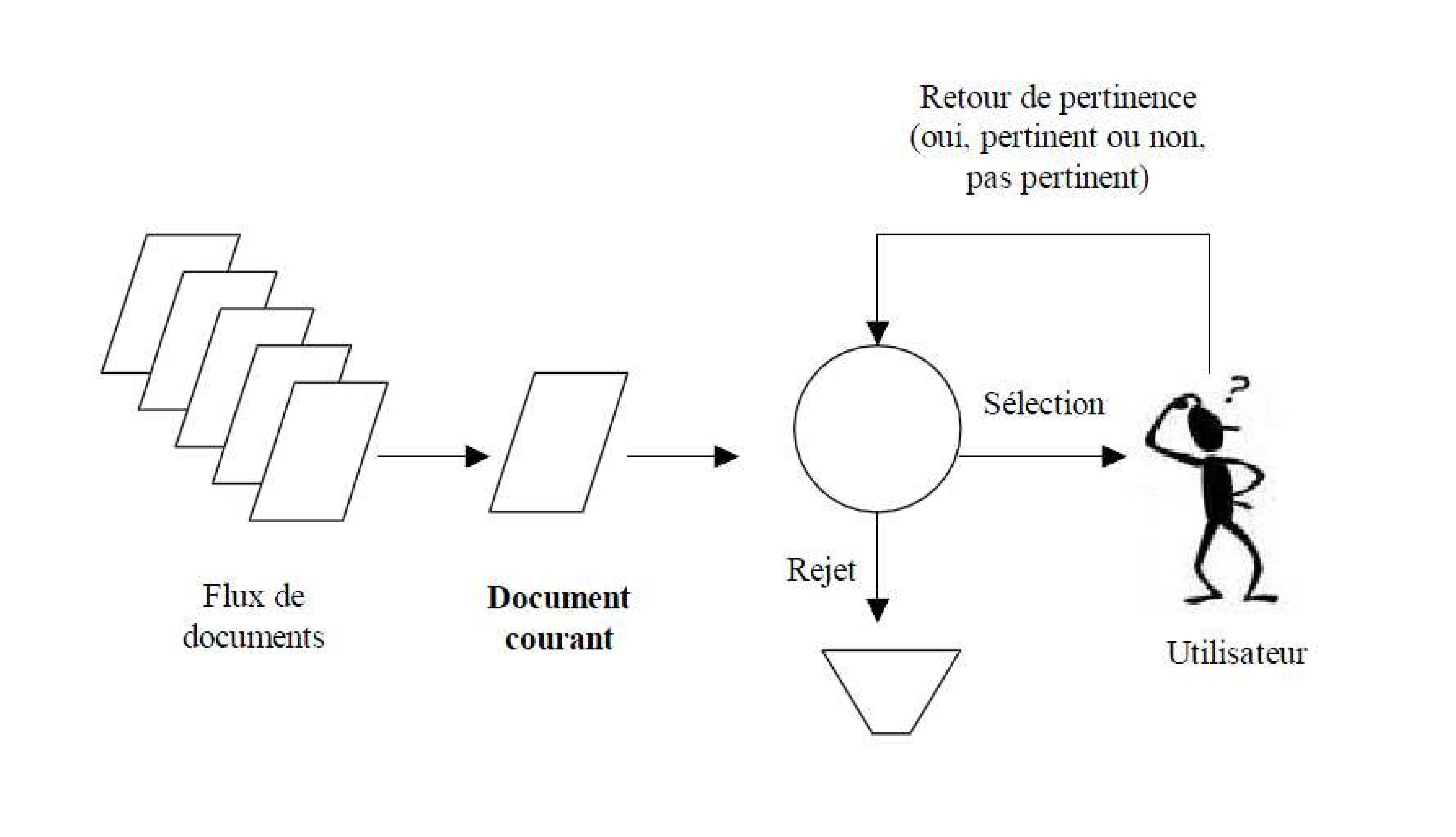, height=3in, width=5.5in}
\caption{Représentation d'un système de filtrage}
\label{FiltringResults}
\end{figure*}

Un système nommé Casper est présenté dans \cite{Bradley2000}. Ce système applique le principe de filtrage dans le but de la recherche des annonces de travail. Ainsi, la requête de l'utilisateur est exécutée en deux étapes:
\begin{enumerate}
  \item Extraction des annonces sur le serveur selon une similarité entre les termes contenus dans chaque annonce et non pas avec un matching exact
  \item Filtrage des annonces retournées par le serveur.
\end{enumerate}
L'avantage de ce type de personnalisation est résumé dans le fait qu'elle ne nécessite aucune modification sur le fonctionnement du moteur de recherche. Cependant, son inconvénient majeur se situe dans le volume des données échangées entre le serveur et le client, ce qui augmente le risque de perte des données pertinentes.

  \item La recommandation des éléments du résultat: ce type de personnalisation consiste à proposer à l'utilisateur des ressources selon ses préférences.
  Une présentation générale des systèmes de recommandation est donnée par \cite{Trousse1999} dans la figure \ref{RecommandationSystems}.
  \begin{figure*}[htbp]
\centering
\epsfig{file=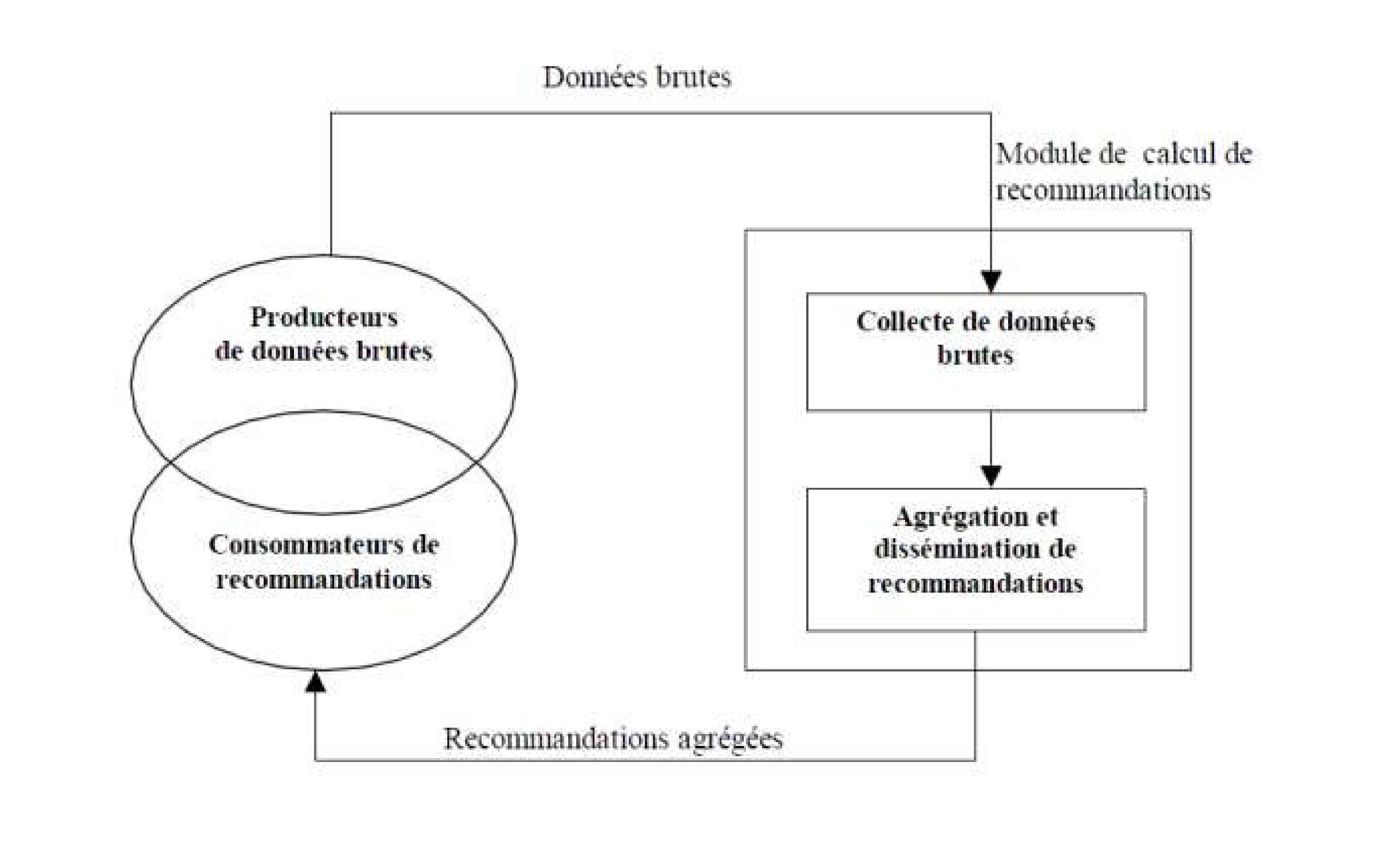, height=3in, width=5.5in}
\caption{Représentation d'un système de recommandation}
\label{RecommandationSystems}
\end{figure*}

  Un exemple de ces systèmes est détaillé dans \cite{Mobasher2000}, où la recommandation n'est faite que par rapport aux pages web précédemment visitées par l'utilisateur, indépendamment des autres utilisateurs. Ce qui suppose que le système a déjà rassemblé suffisamment d'informations sur l'utilisateur.

La recommandation des éléments de réponse est la technique de personnalisation la plus utilisée en pratique puisqu'elle permet de fournir des résultats aux utilisateurs sans limiter leurs choix. Nous avons choisi d'utiliser ce type de personnalisation dans notre approche permettant d'étendre les cercles sociaux des utilisateurs.

  \item Enrichissement des requêtes utilisateur: cette technique de personnalisation consiste à appliquer un prétraitement sur la requête émise par l'utilisateur afin de l'enrichir avec son profil (ses préférences et ses intérêts) ou bien par son contexte (sa localisation, le moment du lancement de la requête, etc.). L'avantage de ce type de personnalisation réside dans le fait qu'elle utilise des techniques qui permettent de réduire la taille du domaine de recherche, en effectuant le minimum de calculs possibles \cite{Latiri2003} \cite{Trabelsi2011} \cite{Trabelsi11}. Nous utiliserons ce type de personnalisation dans notre approche de RI contextuelle, que nous allons proposer dans la suite de ce mémoire.
      Plusieurs définitions ont été données dans la littérature pour définir la technique d'enrichissement de requêtes, nous citons quelques unes dans la suite:
      \begin{definition}
      Selon \cite{Salton1986} l'enrichissement de requêtes est défini comme un processus qui vise à rendre les résultats plus clairs et plus précis en permettant à l'utilisateur de modifier sa requête pour améliorer la pertinence de ses résultats.
      \end{definition}
      \begin{definition}
      \cite{Efthimiadis1996} a également donné la définition suivante: l'enrichissement de requêtes ou l'enrichissement de termes est un processus qui vise à compléter la requête en proposant des termes supplémentaires, et est considéré comme une amélioration de la recherche d'information.
      \end{definition}
\end{itemize}

\section{Recherche d'information contextuelle}
Dans différents domaines, on entend toujours parler du mot contexte. Originellement, la notion du contexte présente le non dit nécessaire pour comprendre un texte donné. En effet, cette notion est de plus en plus utilisée. Selon les études de \cite{Wanner2007}, entre les années 1997 et 2006, le pourcentage des pages web contenant le mot contexte a passé de 5\% à 15\%. Cette croissance s'est multipliée en exponentielle aujourd'hui.

Cependant, cette notion reste très difficile à définir à cause de sa fréquence d'utilisation. En outre, elle devient plus claire quand une tâche possède plusieurs manières d'exécution et d'exploitation. Dans une telle situation, chaque personne choisit en fonction de ses préférences et connaissances la manière qui lui convient en plus pour définir son contexte.

Dans ce qui suit, nous allons donner les principales définitions de la notion du contexte qui existent dans la littérature.

\subsection{Définition du contexte}
Plusieurs définitions du contexte ont été proposées dans la littérature de la RI contextuelle, ces définitions diffèrent essentiellement au niveau des éléments permettant de le construire. Les facteurs contextuels les plus exploités sont les centres d'intérêts de l'utilisateur, la tâche de recherche en cours, les préférences de recherche liées à la qualité de l'information, le temps de soumission de la requête, la localisation géographique de l'utilisateur, l'environnement de recherche, etc. Ainsi, selon Brézillon en \cite{Brezillon2006}, Il n'y a pas de contexte sans contexte, \emph{i.e.}, le contexte n'existe pas en tant que tel, et ne peut être utilisé et défini que dans un cadre bien précis.
\begin{definition}
Le contexte est un ensemble d'informations. Cet ensemble est structuré, il est partagé, il évolue et sert l'interprétation \cite{Winograd2001}.
\end{definition}
\begin{definition}
Un contexte multidimensionnel a également été défini par \cite{Kostadinov2003}. Cette définition ajoute des nouvelles caractéristiques liées à l'aspect temporel du besoin en information et au type de recherche demandé. Les trois principales dimensions retenues pour ce type de contexte sont : sociale, applicationnelle et temporelle.
\begin{itemize}
  \item La dimension sociale définit l'appartenance de l'utilisateur : individuel, groupe ou communauté
  \item La dimension applicationnelle définit le but de la tâche accomplie : recherche ad-hoc, résolution d'un problème, etc.
  \item La dimension temporelle permet de définir le contexte temporel du besoin : temps passé, intention à court terme ou intention à long terme.
\end{itemize}
\end{definition}

\subsection{Recherche d'information contextuelle dans un environnement mobile}
Avec l'évolution considérable des appareils mobiles, des navigateurs web et des packs proposés par les fournisseurs d'accès Internet, le web mobile est apparu. En effet, selon des statistiques de la \emph{DGCIS} (Direction Générale de la Compétitivité de l'Industrie et des Services), le nombre d'internautes mobiles a passé de 13.7 millions en 2010 pour atteindre 18.3 millions en 2011, \emph{i.e.}, une augmentation de 34\%.
De même, \emph{Orange} a fait une étude qui montre que le nombre d'Iphones en France en 2006 a dépassé 1.3 million (5\% de la population Française) et 28\% de mobinautes (soit 8.3 millions de Français) se connectent à Internet mobile, via un navigateur ou une application, au moins une fois par mois.

\begin{figure*}[htbp]
\centering
\epsfig{file=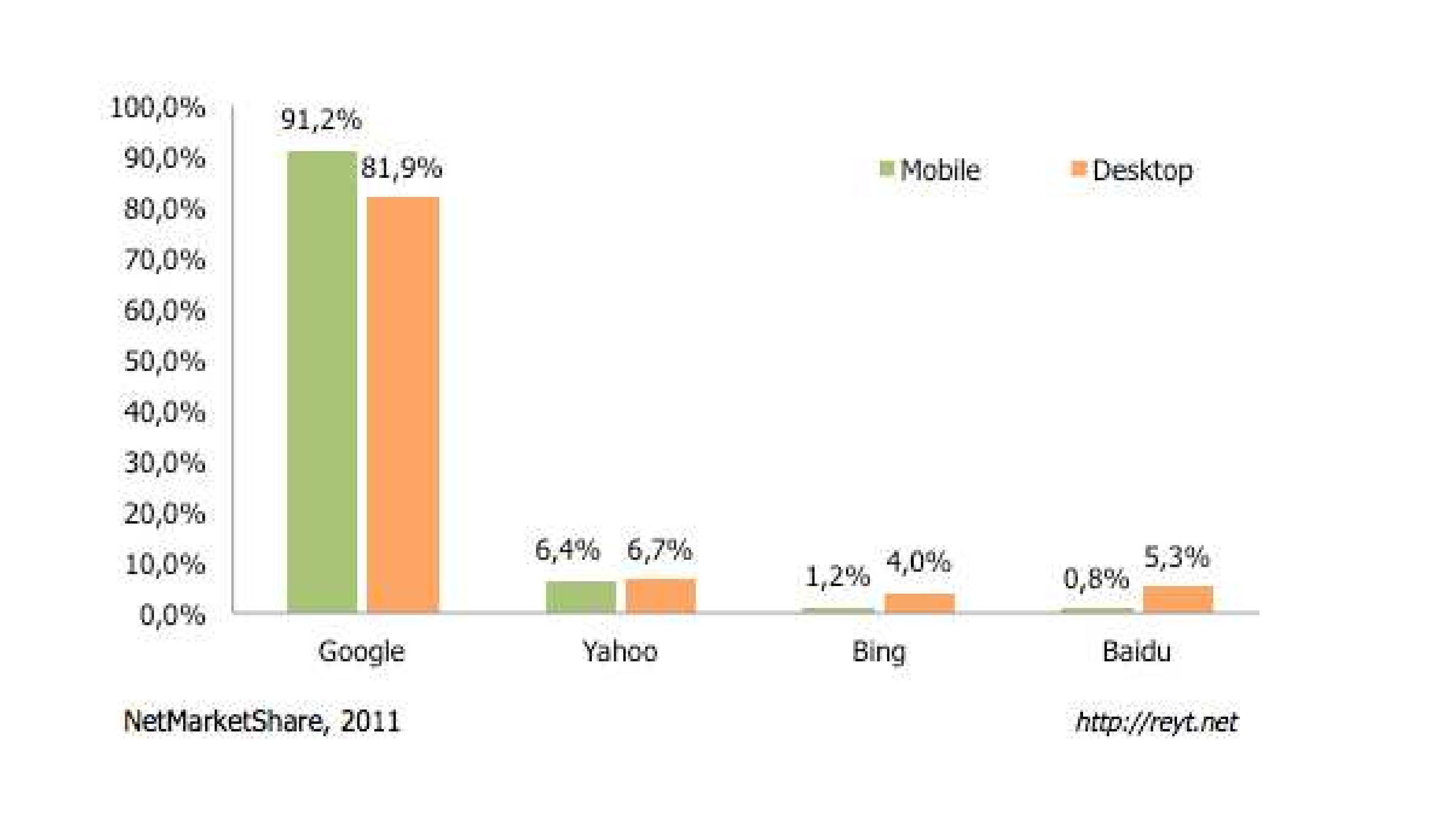, height=3in, width=5in}
\caption{Parts des moteurs de recherche mobiles dans le marché}
\label{marche_mobiles}
\end{figure*}

Par conséquent, le besoin des outils de recherche d'information adaptés aux appareils mobiles ne cesse de cro\^{\i}tre depuis ces dernières années. Afin de répondre à ces demandes, les moteurs de recherche mobile sont apparus pour consolider la position des internautes sur ces appareils nomades. Entre novembre 2010 et septembre 2011, la part de marché des requêtes mobiles sur les recherches globales a doublé en passant de 2.95\% à 5.98\%, selon les chiffres de NetMarketShare \footnote{~\url{http://netmarketshare.com}}. Selon ce site, Google \footnote{~\url{https://google.com}} est indéniablement le moteur de recherche le plus utilisé sur mobile dans le monde. 91.2\% des requêtes envoyées depuis un appareil nomade, \emph{i.e.}, tablette ou smartphone, ont été traitées par Google en septembre 2011 (\emph{c.f.}, Figure \ref{marche_mobiles}).

D'ailleurs l'avantage d'un téléphone mobile par rapport à un PC, c'est que grâce au premier, une application peut savoir exactement où se situe l'utilisateur. Ce type de détail peut considérablement faciliter la RI. Donc, pourquoi taper une requête telle que "restaurant chinois paris 18"? alors que n'importe quelle application mobile peut identifier La position exacte. Dans ce cas, il suffit que cet utilisateur tape uniquement "restaurant chinois" pour que le moteur de recherche mobile retourne la liste des restaurants choinois se trouvant à sa proximité. Il est même possible, en admettant que le profil de ce dernier contient l'information qu'il aime la cuisine chinoise, qu'il tape uniquement "restaurant" pour avoir la liste des restaurants choinois les plus proches de lui.

Toutes ces particularités techniques et fonctionnelles ont engendré plusieurs problématiques dans le domaine de la RI, à savoir:
\begin{itemize}
  \item La combinaison de plusieurs dimensions dans la définition du contexte des utilisateurs, afin d'assurer une meilleure exploitation des informations pouvant ajouter plus de spécification dans leurs recherches mobiles
  \item  L'amélioration des contraintes physiques des appareils mobiles: adapter les résultats retournés par les différentes recherches aux zones d'affichages limitées de ces terminaux, simplifier la saisie des requêtes, diminuer la consommation de l'énergie, et reduire le trafic réseau pour une meilleure adaptation aux limites des bandes passantes
  \item L'exploitaion du contexte courant des utilisateurs pour garantir la précision des résultats retournés après la personnalisation
  \item L'évaluation de la recherche mobile en prenant en considération le contexte des utilisateurs.
\end{itemize}

%\subsubsection{Définition du contexte dans un environnement mobile}
Selon Dey dans \cite{Dey2000}, le contexte dans le cadre mobile est défini comme suit:
\begin{definition}\label{def}
    Le contexte est toute information pouvant être utilisée pour caractériser une situation ou une entité. Une entité est une personne, un endroit ou un objet considéré comme pertinent dans l'intéraction entre un utilisateur et une application, incluant l'utilisateur et l'application eux mêmes.
    Un système est dit context-aware s'il utilise le contexte pour fournir les données pertinentes et/ou les services à l'utilisateur, où la pertinence dépend de la tâche de ce dernier.
\end{definition}

%\subsubsection{Représentation du contexte mobile}
%Dans cette section nous allons cité les principaux modèles qui ont été définit dans le but de donner une représentation unifié du contexte.
%\begin{itemize}
%  \item Modèle hiérarchique
%  \item Modèle de représentation basé sur l'historique de recherche de l'utilisateur
%  \item
%\end{itemize}

\section{Systèmes de recommandation}
De nos jours, un grand nombre de systèmes de recommandation sont utilisés dans divers domaines. Le principal but de ces systèmes est de filtrer le flux d'informations de façon à fournir pour chaque utilisateur les ressources qui répondent à son besoin en information. Afin de satisfaire ces contraintes, les moteurs de ces systèmes gèrent les profils des utilisateurs pour choisir les ressources à transmettre à chacun, et mettre à jours ces profils en fonction des retours et des intéractions des utilisateurs.

\subsection{Définition des systèmes de recommandation}
Les systèmes de recommandation peuvent être définis de plusieurs manières, selon le type de ressources à recommander. En particulier, Burke dans \cite{Burke2002}, a défini les systèmes de recommandation comme étant:
\begin{definition}
\label{recommender_systems}
Tout système permettant de fournir des recommandations personnalisées et de guider l'utilisateur vers des résultats pertinents au sein d'un large espace de données.
\end{definition}
En pratique, la majorité des systèmes de recommandation consistent à des application web permettant de proposer un ensemble de ressources à des utilisateurs. Ces ressources peuvent correspondre aux différents types de données tels que la musique \cite{Passant2008} \cite{Su2010} \cite{Kazuyoshi2006}, les films \cite{Golbeck2009}, les restaurants \cite{Park2008}, les livres \cite{Soledad2010}, etc.

\subsection{Personnalisation des systèmes de recommandation}
Comme l'indique Chris Anderson dans son ouvrage "\emph{the long tail}", il paraît que "\emph{nous quittons l'ère de l'information pour entrer dans celle de la recommandation}". Les systèmes de recommandation sont des composants logiciels dont le but est de faciliter aux utilisateurs la prise des décisions à partir de leurs intéractions avec leurs cercles sociaux et informationnels. Les systèmes de recommandation ont donc comme vocation à nous aider à traiter des informations dont le volume et la complexité sont en croissance exponentielle.

Pour atteindre ces objectifs, les technologies de recommandation font face à des défis scientifiques majeurs, \emph{i.e.}, Comment intégrer l'hétérogénéité des sources d'information pour modéliser les intérêts des utilisateurs? Comment se servir du contexte courant des utilisateurs pour enrichir le processus de recommandation? Et comment prendre en compte les données provenant des cercles sociaux des utilisateurs dans la recommandation?

Dans le but de répondre à ces défis, l'idée de la personnalisation des systèmes de recommandation est née. En effet, l'utilité et l'importance de ce domaine a encouragé les chercheurs à travailler sur cet axe et à intégrer des informations externes aux systèmes dans le processus de recommandation, \emph{e.g.}, les données contextuelles (localisation, temps, etc.), les relations sociales, les intérêts, etc.
Les différéntes approches ayant travaillé et développé cette idée seront citées dans le reste de ce mémoire (\emph{c.f.} Chapitre \ref{ch2}).

%**********************************
\section{Conclusion}
Considéré comme un facteur indispensable dans le domaine de personnalisation de la RI, le contexte joue un rôle très important dans la gestion du flux énorme d'information disponible sur le web. En effet, le contexte d'un utilisateur donne une idée claire sur ses intentions de recherche et sur les ressources qu'il souhaite recevoir comme résultat à sa requête ou d'un processus de recommandation. Plus précisément, deux utilisateurs ayant deux besoins différents peuvent lan\c{c}er une même requête pour décrire leurs besoins distincts. A l'aide de leurs contextes courants, cette requête peut être desambigu\"{\i}sée pour mieux répondre à leurs besoins. De plus, ce même contexte peut être utilisé pour déterminer l'ensemble de personnes pouvant être recommandées comme nouveaux amis à un utilisateur donné.

Plusieurs travaux ont traité ces deux problématiques en donnant des solutions plus-ou-moins efficaces vis-à-vis la grande masse d'information disponible sur le web. Ainsi, après avoir présenté les principales notions de base liées à ce domaine, nous citons dans le chapitre suivant les principales approches qui ont été proposées dans l'état de l'art pour la personnalisation de la RI et des systèmes de recommandation. Nous donnons aussi, une revue critique de ces différents travaux.
\label{conclusion} 
%%%%%%%%%%%%%%%%%%%%%%%%%%%%%%%%%%%%%%%%%%%%%%%%%%%%%%%%%%%%%%%%%%%%%%
%   Fichier : ch2.tex
%%%%%%%%%%%%%%%%%%%%%%%%%%%%%%%%%%%%%%%%%%%%%%%%%%%%%%%%%%%%%%%%%%%%%%
\chapter{Personnalisation de la recherche d'information \&  des systèmes de recommandation: \'{E}tat de l'art}
\markboth{Personnalisation de la RI \&  des systèmes de recommandation}{\'{E}tat de l'art}
\label{ch2} \vspace*{3cm}
\section{Introduction}
Avec le déploiement à large échelle de la 3G/WiFi, la popularité des smartphones et la performance des téléphones mobiles, le domaine de la RI statique (basé sur les PC de bureau) s'est évolué vers un environnement mobile et dynamique. En effet, l'accès à l'information disponible sur le web est devenu possible à tout moment et de n'importe quel emplacement.
Ces nouvelles contraintes; caractérisées par la dynamique de l'environnement, les particularités des appareils mobiles (côté affichage et saisie), ainsi que la taille réduite et l'ambig\"{u}ité des requêtes de l'utilisateur; ont rendu les SRI traditionnels incapables de répondre convenablement aux besoins des utilisateurs.
Ces limites, imposées par le cadre mobile, ont obligé les nouvelles technologies à intégrer l'exploitation du contexte des utilisateurs dans le processus de selection des informations susceptibles d'être utiles pour un utilisateur donné dans le but d'améliorer la précision des ressources qui lui sont retournées.
Ainsi, plusieurs travaux en RI contextuelle ont essayé de donner une modélisation unifiée du contexte. Pour définir une telle représentation les auteurs ont utilisé plusieurs sources de données à savoir: les données collectées par acquisition explicite, \emph{i.e.}, les utilisateurs sont invités à introduire manuellement leurs centres d'intérêt \cite{Glover99} \cite{Sohn2008}; les données collectées par acquisition implicite \cite{Santos2010} \cite{Bellotti2008}, \emph{i.e.}, les informations utilisées dans la construction du contexte de l'utilisateur sont obtenues en observant son comportement et ses réactions vis-à-vis du système; le système de positionnement mondial GPS (Global Positionning System) permettant aux appareils mobiles de déterminer la position physique et la vitesse de mouvement de l'utilisateur, \emph{i.e.}, les systèmes SnapToTell \cite{Lim05} et Magitti \cite{Bellotti2008} se basent sur ce type de positionnement pour déterminer la localisation de l'utilisateur; l'horloge système utilisée pour déterminer l'heure exacte d'une façon automatique; les logs de requêtes, \emph{i.e.}, ces fichiers sont analysés pour déterminer le type du besoin de l'utilisateur à partir de ses requêtes précédentes; et les réseaux sociaux permettant d'avoir une idée sur la description sociale de l'utilisateur.

En se basant sur ces sources de données, différentes approches ont formalisé le contexte de plusieurs manières afin de proposer des modélisations uniformes de ce concept. En effet, ces modèles peuvent être classés en: modèles basés sur l'historique de recherche de l'utilisateur \cite{White09}; modèles attribut/valeur: dans \cite{Bouidghaghen2010} la situation de l'utilisateur est décrite par un vecteur à quatre dimensions contenant des données physiques liées à sa localisation et au temps; modèles en couche: un modèle en couche est proposé dans \cite{beach09}. Ce modèle est composé de trois couches, à savoir: une première couche responsable de la collecte des informations contextuelles à partir des capteurs, des appareils mobiles et des réseaux sociaux, une deuxième couche permettant la fusion de ces données, et une troisième couche au niveau de la couche applicative permettant de filtrer les données à utiliser pour modéliser le contexte selon le cas d'utilisation courant; modèles sémantiques: ces modèles se basent principalement sur les ontologies pour définir le contexte d'un utilisateur \cite{Daoud2008} \cite{Pretschner1999}.

Dans ce chapitre, nous présentons un survol des principales approches ayant intégré le contexte dans leurs travaux de recommandation et de RI.
Ainsi, nous catégorisons les approches de RI en trois groupes: les approches qui ont exploité le contexte cognitif de l'utilisateur, les approches qui ont intégré le contexte dans la RI dans le cadre d'un environnement mobile et les approches qui se sont focalisées sur le contexte social de l'utilisateur. D'autre part, nous citons les principaux travaux qui ont utilisé le contexte mobile et social dans leurs systèmes de recommandation.

\section{Principales approches de personnalisation de la RI}
La RI est une branche en informatique qui s'intéresse à l'acquisition, l'organisation, le stockage et la recherche des informations.
C'est l'ensemble de procédures et de techniques permettant de sélectionner, parmi un ensemble de documents, les informations (documents ou parties de documents) pertinentes en réponse à un besoin en information exprimé par l'utilisateur à travers une requête \cite{Boubekeur2008}. Les modèles classiques en RI sont basés sur une approche généraliste, qui répond invariablement aux utilisateurs en renvoyant une même liste de résultats pour deux utilisateurs ayant émis la même requête et ayant pourtant des besoins en information et des préférences de recherche différentes. Cette problématique est d'autant plus aigue suite à la prolifération des technologies mobiles d'accès à l'information (les agendas personnels, la téléphonie mobile, les smartphones, etc.) qui font de la localisation géographique de l'utilisateur un élément à prendre en compte dans le processus de sélection de l'information.

\subsection{Approches basées sur le contexte de l'utilisateur}
Compte tenu des limites de la RI classique, les approches en RI se sont orientées vers le développement d'une nouvelle génération de moteurs de recherche basée sur l'accès contextuel à l'information. L'objectif de la RI contextuelle est de mieux répondre aux besoins en information de l'utilisateur tout en intégrant son contexte de recherche dans la chaîne d'accès à l'information. La recherche d'information contextuelle consiste à exploiter le contexte de l'utilisateur ainsi que les connaissances liées à sa requête dans le but de mieux répondre à ses besoins en information.

\begin{figure*}[htbp]
\centering
\epsfig{file=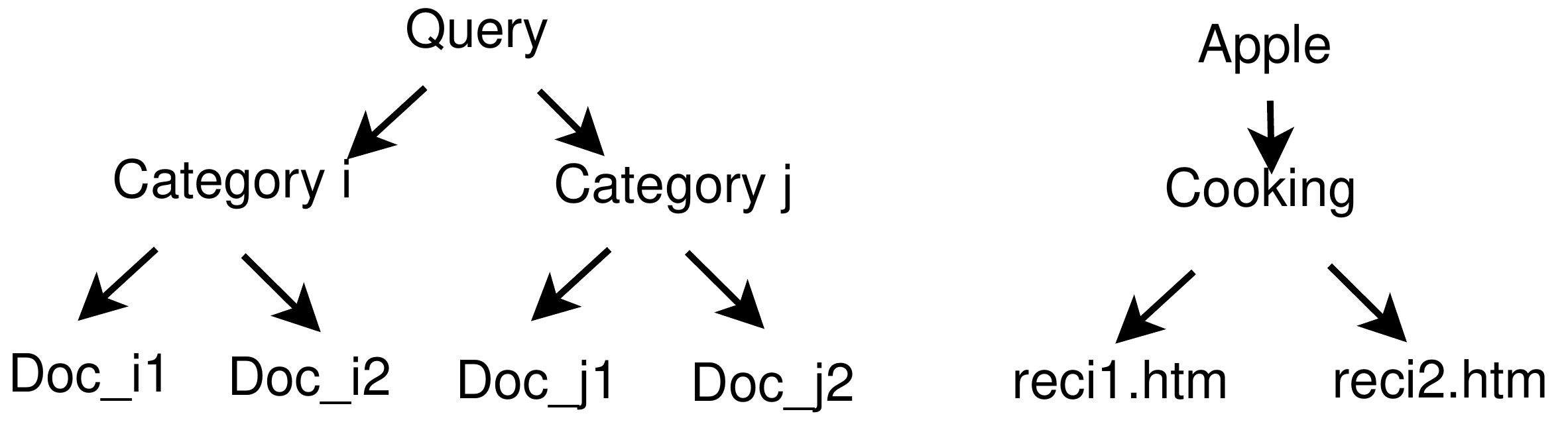, height=1.5in, width=3in}
\caption{Un modèle sous forme de tree représentant l'historique de recherche d'un utilisateur}
\label{Liuapproach}
\end{figure*}

\cite{Liu2004} ont proposé une nouvelle stratégie dans la personnalisation de la RI. Cette approche comprend deux grandes phases, la première permet de faire le mapping de la requête de l'utilisateur en un ensemble de catégories présentant à la fois ses intentions de recherche et le contexte de la requête qu'il a émise. Dans ce travail, le contexte est présenté par l'ensemble des intérêts des utilisateurs précisant leurs intentions de recherche. Plus précisément, ce contexte est formé d'un ensemble de catégories, où chaque catégorie est décrite par des mots clés pondérés par des poids. Souvent, les informations requises de l'utilisateur ne sont pas suffisantes pour décrire ses intentions, c'est la raison pour laquelle ce système utilise des informations externes extraites de l'ODP \footnote{~\url{http://www.dmoz.org}}.
Ainsi, la deuxième phase permet de combiner le contenu de la requête et son contexte dans la recherche des pages web. Un modèle de tree est proposé pour enregistrer l'historique de recherche de l'utilisateur afin de pouvoir générer ses intérêts à long terme. Un exemple de modèle de tree est décrit dans la figure (\emph{c.f.}, Figure \ref{Liuapproach}).

Wilson, dans \cite{Wilson1996}, a proposé un modèle interdisciplinaire général du comportement humain par rapport à l'information. Son modèle comprend plusieurs éléments caractérisant un utilisateur tel que: son caractère, son besoin d'information, des variables intermédiaires, \emph{i.e.}, psychologique, démographique, etc.
En effet, cet auteur place l'utilisateur dans le processus de la RI et identifie son contexte de recherche en analysant le processus de la recherche d'information et ses besoins en fonction de ses besoins personnels (psychologiques, cognitifs, etc.), de son rôle social (travail et rendement) et de son environnement (physique, social, de travail, etc.).

Une autre vision du contexte a été présentée dans \cite{White09}, qui se base sur le principe de poly-représentation, qui exploite le chevauchement des différents éléments de ce contexte implémentés en utilisant des recherches côté serveur (les activités, les collections et les informations sociales) et le code côté client (les interactions et l'historique des recherches). De ce fait, le contexte a été décrit à l'aide de cinq dimensions à savoir:
\begin{itemize}
  \item Les informations sociales: qui présentent les intérêts des utilisateurs qui ont visité, précédemment, la page actuelle de l'utilisateur
  \item L'historique des recherches: décrite par les intérêts à long terme de l'utilisateur
  \item Les activités: décrivant les pages reliées à la page actuelle de l'utilisateur par sa requête courante
  \item La collection: décrivant les pages reliées à la page actuelle de l'utilisateur par des liens hypertextes
  \item Les intéractions: déterminées par les dernières intéractions avec la page actuelle de l'utilisateur.
\end{itemize}
Ce contexte est utilisé dans la prédiction des intérêts (à court, moyen ou long terme) de l'utilisateur, et évalué à travers le pouvoir des différentes dimensions de prédire les futurs intérêts de l'utilisateur. Ces intérêts sont utilisés par la suite dans la personnalisation des recherches des utilisateurs.

Ultérieurement, les mêmes auteurs ont amelioré leur définition du contexte des utilisateurs, dans \cite{White10}, en incluant d'autres sources de données dans la phase de construction. En effet, la nouvelle description du contexte, inspirée des catégories de l'hiérarchie ODP, est utilisée dans la prédiction des intérêts de l'utilisateur. Chaque intérêt est présenté par trois modèles:
\begin{itemize}
  \item La requête: présentant la requête courante de l'utilisateur
  \item Le contexte: présentant la requête courante de l'utilisateur, ainsi que les pages qu'il a précédemment visitées
  \item L'intention: définissant une combinaison entre la requête de l'utilisateur et son contexte.
\end{itemize}

Par ailleurs, Kraft \textit{et al.} ont proposé dans \cite{Kraft05} une nouvelle approche d'enrichissement des requêtes utilisateurs en utilisant leurs contextes, \emph{i.e.}, un bout de texte (un mot, une phrase, etc.) rédigé par quelqu'un. Dans ce travail, un SRI contextuel nommé Y!Q est développé permettant de résoudre deux types de problèmes: (\textit{i}) Comment choisir les informations à utiliser pour construire le contexte de l'utilisateur? et (\textit{ii}) Comment utiliser ce contexte de façon à améliorer les résultats des recherches de l'utilisateur?

\begin{figure*}[htbp]
\centering
\epsfig{file=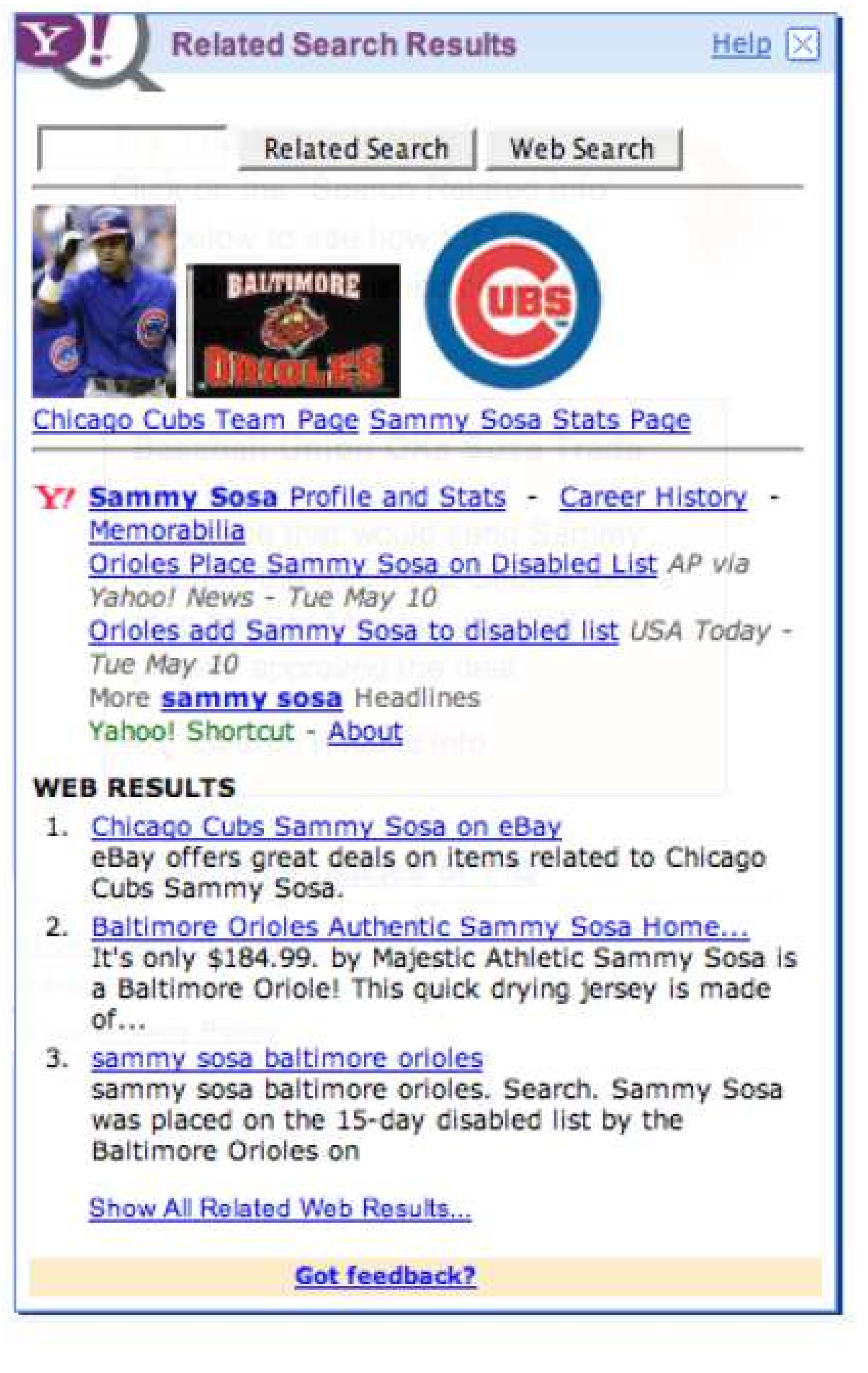, height=4.5in, width=3in}
\caption{Y!Q: Interface d'intéraction avec l'utilisateur}
\label{Y!Q}
\end{figure*}

Pour résoudre ces problèmes, Y!Q (\emph{c.f.}, Figure \ref{Y!Q}) utilise une combinaison des widgets qui détectent le contexte courant de l'utilisateur et le web sémantique dans l'analyse du contexte déjà généré. Ce système peut être divisé en trois composantes: (\textit{i}) Analyseur du contenu: permet de construire le contexte de l'utilisateur utilisé dans la désambiguïsation de sa requête; (\textit{ii}) Planifieur et reformulateur de requête: permet d'enrichir la requête avec le contexte de l'utilisateur; et (\textit{iii}) Classifieur contextuel: s'occupe du reclassement sémantique du résultat retourné à l'utilisateur.

\subsection{Approches basées sur le contexte dans un environnement mobile}
De nos jours, les appareils mobiles ont plus de fonctionnalités que les ordinateurs, incluant les informations de localisation, les caméras, les réseaux sociaux, etc. Nous pouvons utiliser les informations provenants de ces fonctionnalités supplémentaires pour créer de nouveaux domaines de recherche.
La RI et la fouille de données des nouvelles informations nécessitent des nouvelles technologies pouvant traiter ce type d'information: heure, localisation et sémantique. Plus précisement, dans les systèmes contexte-aware, l'information supplémantaire de l'utilisateur mobile présente une grande opportunité dans la personnalisation de la RI.
En effet, le contexte-aware peut être vu comme: "\textit{La perception des éléments dans l'environnement au sein d'un volume de temps et d'espace, la compréhension de leurs significations et la projection de leurs statuts dans un avenir proche}" \cite{Endsley1988}.

\begin{figure*}[htbp]
\centering
\epsfig{file=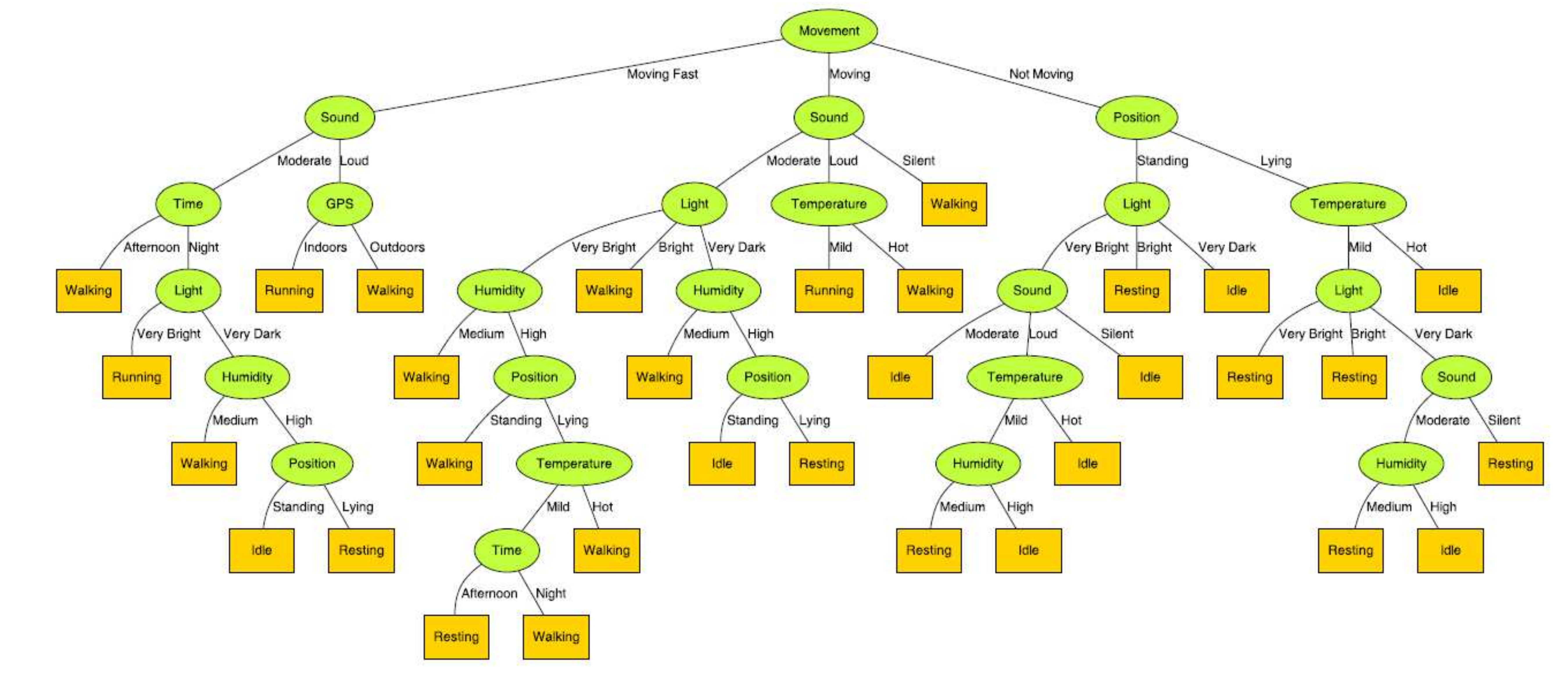, height=3.9in, width=6.5in}
\caption{Exemple d'un arbre de décision généré à partir de la phase d'apprentissage}
\label{Santosapproach}
\end{figure*}

Une nouvelle approche a été présentée par \cite{Santos2010} utilisant différents types de capteurs (capteurs de son, capteurs GPS, capteurs de température, etc.). Ces capteurs permettent de produire différentes catégories de signaux, traités par la suite pour déterminer le contexte de l'utilisateur (se déplace ou non, température forte ou basse, etc.).
Durant la phase d'apprentissage, le contexte de l'utilisateur est fourni manuellement afin de collecter un ensemble d'information, qui va aider par la suite à inférer le contexte de l'utilisateur en utilisant des techniques de classification.
Ce travail se base sur une étape d'inférence permettant de rassembler les données collectées et de déterminer le contexte de l'utilisateur. En effet, cette phase permet de déterminer la relation qui relie l'activité actuelle de l'utilisateur à son environnement. De sorte que si les mêmes conditions se reproduiront une autre fois, l'appareil arrive automatiquement à connaître son contexte.
Ce contexte d'inférence utilise la modélisation à base d'arbre de décision formé lors de la phase d'apprentissage et mis à jour à chaque utilisation du système. Cette structure est détaillée dans la figure \ref{Santosapproach}.

Le contexte généré est utilisé par la suite, dans une phase de publication, dans le cadre des applications sociales, permettant à l'utilisateur de mettre à jour: son statut, les activités ayant une relation avec son contexte actuel et de tagger des resources avec son contexte. L'exemple de Twitter \footnote{~\url{https://Twitter.com}}, où le contexte détérminé permet de répondre à la question "Que fais-je?".

\begin{figure*}[htbp]
\centering
\epsfig{file=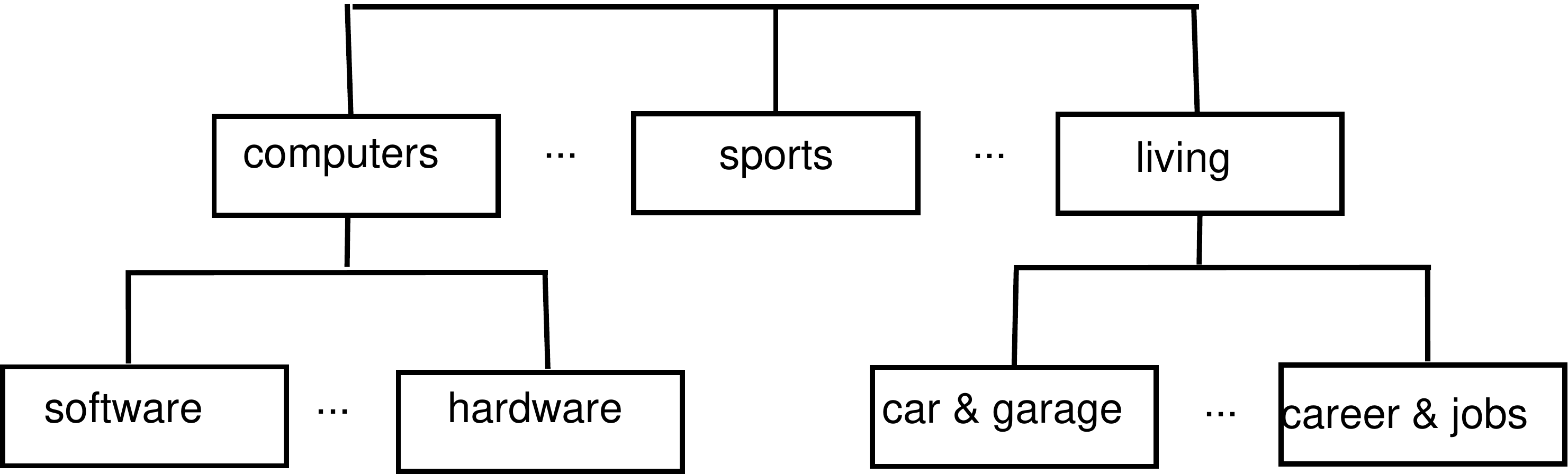, height=1.5in, width=3in}
\caption{Exemple de structure d'une taxonomie}
\label{Taxonomyapproach}
\end{figure*}

Dans le même cadre de recherche, \cite{Cao2010} s'est basé sur la classification des requêtes de l'utilisateur (QC) en considérant son contexte, pour déterminer ses intentions de recherche. En effet, la classification est faite à l'aide du modèle \textbf{Conditional Random Field} (\textbf{CRF}). Le principe de la classification des requêtes est de classer une requête $q_{T}$ dans une liste triée de K catégories $c_{T1}$, $c_{T2}$, ..., $c_{Tk}$, parmi les \{$c_{T1}$, $c_{T2}$, ..., $c_{Tk}$\} de la taxonomie $\Upsilon$, la figure \ref{Taxonomyapproach} décrit un exemple de cette taxonomie.

Le modèle utilisé dans cette approche est un modèle d'apprentissage séquentiel qui n'a pas besoin d'information à priori et qui permet de collecter des informations contextuelles à partir des requêtes de l'utilisateur. Ce modèle est utilisé pour modéliser le contexte de l'utilisateur pour chacune de ses requêtes en utilisant deux types de caractéristiques: locales (\emph{i.e.}, les termes de la requête, le pseudo-feedback et le feedback implicite) et contextuelles.

Cependant, \cite{Sohn2008} ont analysé le besoin en information des utilisateurs mobiles, en effectuant une étude journalière de deux semaines. Durant cette période les utilisateurs, d'âges et de sexes différents, sont menés à retourner un journal contenant leurs recherches. Un ensemble de 421 requêtes est collecté avec une moyenne de 21.1 journal par personne.
Cette étude inclut les types d'information demandés par les participants, les stratégies et les méthodes utilisées pour exprimer leurs besoins, ainsi que les facteurs contextuels qui expliquent chaque besoin et la manière de le modéliser.
Les entrées recueillies sont divisées en 16 catégories de besoins (shopping, horaires des films, actualité, etc.) en se basant sur les journaux des participants et leurs feedbacks durant la période des interviews.
Dans cet article, la méthode Experience Sampling Method (ESM) avec une technique similaire à la technique \emph{snippet} sont utilisées afin de capturer les informations de l'appareil mobile de l'utilisateur. Ces informations sont utilisées dans l'extraction des intérêts des utilisateurs qui vont servir dans la personnalisation de leurs recherches.

\begin{figure*}[htbp]
\centering
\epsfig{file=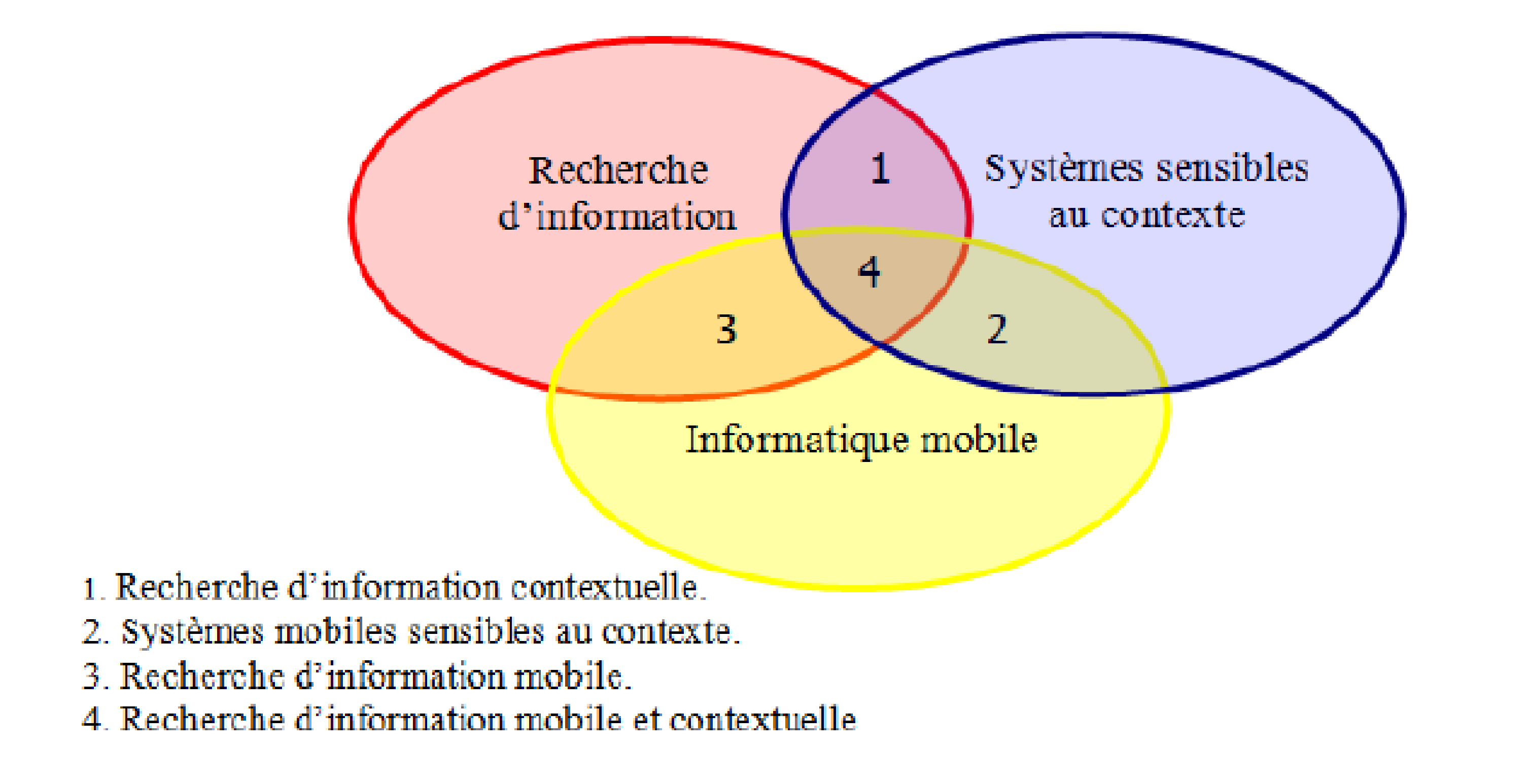, height=2.3in, width=5in}
\caption{Principales composantes du domaine de la RI mobile}
\label{recherchecontextuelle}
\end{figure*}

Récemment, les auteurs dans \cite{Bouidghaghen2010} ont donné une description du positionnement du contexte mobile par rapport au domaine de la RI (\emph{c.f.}, Figure \ref{recherchecontextuelle}). Toutefois, ils ont proposé un système de recherche mobile qui se base sur le contexte mobile de l'utilisateur dans le processus de personnalisation en utilisant un modèle \textbf{Case Based Reasoning} (\textbf{CBR}) permettant de sauvegarder l'historique des recherches des utilisateurs. Le CBR est une technique qui permet de résoudre un problème donné en utilisant les solutions des problèmes similaires déjà résolus, où un cas est présenté par une paire $\langle premisse,valeur\rangle$, prémisse: la description des caractéristiques du cas, et valeur: le résultat du raisonnement basé sur le champ prémisse. Dans cette approche un cas est décrit comme suit:
Case=(S,G), où S: la situation décrivant le contexte actuel de l'utilisateur et G: son profil lié à ce contexte.
En effet, à chaque fois où une nouvelle situation S est construite, les n{\oe}uds du CBR sont visités pour vérifier l'existence d'une situation similaire à S et extraire son intérêt. Dans le cas contraire, où aucune situation similaire n'est extraite, S est ajoutée à la base d'apprentissage.

Ainsi, cette approche peut être modélisée par quatre étapes:
\begin{enumerate}
  \item Identification de la situation actuelle de l'utilisateur $S^{*}$=($X_{l}^{*}$,$X_{u}^{*}$,$X_{v}^{*}$,$X_{w}^{*}$): cette situation est construite en utilisant les capteurs GPS et l'horloge système. Elle est mappée, par la suite, via des ontologies spatiales et temporelles, en un contexte sémantique,
  \item Recherche du cas le plus similaire: une situation $S^{opt}$ est considérée similaire à S si elle satisfait la formule (\ref{similarity})
  \begin{center}
  \begin{equation}
  \label{similarity}
   $$ $S^{opt}$=argmax($\sum_{j}$ $\alpha_{j}$.sim($X_{j}^{*}$,$X_{j}^{i}$))$$
  \end{equation}
  \end{center}
  Où $X_{j}^{*}$ (resp. $X_{j}^{i}$) la valeur de la $j^{\grave{e}me}$ (resp. $i^{\grave{e}me}$) composante du vecteur situation $S^{*}$, $1\leq i\leq n$ et n: le nombre des situations précédentes, $sim_{j}$ la métrique de similarité liée à la $j^{\grave{e}me}$ composante du vecteur situation ayant $\alpha_{j}$ comme poids,
  \item Réutilisation du cas: réordonnancement du résultat de recherche. Afin d'améliorer le résultat de la recherche, la personnalisation n'est effectuée que si: sim($S^{*}$,$S^{opt}$)$\geq$ $\beta$ avec $\beta$ un seuil fixé. Le profil extrait $G^{opt}$ est utilisé pour la phase du réordonnancement des résultats de la recherche de l'utilisateur,
  \item Révision de la solution proposée et/ou le maintien du cas: deux situations sont possibles dans cette phase
  \begin{itemize}
    \item sim($S^{*}$,$S^{opt}$)$\neq$ 1: un nouveau cas <$S^{*}$,$G^{*}$> est ajouté au modèle CBR
    \item sim($S^{*}$,$S^{opt}$)= 1: le profil $G^{*}$ est utilisé dans la personnalisation de la recherche de l'utilisateur.
  \end{itemize}
\end{enumerate}

\subsection{Approches basées sur le contexte social de l'utilisateur}
Malgré l'efficacité de la recherche obtenue par les techniques présentées précédemment, elles sont généralement limitées au niveau de l'exploitation du réseau social de l'utilisateur : amis, relations, etc. D'une manière générale, ces informations sociales peuvent être exploitées pour incorporer le domaine d'intérêts de l'utilisateur dans la RI sociale.

Dans \cite{Zhou2008} les annotations de l'utilisateur sont utilisées dans la personnalisation du SRI. Cette approche permet de: (\textit{i}) définir les intérêts des utilisateurs à partir de leurs annotations sociales; et (\textit{ii}) catégoriser les utilisateurs par domaines en se basant sur leurs annotations sociales; et (\textit{iii}) combiner les modèles des termes de langage avec ceux obtenus à partir des sujets des documents.
Et se compose de deux phases:
\begin{enumerate}
  \item Découvrir des sujets à partir des annotations et des contenus des documents tout en catégorisant les utilisateurs par domaines d'intérêt. En effet, un modèle probabiliste génératif est développé pour la génération des contenus des documents, ainsi que l'ensemble des tags associés
  \item Améliorer les modèles des documents, ainsi que le langage des requêtes par l'intégration des domaines d'intérêt. Dans cette étape, un framework de RI, basé sur la minimisation des risques, est développé. Ce framework est fondé sur la théorie de la décision Bayésienne permettant l'amélioration des modèles de langage \cite{Lafferty2001}, des requêtes et des documents.
\end{enumerate}

Dans \cite{BenJabeur2010}, un modèle de RI est proposé visant à associer la pertinence des ressources bibliographiques à l'importance sociale de leurs auteurs.
Dans cette approche, la pertinence sociale des documents, en particulier les ressources bibliographiques, est estimée à l'aide d'un calcul qui mesure l'importance sociale d'un auteur $a_{i}$ d'une ressource $C_{G}(a_{i})$. Ce calcul est fait à l'aide de la formule suivante:
  \begin{center}
  \begin{equation}\label{eq2}
    $$$Imp_{G}$(d)=$\sum_{i=1}^m$ (w($a_{i}$,d)$C_{G}$($a_{i}$))$$
  \end{equation}
  \end{center}

Par la suite, cette mesure est combinée avec une métrique de la RI traditionnelle pour donner à la fin:
  \begin{center}
  \begin{equation}\label{eq3}
    $$R(q,d,G)=$\alpha$ RSV(q,d)+$(1-\alpha)$$Imp_{G}$(d)$$
  \end{equation}
   \end{center}

Où: $\alpha \in [0,1]$, RSV(q,d) une mesure de similarité thématique entre une requête q et un document d, et $Imp_{G}$(d) l'importance sociale du document d dans le réseau social G.

Ainsi, la pertinence d'un document est estimée par la combinaison de la pertinence thématique et de la pertinence sociale, qui est à son tour dérivée de l'importance sociale des auteurs associés. Dans la phase de validation, les auteurs ont utilisé une collection d'articles SIGIR, dont les annotations sociales sont extraites à partir du réseau social académique "CiteULike.org".

\begin{figure*}[htbp]
\centering
\epsfig{file=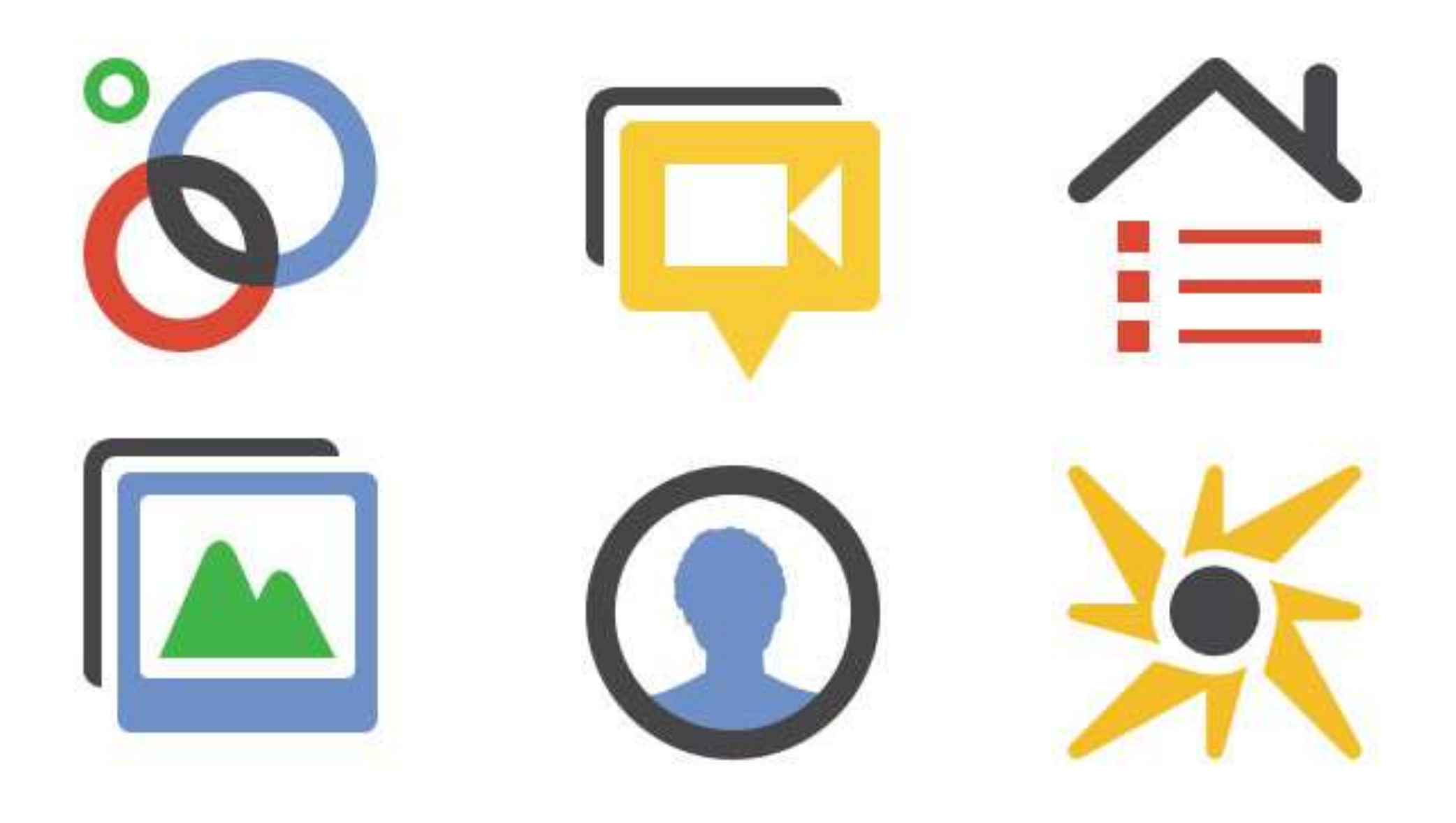, height=1.5in, width=3in}
\caption{Les dimensions du contexte social proposé par Google+}
\label{Googleapproach}
\end{figure*}

Par exemple, en 2011, Google a lancé Google+ \footnote{~\url{https://plus.google.com}} son dernier effort en web social.
Cette initiative a rassemblé différentes idées présentées dans la figure \ref{Googleapproach}.

Parmi les dimensions utilisées dans la description du contexte social de l'utilisateur nous citons: (\textit{i}) Les cercles sociaux présentant l'ensemble des amis de l'utilisateur organisés par sujet (amis, famille, collègues, etc.). Ce type d'organisation est efficace de point de vue communication et partage des ressources entre les amis; et (\textit{ii}) Les déclics (ou Sparks) permettant de retourner automatiquement des documents aux utilisateurs en fonction de leurs intérêts.

\subsection{\'{E}tude comparative et discussion}
Dans ce qui suit, nous allons présenter une synthèse des approches de la RI contextuelle, présentées précédemment, à travers deux tableaux qui détaillent les principales caractéristiques et principes de chacune d'elles (\emph{c.f.}, Tableau \ref{RI1} et Tableau \ref{RI2}).
En effet, ces approches sont appuyées sur trois types de contextes exploités dans la phase de la personnalisation, à savoir:
\begin{itemize}
  \item Le contexte lié à l'utilisateur et à la requête qu'il a émise,
  \item Le contexte mobile de l'utilisateur lié aux contraintes spatio-temporelles présentes lors de l'émission de sa requête,
  \item Le contexte social lié au cercle social de l'utilisateur.
\end{itemize}

Les tableaux \ref{RI1} et \ref{RI2} donnent une comparaison entre les approches présentées ci-dessus, en précisant la spécificité de chacune d'elles sur les axes suivants:
\begin{enumerate}
  \item \textbf{Type du contexte}: spécifie le type du contexte exploité lors de la personnalisation.
   \item \textbf{Dimensions du contexte}: indique les dimensions utilisées pour décrire le contexte des utilisateurs. Ces informations peuvent être détectées à l'aide des capteurs (de son, de luminosité, GPS, etc.), par l'intervention des utilisateurs, par l'historique de leurs recherches, etc.
  \item \textbf{Prédiction des intérêts}: précise si l'approche en question permet d'aller au delà de la personnalisation des recherches des utilisateurs, pour prédire leurs intérêts qui peuvent reflèter leurs intentions derrière les requêtes émises.
  \item \textbf{Technique utilisée}: différentes techniques peuvent être appliquées afin d'employer et de tirer profit du contexte des utilisateurs dans la personnalisation des résultats de leurs recherches.
  \item \textbf{Source de données}: cette caractéristique définit la source de données utilisée par l'approche en question pour la construction du contexte des utilisateurs lors du lancement de leurs requêtes.
\end{enumerate}

Par ailleurs, les tableaux \ref{RI1} et \ref{RI2} présentent une étude comparative entre les différentes approches de l'état de l'art ayant traitées le problème de personnalisation de la RI en exploitant trois types de contextes à savoir: local, mobile ou social. Ainsi, nous citons les critiques suivantes:
\begin{itemize}
  \item Aucune approche, sauf celle de \cite{Bouidghaghen2010}, n'a mis en relief la relation qui existe entre les intérêts des utilisateurs et leurs mobilités. En effet, une étude comparative va être détaillée dans le chapitre \ref{ch4}, permettant de valoriser et citer les avantages de la technique que nous avons utilisée dans ce mémoire pour faire l'association situation/intérêt par rapport au modèle CBR.
  \item Aucune approche n'a trouvé une technique efficace pour la prédiction des intérêts des utilisateurs par rapport à l'état ubiquitaire ou mobile de ces derniers. Par contre, nous trouvons quelques travaux qui ont visé à exploiter les intérêts des utilisateurs, mais l'extraction de ces préférences est conditionnée par l'intervention de ces derniers.
  \item Concernant le nombre énorme de situations qui peuvent être détectées pour chaque utilisateur à cause de sa mobilité, aucune approche n'a tenté de minimiser le nombre de ces situations pour diminuer le temps de calcul et faciliter la prédiction des intérêts.
\end{itemize}

\begin{landscape}
\begin{table}
\centering
\begin{tabular}{|p{4cm} |p{2cm} |p{4cm} |p{2cm} |p{3cm} |p{3cm} |p{2.5cm}|}
\hline\hline
\textbf{Approche} & \textbf{Type du contexte} & \textbf{Dimensions du contexte} & \textbf{Prédiction des intérêts} & \textbf{Technique utilisée} & \textbf{Source de données}\\
  \hline\hline
   Liu \textit{et al.} (2004) & local & intérêts & non & désambiguïsation de requêtes & données réelles\\
 \hline
   White \textit{et al.} (2009) & local \& social  &  social, historique, activité, collection et intéraction & oui & catégorisation des intérêts & fichiers logs\\
   \hline
   White \textit{et al.} (2010) & local & requête et pages visitées & oui & catégorisation des intérêts & fichiers logs\\
   \hline
   Kraft \textit{et al.} (2005) & local & phrase rédigée par l'utilisateur & non & enrichissemnt de requêtes & données détectées avec Y!Q\\
  \hline
  Santos \textit{et al.} (2010) & mobile & mouvement, son, temps, localisation, temperature, humidité et luminosité & non & arbre de décision & capteurs \\
   \hline
   Cao \textit{et al.} (2010) & mobile & requêtes précédentes et pages visitées & non & CRF & moteur de recherche commercial\\
   \hline
   Sohn \textit{et al.} (2008) & mobile & activité, localisation, temps et conversation & non & ESM & étude journalière\\
   \hline
   Bouidghaghen \textit{et al.} (2010) & mobile & temps et localisation & non & CBR & étude journalière\\
   \hline
\end{tabular}
\caption{Tableau récapitulatif des approches de personnalisation de la RI}
\label{RI1}
\end{table}
\end{landscape}

\begin{landscape}
\begin{table}
\centering
\begin{tabular}{|p{4cm} |p{2cm} |p{4cm} |p{2cm} |p{3cm} |p{3cm} |p{2.5cm}|}
\hline\hline
\textbf{Approche} & \textbf{Type du contexte} & \textbf{Dimensions du contexte} & \textbf{Prédiction des intérêts} & \textbf{Technique utilisée} & \textbf{Source de données}\\
  \hline\hline
   Zhou \textit{et al.} (2008) & social & annotations & oui & modèle probabiliste génératif \& minimisation des risques & del.icio.us\\
   \hline
   Jabeur et Tamine-Lechani (2010) & social & importance sociale et annotations & non & modèle de pondération & SIGIR \& "CiteULike.org"\\
  \hline
\end{tabular}
\caption{Tableau récapitulatif des approches de personnalisation de la RI (suite)}
\label{RI2}
\end{table}
\end{landscape}

\section{Principales approches de recommandation d'amis}
Chaque année, le web voit arriver une nouvelle mode ou une nouvelle technologie qui semble venir le révolutionner en entier.
Ces dernières années avec l'apparition du web social, le nombre des systèmes de recommandation s'est multiplié afin de pouvoir gérer le grand volume d'information contenu dans le web social, pour un utilisateur donné, et ainsi lui faciliter le choix du contenu qui lui semble intéréssant à partir d'un catalogue trop vaste.
Ce domaine représente une grande similarité par rapport au domaine de la RI. En effet, les techniques utilisées sont les mêmes avec la différence qu'il n'y a pas de requête explicite exprimée par l'utilisateur.

Les systèmes de recommandation peuvent être classés de différentes manières. La classification qu'on va suivre dans ce mémoire permet de définir deux catégories de recommandation: approches basées sur le contexte social de l'utilisateur et approches basées sur le contexte mobile de l'utilisateur.
Dans ce mémoire nous allons nous intéresser plus aux systèmes de recommandation d'amis.

\subsection{Recommandation basée sur le contexte social de l'utilisateur}
La recommandation basée sur le contexte social consiste à analyser le cercle social, \emph{e.g.}, amis, intérêts, etc. dans le but de déterminer les ressources, \emph{i.e.}, dans notre cas d'étude des amis, succeptibles à être intéressantes pour un utilisateur appartenant à une communauté donnée.

Dans \cite{Silva2010}, les auteurs ont proposé une nouvelle approche de recommandation d'amis basée sur les relations sociales des utilisateurs. La procédure de recommandation est appuyée sur la typologie des relations entre les n{\oe}uds d'un réseau et opère en deux étapes: (\textit{i}) Filtrage: cette phase permet d'extraire les amis de $2^{eme}$ degré de l'utilisateur, \emph{i.e.}, les amis de ses amis. Les n{\oe}uds du graphe sont filtrés de façon à garder que les n{\oe}uds adjacents à chaque noeud connecté au noeud central du processus de recommandation; et (\textit{ii}) Ordonnancement: dans cette phase un coefficient est associé à chaque noeud du graphe. Cette valeur correspond au degré d'intéraction entre un n{\oe}ud et le n{\oe}ud central autours duquel tourne le processus de recommandation.
Une limite de cette approche réside dans sa restriction aux relations d'amitié dans l'étape de filtrage. Cette limite peut être vue plus clairement avec un passage à l'échelle, \emph{i.e.}, vers des réseaux plus larges comme Facebook \footnote{~\url{https://www.facebook.com}} ou Orkut \footnote{~\url{http://www.orkut.com}}, où le nombre d'amis d'une personne dépasse souvent 150 personnes.

Ainsi, dans \cite{Xin2009} les auteurs ont proposé un framework Multi-scale Continuous Conditional Random Fields (MSCCRF), où ils ont modélisé les dépendances du réseau social par des propriétés du modèle de Markov. Dans cette approche, le processus de recommandation est fait sur la base de similarité entre les utilisateurs indépendamment des informations contenues dans les réseaux sociaux.

Par ailleurs, Konstas \textit{et al.} ont utilisé le réseau social "last.fm" \cite{Konstas2009} dans l'évaluation de leur modèle de recommandation collaboratif qui inclut à la fois les relations d'amitié et l'ensemble d'annotations des utilisateurs.
En effet, cette approche se base sur le modèle Random Walk with Restarts (RWR) \cite{Lovasz1996} dans la présentation des réseaux sociaux.
Ce travail se compose de trois étapes:
\begin{enumerate}
    \item Analyse du réseau social "last.fm" pour construire un graphe d'utilisateurs, les relations qui les lient, ainsi que leurs annotations sociales
    \item Utilisation du graphe généré dans l'étape précédente dans la validation de la nouvelle variante du modèle RWR proposée, et dans la démonstration de son impact dans les systèmes de recommandation. En effet, le lien de parenté entre deux noeuds du graphe est mesuré, gr\^{a}ce à la théorie RWR, avec la formule suivante:
        \begin{center}
        \begin{equation}\label{eq1}
        $$$\textbf{p}^{(t+1)}$=(1-a)\textbf{S}$\textbf{p}^{(t)}$+a\textbf{q}$$
        \end{equation}
         \end{center}
        Où: $\textbf{p}^{(t)}$ est un vecteur colomne avec $p_{i}^{(t)}$ la probabilité que le noeud courant soit le noeud $i$, \textbf{q} est un vecteur colomne de 0 où l'élément correspendant au noeud de début vaut 1, \textbf{S} est la table des probabilités où chaque élément $S_{i,j}$ présente la probabilité que $j$ sera le noeud suivant à partir du noeud courant $i$, et a est la probabilité pour retourner au noeud de début $x$
    \item Preuve de l'utilité du modèle RWR dans l'extraction des informations utiles à partir des réseaux sociaux.
\end{enumerate}
La principale limite de ce travail réside dans le manque de la dynamicité dans l'information utilisée durant le processus de recommandation.

Plus récemment, \cite{Ma2011} ont tenté d'inclure les matrices de factorisation dans leur définition du système de recommandation. Deux types d'information sont utilisés: les réseaux des amis ainsi que les préférences des utilisateurs. En effet, les auteurs se basent sur l'hypothèse que les décisions d'une personne dépendent fortement de ces préférences et des recommandations de ses amis, pour proposer un framework probabiliste de la matrice de factorisation. Ce modèle fusionne les matrices utilisateur/ressource et les cercles des utilisateurs dans le processus de recommandation.

\begin{figure}
    \centering
    \subfigure[]{\label{sub1} \includegraphics[width=7cm]{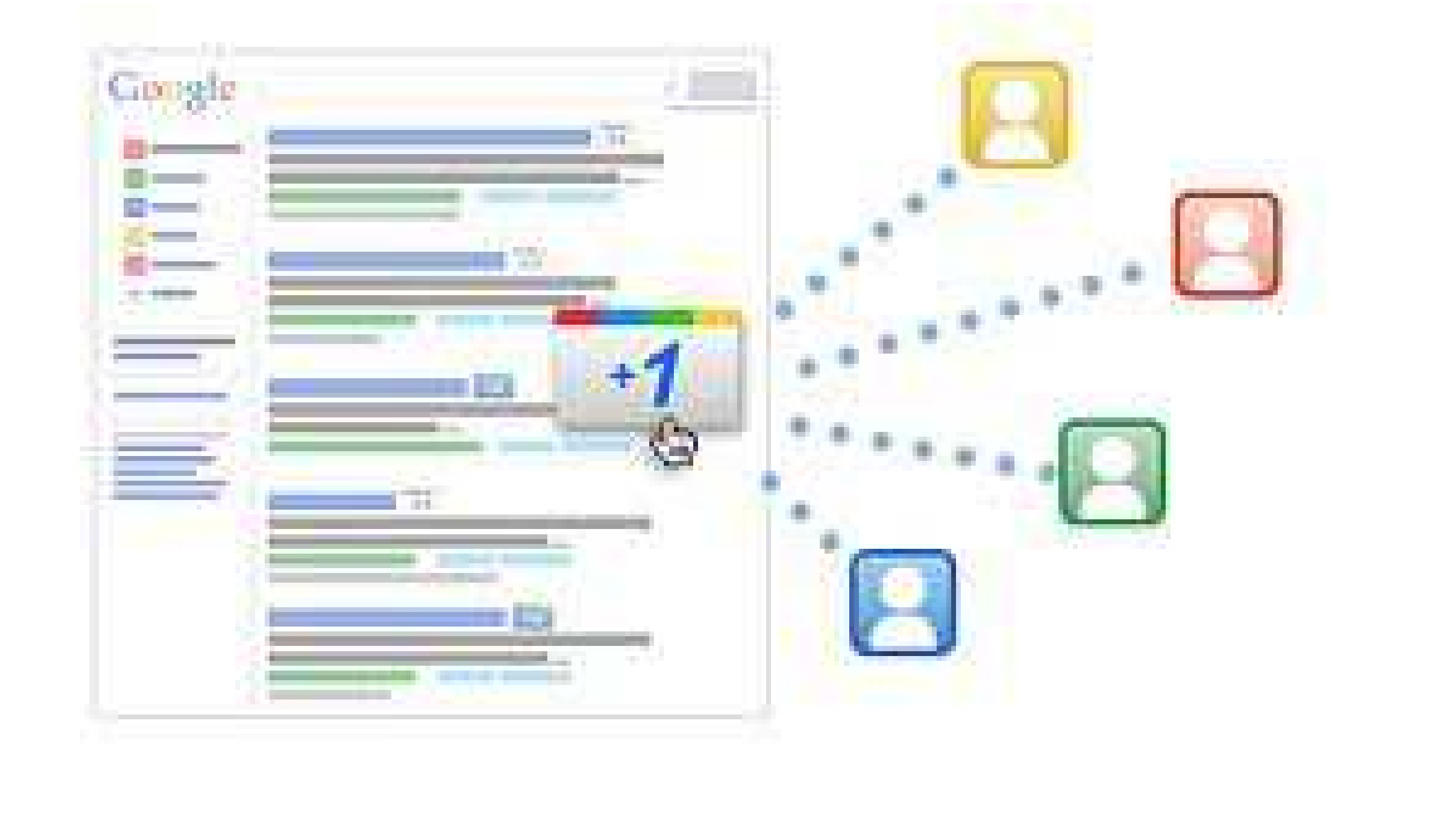}}
    \subfigure[]{\label{sub2} \includegraphics[width=7cm]{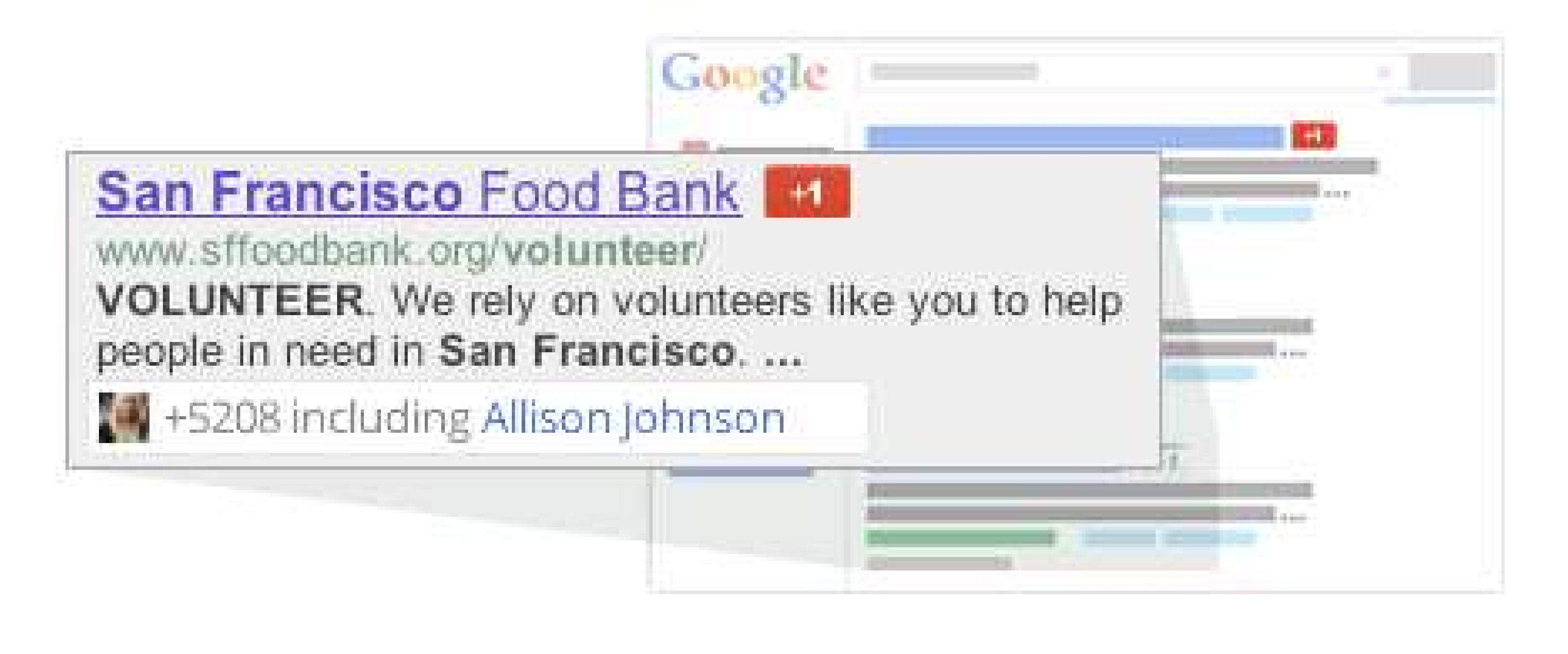}}
    \subfigure[]{\label{sub3} \includegraphics[width=7cm]{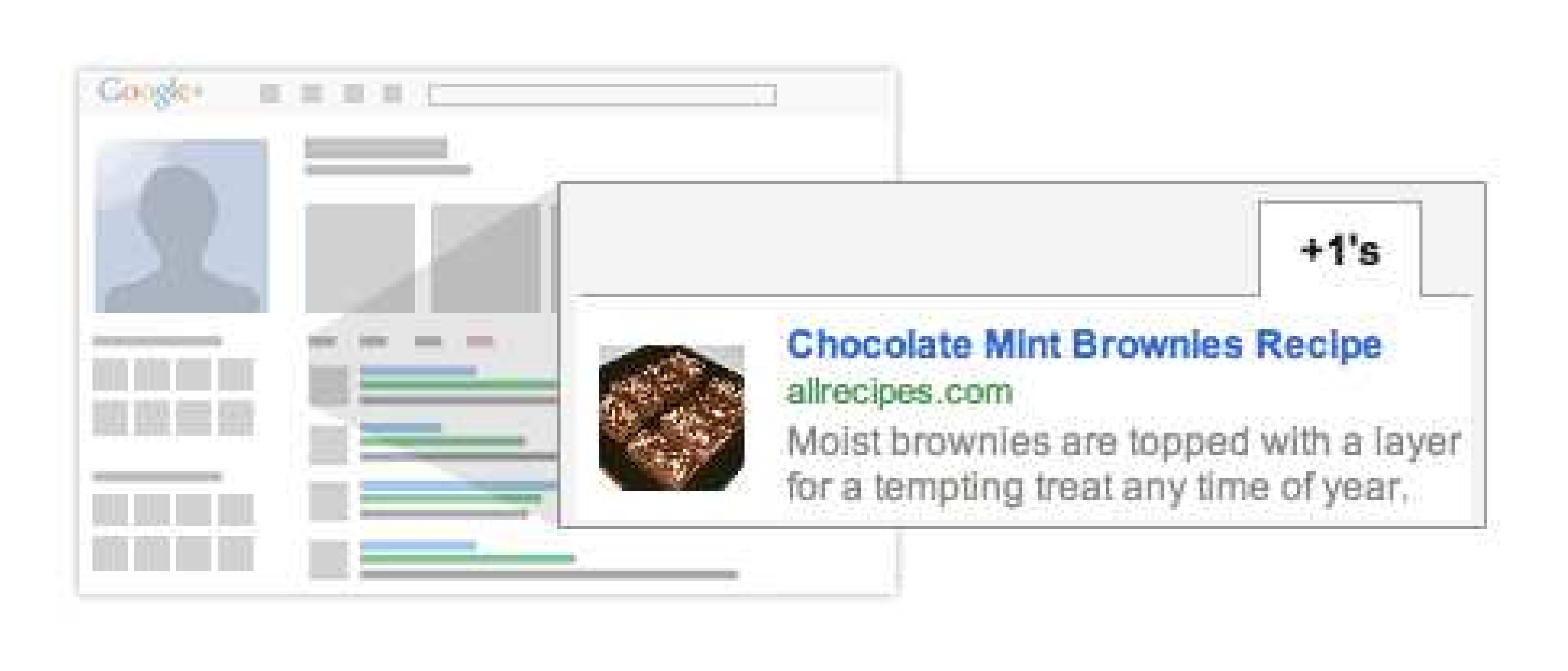}}
    \caption{Le système de recommandation de Google "\textbf{+1}". \subref{sub1} Le bouton "+1". \subref{sub2} Publication des recommandations avec "\textbf{+1}". \subref{sub3} Recommandation des intérêts avec "\textbf{+1}".}
    \label{google+1}
\end{figure}

Google, par exemple, a lancé en 2011 son système de recommandation "\textbf{Google+1}" \footnote{~\url{http://www.google.com/+1/button}}. Le bouton "+1" (\emph{c.f.}, Figure \ref{google+1}) permet aux utlisateurs de Google de signaler un contenu donné comme contenu recommandable pour qu'il soit visible dorénavant dans le résultat de leurs recherches. A travers cette fonction baptisée "+1", Google tente une nouvelle incursion dans l'univers des systèmes de recommandation et des réseaux sociaux. Plus précisément, il s'agit d'un bouton grâce auquel les utilisateurs peuvent recommander à leurs amis, utilisateurs de Google, un contenu qu'ils jugent intéressant.

\subsection{Recommandation basée sur le contexte mobile de l'utilisateur}
Plusieurs études ont essayé d'inclure le contexte des utilisateurs dans la définition de nouveaux systèmes de recommandation \cite{Adomavicius2011}, \cite{AdomaviciusTuzhilin2011}.

Zheng et \emph{al.}, dans \cite{Zheng2010} ont proposé l'utilisation des coordonnées GPS pour répondre à deux différentes questions de la vie réelle:
\begin{enumerate}
  \item Si nous voulons faire une activité, tel que le tourisme dans une ville comme Pékin, où allons-nous?
  \item Si nous avons déjà visité certains endroits pour faire une activité, tel que la construction des Nids d'oiseaux au Parc Olympique à Pékin, que pouvons faire d'autre?
\end{enumerate}
Pour répondre à ces questions les auteurs ont utilisé les trajectoires GPS dans la recommandation du comportement des utilisateurs ou même la recommandation de leurs futures actions, en modélisant leurs historiques de localisations et d'activités. Cet ensemble de données a été collecté grâce à 162 utilisateurs et 4 millions de points GPS durant une période de 2.5 ans.

Ces mêmes auteurs ont traité la même problématique, de recommandation de localisation et d'activité, d'une autre manière. Ainsi, ils ont proposé dans \cite{ZhengC10} un système de filtrage collaboratif de localisation et d'activité User Collaborative Location and Activity Filtering (UCLAF) basé sur une représentation en tenseur 3-D.
En particulier, un tenseur $\mathcal{A}$ est défini par $\mathcal{A}$ $\in$ $\mathcal{R}^{m\times n\times r}$, o\`{u} m est le nombre des utilisateurs, n le nombre des localisations et r le nombre des activités.
Cependant, ce système ne considère pas les relations entre les utilisateurs pouvant donner plus de précision et de dynamicité aux résultats de la recommandation.

Dans le même axe de recherche, \cite{Hariri2011} ont présenté une nouvelle approche de fouille de données contextuelle à partir des revues des utilisateurs. Ces auteurs précisent la différence entre le "rang" qu'un utilisateur peut attribuer à une ressource et "l'utilité" qui peut être gagnée en le choisissant. Dans ce travail, le contexte est utilisé dans le but de définir une fonction d'utilité qui reflète à quel point une ressource est préférée par un utilisateur.

Le plus grand inconvénient des approches précédemment détaillées, réside dans le manque de dynamicité dans le processus de recommandation.
Afin de résoudre ce problème, un framework nommé SocialFusion a été proposé par \cite{beach09} visant à intégrer le contexte mobile dans la recommandation à partir de la fusion de la mobilité, les capteurs et l'information sociale. Ce framework fonctionne suivant trois étapes: (\textit{i}) Collecte des données à partir des réseaux sociaux, les téléphones mobiles et différents types de capteurs; (\textit{ii}) Les données collectées sont fusionnées sous forme d'indices contextuels appelés descripteurs; et (\textit{iii}) En se basant sur les données rassemblées et l'ensemble des descripteurs, une action contextuelle est recommandée à l'utilisateur.
Un facteur important présentant les intérêts des utilisateurs a été négligé dans la recommandation. En effet, une telle information pourra donner une idée sur leurs préférences, donc plus de précision sur le type de personnes à recommander.

Une autre approche de recommandation mobile et dynamique d'amis a été introduite dans \cite{Qiao2011}, qui applique la technique de clustering dans le processus de découverte des communautés. L'approche proposée combine le contexte mobile et la modélisation d'ontologies pour l'extraction des données utilisées dans la recommandation. Ces données sont restreintes aux relations d'amitié entre les utilisateurs en négligeant la description de leurs intérêts et préférences. Une autre limite de ce travail réside dans l'utilisation de l'algorithme \textbf{GN} \cite{Girvan2002} à cause de sa complexité élevée $(O(m^{2}n))$, où $n$ est le nombre des utilisateurs et $m$ le nombre des relations qui les relient. Une étude comparative entre cette approche et notre approche de recommandation sera détaillée dans le chapitre \ref{ch4}.

\subsection{\'{E}tude comparative et discussion}
La plupart des systèmes de recommandation présentés ci-dessus, exploitent des réseaux sociaux connus. En effet, après l'étude de ces approches, la classification en approches basées sur le contexte social des utilisateurs et celles basées sur leur contexte mobile demeure automatique. Ainsi, la première famille de travaux permet d'extaire les relations reliant les utlisateurs d'un réseau donné pour les appliquer dans le processus de recommandation. Bien que la deuxième famille visait à recommander des nouvelles ressources aux utilisateurs par rapport à leurs contextes mobiles.

Le tableau \ref{SR} présente une comparaison entre les différentes approches déjà détaillées. Cette comparaison est menée suivant les axes suivants:
\begin{enumerate}
  \item \textbf{Type de recommandation}: décrit le type de ressources peuvant être recommandées par chaque approche, \emph{e.g.}, amis, films, etc.
  \item \textbf{Type du contexte}: précise le type du contexte utilisé dans la personnalisation du système de recommandation proposé par chaque approche. Dans ce mémoire on distingue entre deux types de contexte: social et mobile.
  \item \textbf{Découverte des communautés}: précise si l'approche inclut ou non l'extraction des communautés à partir du réseau des utilisateurs.
  \item \textbf{Technique utilisée}: indique la technique utilisée par le processus de recommandation.
  \item \textbf{Source de données}: précise la base de test utilisée par chaque approche dans la phase de validation.
\end{enumerate}

Ainsi, le tableau \ref{SR} donne une étude comparative des différentes approches proposées dans l'état de l'art des systèmes de recommandation basés sur l'exploitation des réseaux sociaux. Par ailleurs, nous pouvons citer les constatations suivantes:
\begin{itemize}
  \item Aucune approche, à part celle de \cite{Qiao2011}, n'a pu combiner entre le contexte social et mobile de l'utilisateur d'une façon à garder l'aspect dynamique des réseaux sociaux. Dans notre cas d'étude, en admettant que le type des personnes qui peuvent intéresser un usager donné n'est pas statique, la propriété de dynamicité devient très intéressante.
  \item La majorité des approches détaillées dans le tableau \ref{SR} appliquent un algorithme de découverte de communautés d'utilisateurs dans le processus de recommandation. Cependant, ils perdent avec ce découpage du réseau en communautés la dynamicité des liens qui relient entre ses noeuds puisqu'ils utilisent des propriétés statiques dans leurs définitions. \cite{Qiao2011} ont appliqué l'approche \textbf{GN} qui permet de préréserver cet aspect. Seulement, cette approche donne une compléxité élevée en termes de temps d'exécution par rapport à notre approche. L'étude comparative des deux approches est présentée dans le chapitre \ref{ch4}.
\end{itemize}

\begin{landscape}
\begin{table}
\centering
\begin{tabular}{|p{4.5cm}| p{3.5cm} |p{2cm} |p{3.5cm} |p{3.5cm} |p{3cm}|}
\hline\hline
\textbf{Approche} & \textbf{Type de recommandation} & \textbf{Type du contexte} & \textbf{Découverte des communautés} & \textbf{Technique utilisée} & \textbf{Source de données}\\
  \hline
   \hline
   Xin \textit{et al.} (2009) & recommandation des ressources & social & non & Multi-scale Continuous Conditional Random Fields & MovieLens \& Epinions\\
   \hline
   Konstas \textit{et al.} (2009) & recommandation des ressources & social & non & Random Walk with Restarts & "last.fm"\\
   \hline
   Silva \textit{et al.} (2010) & recommandation d'amis & social & oui & clustering & Oro-Aro\\
   \hline
   Ma \textit{et al.} (2011) & recommandation des films & social & non & matrices collaboratives de factorisation & Douban\\
   \hline
   Zheng \textit{et al.} (2010) & recommandation du comportement \& recommandation des futures actions & mobile & non & matrices collaboratives de factorisation & trajectoires GPS\\
   \hline
   Beach \textit{et al.} (2009) & recommendation d'amis & Mobile \& Social & non & filtrage collaboratif & données réelles\\
   \hline
   Hariri \textit{et al.} (2011) &  recommandation de ressources & mobile & non & classification & critiques des clients\\
   \hline
   Qiao \textit{et al.} (2011) &  recommandation d'amis & mobile \& social & oui & clustering & Renren\\
  \hline
\end{tabular}
\caption{Tableau récapitulatif des approches de personnalisation des systèmes de recommandation}
\label{SR}
\end{table}
\end{landscape}

\section{Conclusion}\label{conclusion2}
Dans ce chapitre, nous avons présenté un état de l'art sur la personnalisation de la RI et des systèmes de recommandation via le contexte des utilisateurs resumé dans les tableaux \ref{RI1}, \ref{RI2} et \ref{SR}. Nous avons réalisé, également, une étude comparative permettant de critiquer ces approches et de dégager leurs avantages et leurs inconvénients. Les constats majeurs sont: (\textit{i}) Aucune approche de RI contextuelle n'a trouvé une solution efficace pour réduire le nombre énorme de contextes pouvant être collectés; (\textit{ii}) Aucune approche de RI contextuelle ou de recommandation ne permet de prédire les intentions de recherche de l'utilisateur dans le cadre d'un environnement mobile; et (\textit{iii}) La majorité des approches de recommandation d'amis ne garantissent pas la dynamicité dans leurs représentations des liens entre les utilisateurs.

Dans le but de remédier à ces insuffisances, nous proposons dans le chapitre suivant une nouvelle apporche de RI contextuelle qui se base sur la prédiction des intérêts des utilisateurs dans le processus d'enrichissement de requêtes et l'extension de leurs cercles sociaux à travers le processus de recommandation d'amis.

\chapter{La prédiction des intérêts des utilisateurs pour la RI contextuelle et la recommandation d'amis}
%\markboth{Recherche d'information \& Systèmes de recommandation}{Notions de base}
\label{ch3} \vspace*{3cm}
\section{Introduction}

De nos jours, les appareils mobiles ont évolué pour offrir les meilleures fonctionnalités aux utilisateurs (des écrans en couleurs plus grands, une puissance de traitement accrue et plus rapide, et une connexion Internet à large bande).
Néanmoins, la majorité des sites web et des moteurs de recherche traditionnels sont conçus sur la base des ordinateurs de bureau.
Pour cette raison, l'expérience actuelle de la recherche mobile est loin d'être satisfaisante.
Les analystes des moteurs de recherche, conscients de ce problème, ont conçu des vues orientées mobiles pour fournir les mêmes services avec des interfaces plus petites.
Plusieurs approches basées sur la transformation du contenu informationnel ont été proposées afin de réduire le nombre de pages accessibles par le web mobile. Tels travaux portent seulement sur l'une des limites majeures du Web mobile, à savoir la taille de l'écran, en négligeant le problème des modes de saisie difficiles.
En effet, les utilisateurs se trouvent souvent fatigués et parfois obligés à abondonner devant l'effort énorme qu'ils doivent fournir pour saisir une seule requête. Ce qui nous ramène au problème des requêtes courtes, ambig\"{u}es et incompatibles avec les besoins réels des utilisateurs \cite{Sweeney2006}.
Par conséquent, ces derniers sont obligés de visiter un grand nombre de pages web inutiles et d'analyser des centaines de ressources afin de trouver leur besoin exact. Une telle recherche peut être acceptable avec des ordinateurs de bureau, mais devient une tâche extrêmement ardue dans un environnement mobile.

Ainsi, un nombre important des SRI se sont basés sur le méchanisme d'autocompletion qui permet d'ajouter des termes à la requête initiale de l'utilisateur pour la compléter et mieux répondre à ses besoins. Par ailleurs, une approche sémantique basée sur ce méchanisme peut présenter des difficultés à cause de l'absence de relations sémantiques entre les concepts (seules les relations syntaxiques sont considérées). Cette limite nous a donné l'idée de proposer une approche sémantique qui intégre la situation mobile de l'utilisateur dans la prédiction de ses intérêts.
En effet, dans un environnement mobile, les besoins informationnels de l'utilisateur peuvent changer à tout moment avec le changement de sa situation physique (cercle social, localisation, temps, etc.). Par exemple, considérons un utilisateur qui a lancé la requête "Mona Lisa". Avec une telle requête, ce n'est pas clair si cet utilisateur est intéréssé par le chef-d'{\oe}uvre de \emph{Leonardo Da Vinci} ou par le film de \emph{Julia Roberts}. Ainsi, sans comprendre l'intention de recherche de l'utilisateur, la plupart des approches de l'état de l'art optent à classer cette requête dans les deux catégories "art" et "film". Cependant, si on arrive à deviner que la localisation actuelle de cet utilisateur est "musée", nous pouvons supposer qu'il est intéréssé par l'"art". Inversement, si l'utilisateur est localisé dans une salle de cinema, c'est la catégorie "film" qui est sélectionnée pour décrire son intérêt.

D'un autre côté, les réseux sociaux se sont multipliés en termes de taille et du nombre de services fournis. Avec un tel développement, les systèmes de recommandation sociaux se sont imposés dans le paysage du web social, comme un support d'extension et de renforcement des relations sociales. Un réseau social peut être définit comme "\emph{un graphe d'individus connectés à travers leurs intérêts personnels ou professionnels, leurs localisations ou origines, leurs parcours éducatifs, etc.}". Plus particulièrement, un réseau social est une commmunauté d'individus connectés à l'aide d'un ou plusieurs types de relations, \emph{e.g.}, préférences, amitiés, localisations, affaires, etc. Ces réseaux sont utilisés dans certains systèmes de recommandation afin de personnaliser leurs ressources proposées aux différents utilisateurs, \emph{e.g.}, produits commer\c{c}iaux, films, extraits musicaux, amis, etc.
En effet, les relations dans les réseaux sociaux sont incomplètes et seule une partie du cercle social réel de l'utilisateur est modélisée à travers des connexions virtuelles. Par conséquent, dans le but de rassembler tels utilisateurs, la proposition d'un système de recommandation d'amis devient une exigence.

La majorité des approches existantes qui ont proposé des systèmes de recommandation se concentrent sur le but de recommander des ressources aux utilisateurs en considérant ou bien leurs contextes mobiles (localisation, temps, etc.), ou bien leurs contextes sociaux (intérêts, amis, etc.). Donc, l'idée est de trouver une manière de profiter, à la fois, des intérêts dynamiques des utilisateurs et de leurs localisations pour pouvoir sélectionner les nouvelles personnes pouvant être leurs futurs amis.

Pour résumer et à travers l'état de l'art présenté dans le chapitre précédent, nous avons souligné les principaux travaux ayant proposé des nouvelles approches de RI et de recommandation contextuelles dans le cadre d'un environnement mobile. En particulier, nous avons spécifié les méthodes qui ont été utilisées dans la représentation et l'exploitation de ce contexte.
Un ensemble de défis a été dégagé suite à cette étude: (\textit{i}) vu que les intérêts des utilisateurs sont multiples, hétérogènes et même contradictoires et doivent être compris vis-à-vis de leurs situations courantes, comment présenter une telle situation, \emph{i.e.}, choisir les dimenssions pouvant contribuer à sa construction? Comment utiliser cette situation pour activer les intérêts des utilisateurs et personnaliser leurs résultats de recherche? (\textit{ii}) comment choisir les informations à utiliser pour personnaliser le processus de recommandation d'amis à partir de l'ensemble de propriétés permettant de décrire un utilisateur donné?

\section{\textsc{Dbpedia}: Base de connaissance}
\begin{figure*}[htbp]
\centering
\epsfig{file=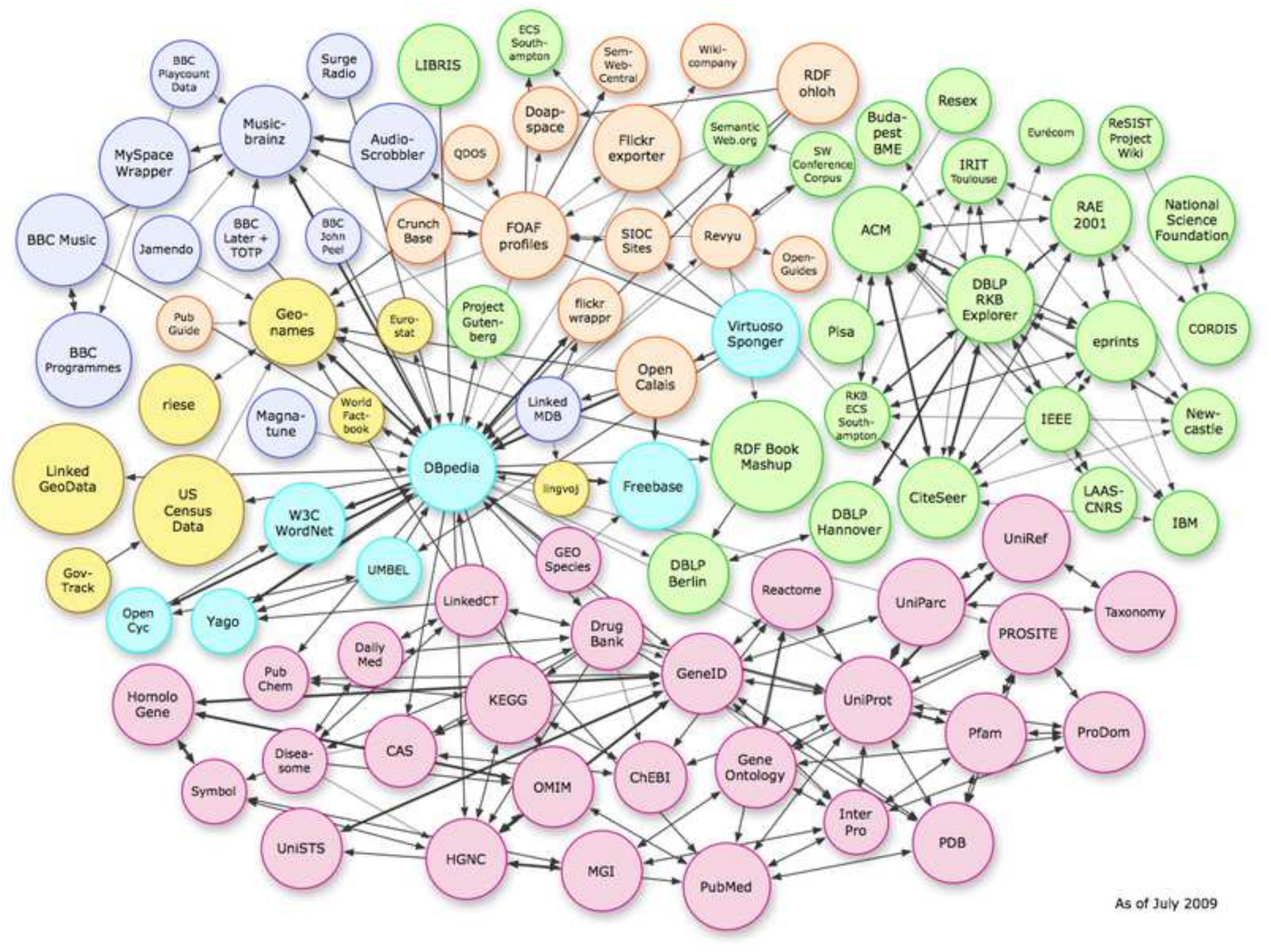, height=4in, width=6in}
\caption{Structure de Linked Open Data Cloud (2009): relation entre \textsc{Dbpedia} et le reste des datasets}
\label{LinkedData}
\end{figure*}

\begin{figure*}[htbp]
\centering
\epsfig{file=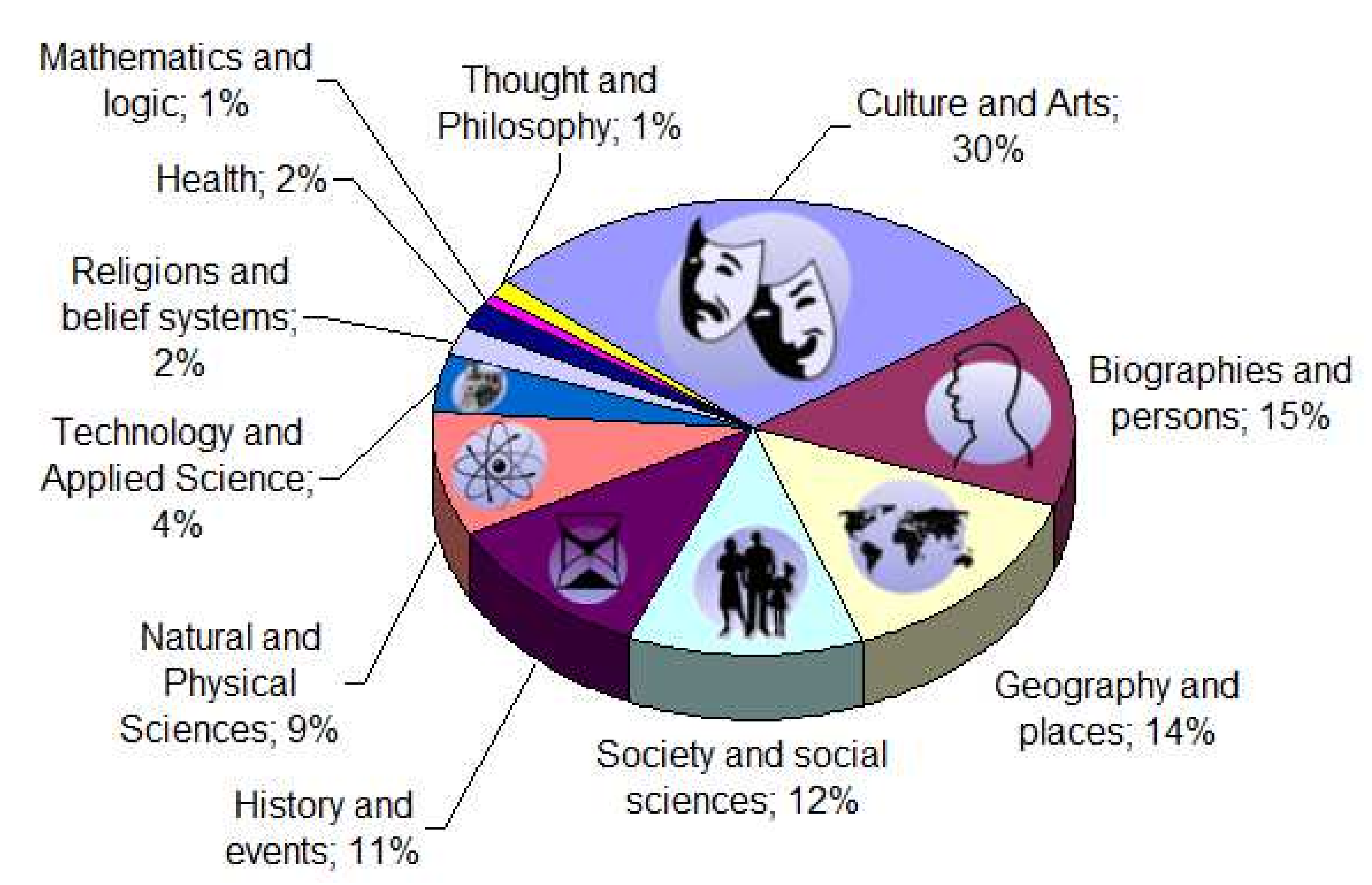, height=3in, width=4in}
\caption{Catégorisation du contenu de \textsc{Dbpedia} par thème}
\label{DbpediaDomains}
\end{figure*}

Depuis quelques années, Le World Wide Web a radicalement modifié la façon dont les internautes partagent leurs connaissances en supprimant les limites et les obstacles qui bloquent l'édition et l'accès aux données, dans le but de mondialiser l'espace d'information.
%Toutefois, le Web a évolué et s'est transformé d'un espace mondial d'information de documents liés entre eux à un espace reliant à la fois, les documents et les données.
A la base, cette évolution est un ensemble de méthodes permettant la publication et la connexion des données structurées sur le Web connu sous le nom de \textbf{Linked Data} \cite{Berners-Lee2009}. En effet, l'utilisation de Linked Data a conduit à l'extension du Web avec un espace mondial de données permettant la connexion de données appartenant à divers domaines tels que les personnes, les communautés en ligne, les livres, les publications scientifiques, les films, les émissions de télévision et de radio, les gènes, les essais cliniques, etc.
Dans ce contexte et vu la diversité et la taille des données contenues dans \textsc{Dbpedia}, ce dataset présente un hub permettant de lier les différentes sources de données des données liées \cite{Auer2007}. \textsc{Dbpedia} permet, comme le montre la figure \ref{LinkedData}, de relier plusieurs datasets \emph{e.g.}, Geonames, Yago, OpenCyc, etc. De plus, ce dataset vise à présenter les articles de \textbf{Wikipedia} d'une manière exhaustive afin de couvrir le maximum de concepts appartenant à plusieurs domaines, \emph{e.g.}, culture et art, géographie, sciences, etc. (\emph{c.f.}, Figure \ref{DbpediaDomains}).

Cet énorme volume de données est stocké à l'aide du modèle de représentation \textbf{RDF} (Resource Description Framework) avec une structure sous forme de triplets <sujet, prédicat, objet>, où le sujet et le prédicat sont des URIs et l'objet est un URI ou une simple chaîne de caractères. Ces données peuvent être consultées en utilisant des requêtes \textbf{SPARQL}.

Nous donnons dans le tableau \ref{tab:dbpedia} un exemple simple d'une requête SPARQL permettant de retourner l'abstract en langue Anglaise de la ressource \textsc{Dbpedia} ayant comme label "mobile\_computing".

\begin{table*}[!h]
\centering\scalebox{0.85}{
\begin{tabular}{|l|}
\hline
"SELECT ?abstract\\
FROM NAMED <http://dbpedia.org>\\
WHERE \{\\
<http://dbpedia.org/resource/mobile\_computing>\\
<http://dbpedia.org/property/abstract> ?abstract.\\
FILTER langMatches (lang(?abstract), 'en')\\
\}\\
\hline
\end{tabular}}
\caption{Un exemple d'une requête SPARQL}
\label{tab:dbpedia}
\end{table*}

\begin{figure*}[htbp]
\centering
\epsfig{file=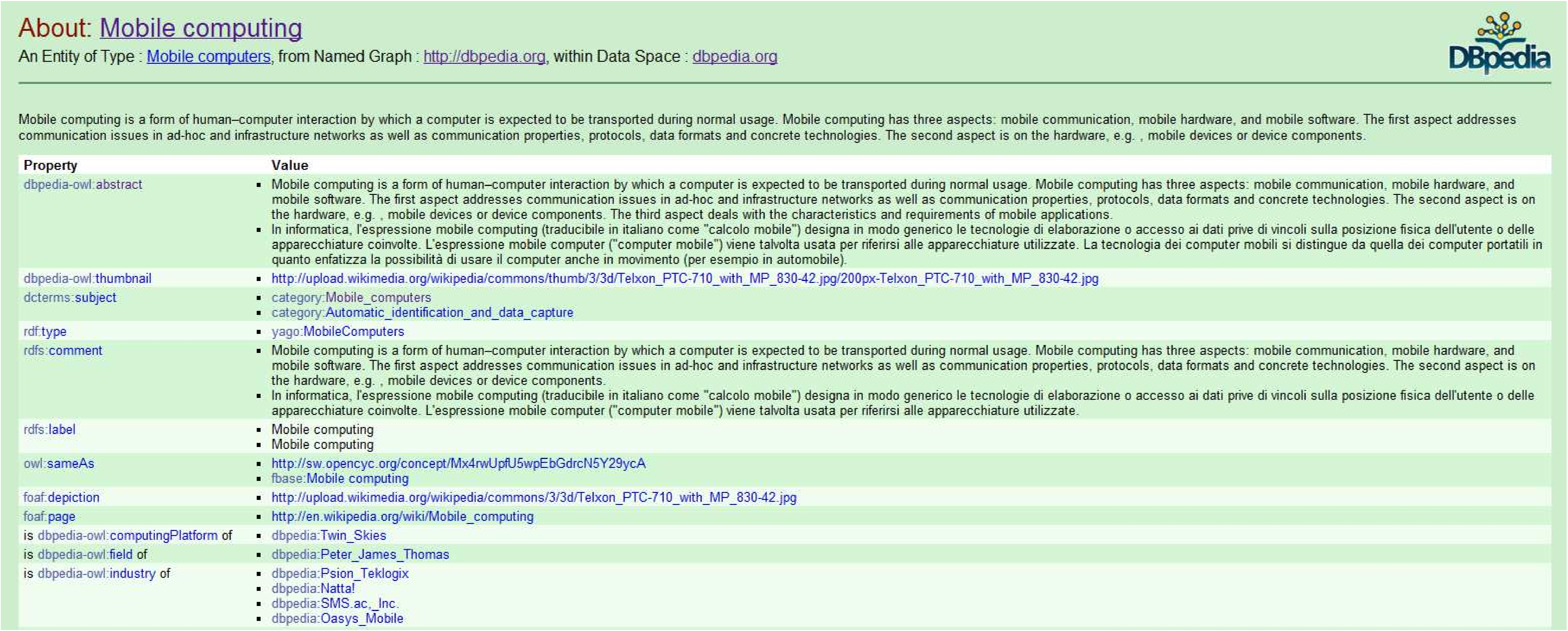, height=3.7in, width=6.2in}
\caption{Exemple d'une page \textsc{Dbpedia} décrivant le concept "mobile\_computing"}
\label{mobile_comp}
\end{figure*}

Dans notre travail, nous avons utilisé \textsc{Dbpedia} pour modéliser plusieurs types de données, \emph{e.g.}, les intérêts, les localisations, etc. En effet, pour chaque donné (chaîne de caractères), on cherche le concept dans \textsc{Dbpedia} ayant comme label cette chaîne. Un exemple d'une page \textsc{Dbpedia} est donné par la figure \ref{mobile_comp} énumérant quelques propriétés du concept "mobile\_computing": "dcterms:subject", "rdf:type", "rdfs:label", etc.

\section{\textsc{FOAF}: Friend Of A Friend}
\begin{figure*}[htbp]
\centering
\epsfig{file=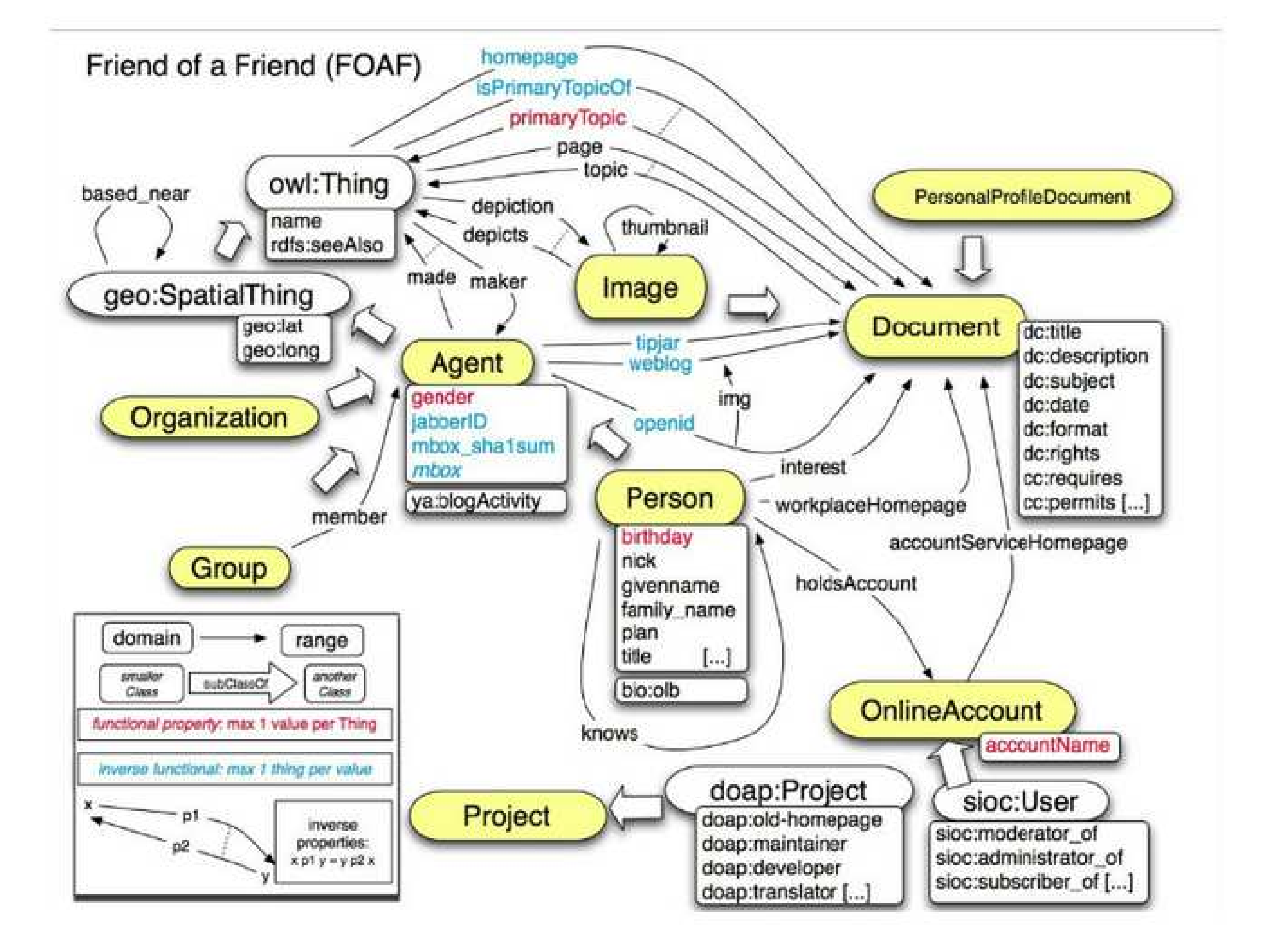, height=3.5in, width=5.5in}
\caption{Spécification de \textsc{FOAF}}
\label{foafspecification}
\end{figure*}

Friend-of-a-Friend est un modèle de structuration de données sur les personnes dont la spécification se retrouve dans la figure \ref{foafspecification}. Il permet de structurer des données sur les individus, de les partager et surtout de faire des liens entre ces données.

Ce projet a été lancé par Dan Brickley et Libby Miller et peut être utilisé dans la création des informations système qui supportent les communautés en ligne.
L'objectif de \textsc{FOAF} est de fournir aux machines une technique pour comprendre le contenu des pages personnelles. \textsc{FOAF} s'appuie, dans la représentation des données, sur deux technologies W3C (World Wild Web Consortium) et RDF (Resource Description Framework).

En effet, une étude a été faite en 2004 par \cite{Ding2005} sur les ontologies les plus publiées dans le web. Cette étude a montré que l'ontologie \textsc{FOAF} a occupé la deuxième position avec un nombre de documents publiés supérieur à 1.129.749 documents.

L'élément le plus important dans un document \textsc{FOAF} est le vocabulaire \textsc{FOAF}, identifié par l'URI \emph{http://xmlns.com/foaf/0.1/}. En effet, ce vocabulaire définit plusieurs classes, \emph{e.g.}, \emph{foaf:Person}, \emph{foaf:Document}, etc. et différentes propriétés \emph{foaf:name}, \emph{foaf:knows}, \emph{foaf:interest}, etc.

Nous avons utilisé cette ontologie pour décrire l'ensemble des utilisateurs appartenant au réseau social, dont nous nous sommes servis pour élaborer notre approche de recommandation d'amis.

\section{Description générale de l'approche de RI et de recommandation d'amis basée sur la situation mobile des utilisateurs}
Dans ce qui suit, nous allons décrire notre approche de prédiction d'intérêts basée sur la situation des utilisateurs pour la personnalisation des résultats de leurs recherches et la recommandation d'amis dans un environnement mobile. Cette approche est constituée de deux parties: (\textit{i}) \emph{Enrichissement de requêtes mobiles basé sur la prédiction des intérêts des utilisateurs}: la situation sémantique de l'utilisateur est générée par un mapping entre sa situation physique et les concepts sémantiques extraits à partir de \textsc{Dbpedia}. Par exemple, si l'utilisateur est localisé au niveau des coordonnées GPS " 36.851111, 10.226944 " et le temps est " Lundi 27 Février 11:00:00 2012 ", donc nous pouvons déduire qu'il a comme situation "musée, hiver, matin". Ensuite, la similarité entre cette situation et les situations précédentes est calculée pour sélectionner celle la plus adéquate au profil utilisateur et l'exploiter dans le processus de personnalisation; et (\textit{ii}) \emph{Recommandation dynamique d'amis basée sur la découverte des communautés}: Cette recommandation vise à étendre le cercle social de l'utilisateur. En effet, le réseau social décrit à l'aide de l'ontologie \textsc{Foaf} est analysé pour découvrir l'ensemble des communautés permettant de proposer à un utilisateur donné les personnes pouvant lui être intéressantes. L'architecture générale de cette approche est décrite par la Figure \ref{approachfig}.

Ces deux parties s'interagissent via les intérêts des utilisateurs. En effet, lorsqu'un nouvel intérêt est prédit selon la situation courante de l'utilisateur, cet intérêt est ajouté à sa description \textsc{Foaf} pour qu'il soit utilisé dans le processus de recommandation. Cette intéraction sera détaillée dans la section \ref{RI&SR}.

\subsection{Enrichissement de requêtes mobiles basé sur la prédiction des intérêts des utilisateurs: SA-IRI}\label{part1}
L'idée de base de l'approche d'enrichissement de requête, proposée dans cette section, répond au fait que les intérêts des utilisateurs sont multiples, dynamiques et même contradictoires et doivent être compris selon le contexte de chaque utilisateur.

Ainsi, nous proposons une nouvelle approche de RI basée sur la prédiction des intérêts des utilisateurs appelée \textbf{SA-IRI} (\textbf{Situation-Aware Information Retrieval based Interest}) \cite{BenSassi2012}. Cette approche permet de définir une représentation dynamique de la situation sémantique d'un utilisateur dans son environnement mobile. Cette situation sémantique est utilisée dans l'activation des différentes règles de classification, de telle sorte à proposer l'intérêt le plus approprié pour enrichir la requête de l'utilisateur.
\begin{exemple}
Supposons qu'un utilisateur a lancé la requête \emph{Mona Lisa}. Dans ce cas, ce n'est pas clair si cet utilisateur est intéréssé par le chef-d'{\oe}uvre de \emph{Leonardo Da Vinci} ou par le film de \emph{Julia Roberts}. Ainsi, sans comprendre l'intention de recherche de l'utilisateur, la plupart des approches existantes optent à classer cette requête dans les deux catégories "arts" et "films". Cependant, si on trouve que la localisation actuelle de l'utilisateur est "musée", c'est comme s'il est intéréssé par le domaine d'"arts". Inversement, si l'utilisateur est localisé dans une salle de cinema, c'est la catégorie "films" qui est proposée pour décrire son intérêt.
\end{exemple}
Par conséquent, cette approche fonctionne suivant trois étapes: (\textit{i}) \emph{Construction de la situation sémantique de l'utilisateur}: les informations sémantiques sont extraites à partir de la situation physique de l'utilisateur (localisation, temps et saison), dans le but de deviner l'intérêt derrière cette situation; (\textit{ii}) \emph{Prédiction des intérêts de l'utilisateur}: l'intérêt associé à la situation de l'utilisateur est prédit à l'aide de \textsc{Dbpedia}; et (\textit{iii}) \emph{Enrichissement de la requête de l'utilisateur}: la requête initiale de l'utilisateur est étendue avec l'intérêt déjà prédit. Ces différentes étapes sont décrites dans ce qui suit.

\subsubsection{Construction de la situation sémantique de l'utilisateur}\label{step1_1}
En se basant sur l'hypothèse que les intérêts des utilisateurs dépendent de la signification sémantique des entités physiques décrivant leurs environnements, nous proposons de modéliser leurs situations en concaténant les différentes dimensions sémantiques de leurs contextes spatial et temporel.
En particulier, les informations de localisation et de temps sont complexes et peuvent être modélisées suivant plusieurs degrés de granularité.
Dans la suite nous allons décrire notre modèle de représentation des informations spatio-temporelles.
\begin{enumerate}
  \item \textbf{Modélisation de la localisation}: la dimension liée à la localisation a généralement une grande importance dans la modélisation du contexte des utilisateurs surtout dans les applications mobiles \cite{Bellotti2008}. En effet, la localisation peut être décrite à travers plusieurs types de données: des coordonnées absolus (des coordonnées GPS), un nom (le Louvre, la Tour-Eiffel, etc.), un type (musée, plage, école, etc.), etc.
      Plusieurs outils en ligne permettent de faire la catégorisation des lieux géographiques, le cas pour \emph{Geonames}
      \footnote{~\url{http://www.geonames.org}}, qui donne le nom d'un emplacement à partir de son adresse en utilisant la technique de \textbf{géocodage inversé}.

      Dans ce mémoire, en se basant sur l'hypothèse que le nom de l'emplacement n'est pas très important pour déterminer l'intention de recherche de l'utilisateur, nous avons choisi de caractériser une localisation par son type.
      Par exemple, le fait que l'utilisateur est dans "le musée de Louvre" ou "le musée Grévin" ne touche pas à son intérêt (donné par la catégorie art), contrairement au fait qu'il est dans un musée. Par conséquent, deux lieux peuvent être regroupés sémantiquement via leur type commun même s'ils n'appartiennent pas au même emplacement physique.

  \item \textbf{Modélisation du temps}: les informations temporelles sont complexes et peuvent être modélisées avec différentes structures et suivant plusieurs types de granularité: fine (heures, minutes, etc.), moyenne (parties de la journée, jours, etc.) ou grande (mois, années, etc.). En effet, plusieurs approches ont travaillé sur cet aspect dans le but d'extraire les informations utiles à partir du contexte temporel des utilisateurs \cite{Artale2006} \cite{Berberich2010}. Dans ce mémoire, nous considérons que la dimension temps peut fortement influencer les activités des utilisateurs (des activités liées à leurs travaux, des activités de divertissement, etc.).

      Par exemple, supposons qu'un utilisateur a lancé la requête "sport" alors qu'il est dans une montagne. Ainsi, deux cas de figures peuvent être distingués:
      \begin{itemize}
  \item Si la saison est l'hiver, nous pouvons restreindre notre recherche sur les "sports d'hiver", \emph{e.g.}, le ski, le snowboard, etc.
  \item Sinon, pour le reste des saisons de l'année, nous pouvons admettre que l'utilisateur est intéréssé par d'autres types de sport, \emph{e.g.}, l'escalade.
      \end{itemize}
      Par conséquent, nous proposons de restreindre la modélisation du temps sur les catégories qui nous semblent intéréssantes et qui peuvent affecter les intentions de recherche des utilisateurs, \emph{e.g.}, matin et automne. Dans ce mémoire, nous présentons le temps par deux dimensions: la partie de la journée et la saison de l'année.
      \begin{itemize}
        \item \textbf{Partie de la journée}: selon le type d'activités pouvant être réalisées par un utilisateur durant la journée, nous pouvons distinguer trois principales parties: matin, midi et soir.
        \item \textbf{Saison}: cette dimension désigne une saison de l'année: automne, hiver, printemps ou été.
      \end{itemize}
\end{enumerate}
Ainsi, en combinant ces différentes dimensions nous définissons notre modèle qui présente la situation de l'utilisateur. Plus précisément, une situation $S$ est définie par:

\begin{definition}
\textsc{Situation sémantique}: une situation (ou situation sémantique) est décrite par un vecteur à trois dimensions: S= ($S_{l}$,$S_{s}$,$S_{t}$) où: $S_{l}$ (resp. $S_{s}$ et $S_{t}$) définit le type de localisation (resp. saison et partie de la journée).
\end{definition}

Par exemple, si l'utilisateur est localisé par les coordonnées GPS suivantes "35.1877778, 8.655" et le temps est "Samedi 29 Janvier 13:00:00 2012", alors nous pouvons décrire sa situation sémantique par le vecteur à trois dimensions suivant: $$S= (S_{l},S_{s},S_{t})= (montagne,hiver,midi)$$
Cette situation est utilisée dans l'étape 2 décrite dans la section \ref{step1_2} pour extraire l'intérêt de l'utilisateur.

\subsubsection{Prédiction des intérêts de l'utilisateur}\label{step1_2}
En se basant sur la nature de l'environnement mobile, les intérêts des utilisateurs ont tendance à changer selon leurs situations (localisation, temps, etc.). Par conséquent, nous avons gardé notre concentration sur les techniques dynamiques permettant d'associer les intérêts des utilisateurs à leurs situations actuelles. Ainsi, nous avons fusionné une technique de classification avec un traitement sémantique dans la formulation du processus de la prédiction des intérêts des utilisateurs. Ce processus peut être décrit à travers deux grandes étapes, à savoir: la génération des règles associatives construisant un classifieur, et la comparaison de la situation de l'utilisateur avec la prémisse de chacune de ces règles pour extraire l'intérêt de l'utilisateur (la conclusion de la règle choisi ou le concept extrait de \textsc{Dbpedia}).
Ainsi, ces deux étapes sont décrites comme suit:
\begin{enumerate}
  \item \textbf{Génération des règles de classification}: Notre solution consiste à utiliser la technique de classification dans la génération des relations de type situation/intérêt. Chaque règle contient comme prémisse la totalité ou une partie de la situation de l'utilisateur, tandis que la conclusion est constituée par son intérêt, \emph{e.g.}, S$_t$ $\wedge$, \ldots,$\wedge$ S$_l$ $\Rightarrow$ Intérêt$_{i}$. Lorsque la situation de l'utilisateur est construite, l'ensemble des règles générées est visité pour choisir celle ayant comme prémisse la situation la plus similaire à sa situation actuelle.

Toutefois, l'ensemble surdimensionné de règles pouvant être générées, constitue un obstacle sérieux devant la précision de la classification \cite{Gasmi2007}. Pour surmonter ce problème, nous avons utilisé un algorithme de classification basé sur les bases génériques de règles d'association \cite{Gasmi2006} \cite{Gasmi2009}.

Dans notre cas, la base générique $\mathcal{IGB}$ présente un intérêt particulier puisqu'elle donne le meilleur rapport à savoir compacité/informativité \cite{BenYahia2004} \cite{Hamrouni2006}, et son choix est justifié par les raisons suivantes:
\begin{itemize}
  \item \textbf{Génération des règles à large couverture}: Les règles d'association de la base $\mathcal{IGB}$ couvrent le maximum d'information utile. En effet, ces règles sont basées sur les itemsets fermés fréquents et ont des prémisses minimales (en terme de nombre d'items) présentées par les plus petits générateurs minimaux satisfaisant le seuil de confiance
  \item \textbf{Informativité}: La base $\mathcal{IGB}$ est dite sans perte d'information grâce à l'axiome de dérivation de \cite{Gasmi2009}:
  \begin{itemize}
    \item \emph{A1. Augmentation}: Si X$\Rightarrow$$Y$ $\in$ $\mathcal{IGB}$ alors X $\cup$ Z$\Rightarrow$$Y-\{$Z$\}$ $\in$ $\mathcal{AR}$\footnote{$\mathcal{AR}$ est considéré comme un ensemble de règles d'association.} ,Z $\subset$Y.
    \item \emph{A2. Decomposition}: Si X$\Rightarrow$$Y$ $\in$ $\mathcal{AR}$ alors X $\Rightarrow$$Z$ $\in$  $\mathcal{AR}$, Z $\subset$ Y $\wedge$ $\omega(XZ)$ = $XY$.
\end{itemize}
  \item \textbf{Compacité}: Des résultats obtenus en utilisant des benchmark, ont montré que la base $\mathcal{IGB}$ présente des bénéfices importants en terme de compacité des bases génériques.
\end{itemize}

Afin de simplifier les avantages précédemment cités, nous rappelons dans ce qui suit quelques définitions liées au domaine de la fouille de données:

\begin{definition}
\textsc{Itemset fermé fréquent}: \cite{Pasquier1999} Un itemset I $\subseteq  \mathcal{I}$ est dit fermé si I = $\omega(I)$. I est dit fréquent si son support relatif, Support(I)= $\frac {\mid\Psi(I)\mid}{\mid\mathcal{O}\mid}$, dépasse le seuil predéfini par l'utilisateur, noté \emph{minsup}.
\end{definition}

\begin{definition}
\textsc{Générateur minimal}: \cite{Hamrouni2008} Un itemset g $\subseteq \mathcal{I}$ est dit un générateur minimal d'un itemset fermé \emph{f}, si et seulement si $\omega(g) = \emph{f}$ et $\nexists   g_1\subset g$ avec $\omega(g_1) = \emph{f}$.
L'ensemble $\mathcal{G}_\emph{f}$ des générateurs minimaux de f est: $\mathcal{G}_\emph{f}$ = \{g $\subseteq \mathcal{I} \mid \omega(g) = \emph{f} \wedge \nexists g_1 \subset g$ avec $\omega(g_1) = \emph{f}$\}.\\
\end{definition}

\begin{definition}
\textsc{Base générique $\mathcal{IGB}$}: \cite{Gasmi2009} \emph{Admettons que $\mathcal{FCI_K}$ est l'ensemble des itemsets fermés fréquents et $\mathcal{G}_f$ est l'ensemble des générateurs minimaux de tous les itemsets fréquents contenus dans l'itemset fermé fréquent f.}
\begin{center}
$\mathcal{IGB} = \{\mathcal{R} = g_s \Rightarrow (\mathcal{I} - g_s)/\mathcal{I}\in \mathcal{FCI_K} \wedge(\mathcal{I} - g_s)\neq \varnothing \wedge g_s \in \mathcal{G}_f, f \in \mathcal{FCI_K} \wedge f \subseteq \mathcal{I} \wedge confidence(\mathcal{R})\geq minconf \wedge\nexists g' \subset g_s / confidence(g' \Rightarrow\mathcal{I}- g')\geq minconf\}$
\end{center}
\end{definition}

De ce fait, le choix du classifieur \textsc{GARC} \cite{Bouzouita2006} est expliqué par la taille indésirable des règles associatives de classification générées par les autres types de classifieurs. Ce choix est fait aussi grâce à une comparaison entre la technique de classification et celle du CBR, qui a prouvé les limites suivantes: (\textit{i}) Détection du bruit: Les expériences précédentes des utilisateurs sont utilisées sans être validées dans la situation actuelle; et (\textit{ii}) Détection des cas généraux: La maintenance de la base des cas devient de plus en plus difficile avec l'augmentation de sa taille, ce qui mène au problème de redondance des règles. Les résultats de cette comparaison sont détaillés dans le chapitre \ref{ch4}.

\item \textbf{Découverte de l'intérêt de l'utilisateur}: En se basant sur les règles de classification, générées précédemment, la situation actuelle de l'utilisateur est comparée à chacune des prémisses de ces règles afin de trouver une situation, qui maximise la relation de similarité. Deux cas de figures sont possibles:
    \begin{itemize}
      \item Une ou plusieurs règles contenant une situation similaire est extraite: dans ce qui suit, nous admettons que la similarité entre deux situations est donnée par:
          \begin{definition}
          Deux situations $S1=(S_{l1},S_{s1},S_{t1})$ et $S2=(S_{l2},S_{s2},S_{t2})$ sont similaires si et seulement si elles partagent au minimum deux dimensions en commun.
          \label{sim}
          \end{definition}
          Dans ce cas, si une seule règle est extraite alors sa conclusion est affectée automatiquement à l'intérêt de l'utilisateur. Sinon, si plusieurs règles sont extraites alors la conclusion de la règle ayant les plus grandes valeurs de support et de confiance est choisie pour définir l'intérêt de ce dernier.
      \item Aucune règle n'est extraite: dans ce cas, nous éffectuons un traitement sémantique permettant de découvrir l'intérêt de l'utilisateur.
      L'idée derrière ce traitement est de trouver les chevauchements entre sa requête et sa localisation.
      Plus précisément, la requête de l'utilisateur est mappée en concept par le biais de \textsc{Dbpedia}. Ainsi, chaque concept utilisé est identifié par la propriété \textbf{rdfs:label} (seuls les concepts ayant une description en triplet RDF dans \textsc{Dbpedia}). Un tel concept définit le sujet, \emph{i.e.}, \textbf{rdfs:label}, et l'objet est une chaîne de caractères qui présente exactement la requête émise par l'utilisateur. Deux cas sont distingués dans ce traitement:
      \begin{enumerate}
        \item La requête de l'utilisateur définit une catégorie dans \textsc{Dbpedia} par exemple "sport", "art", etc. Dans ce cas, nous extrayons toutes les sous-catégories ayant une relation sémantique avec le type de la localisation de l'utilisateur (formant lui même un concept dans \textsc{Dbpedia}). Cette relation est définie par la propriété \textbf{skos:broader}.
            \begin{exemple}
            Prenons le cas d'un utilisateur dans la plage, qui a lancé la requête "sport". Dans ce contexte, les sous-catégories retournées par notre système sont liées aux activités sportives de plage, \emph{e.g.}, beach soccer, beach polo, rowing, beach volleyball, etc. Ces sous-catégories sont les condidats pour présenter l'intérêt de cet utilisateur.
            \end{exemple}
        \item La requête de l'utilisateur définit une sous-catégorie dans \textsc{Dbpedia} par exemple "le titre d'un film", "le nom d'une peinture", etc. Dans ce cas, l'idée est de désambiguïser cette requête. Par conséquent, nous nous intéressons aux super-classes ayant une relation avec le type de la localisation de l'utilisateur, \emph{e.g.}, peinture, film, etc. Cette relation est définie par la propriété \textbf{dcterms:subject}.
            \begin{exemple}
            Supposons qu'un utilisateur a lancé la requête "Mona Lisa". Ce n'est pas évident s'il soit intéressé par le chef-d'{\oe}uvre de \emph{Leonardo Da Vinci} ou par le film de \emph{Julia Roberts}. Dans ce cas, en admettant qu'il est localisé dans une musée, nous devinons qu'il est intéressé par l'art.
            \end{exemple}
      \end{enumerate}
      Dans le cas où plusieurs concepts sont extraits, le concept le plus fréquent (en terme de nombre d'apparition) est retenu comme le candidat le plus probable pour modéliser l'intérêt de l'utilisateur.

      Finalement, après l'identification de l'intérêt de l'utilisateur, une nouvelle transaction ayant la forme (situation, intérêt) est ajoutée à la base d'apprentissage. Cette mise à jour va permettre d'augmenter le nombre de règles extraites dans la prochaine exécution du système.
    \end{itemize}
\end{enumerate}

\subsubsection{Enrichissement de la requête de l'utilisateur}\label{step1_3}
Face au grand volume d'information disponible sur le web et des contraintes des appareils mobiles, \emph{i.e.}, la taille de la zone d'affichage, la difficulté de saisie des requêtes, etc. ces dernières sont plus courtes et plus ambig\"{u}es. Pour répondre à ces limites nous avons choisi d'enrichir la requête initiale de l'utilisateur avec l'un de ses intérêts.

Dans ce mémoire, l'enrichissement de requête consiste à rajouter à la requête initiale de l'utilisateur l'intérêt extrait dans l'étape précédente décrite dans la section \ref{step1_2}. Ainsi, ce processus se limite à la concaténation des deux chaînes de caractères, \emph{i.e.}, les deux cha\^{\i}nes de caractères définissant la requête de l'utilisateur et l'intérêt prédit.

Dans la suite, nous présentons l'algorithme \textit{QueryEnrichment} décrit par le pseudo-code de l'algorithme \ref{algoSAIRI}, qui énumère les différentes étapes du processus d'enrichissement de requêtes. Les notations utilisées dans cet algorithme sont résumées dans le tableau \ref{tab:notations1}.

\begin{algorithm}[!h]
  {
     \Donnees{
 Q: Une nouvelle requête.
   }
     \Res{$Q_{e}$: Une requête enrichie}
 \Deb{
            S= \textsc{situationConstruction()}\;
            \{r\}= \textsc{situationClassification(S)}\;
            \Si {$\{r\}\neq\emptyset$}
                  {\Si {\{r\}.size=1}
                  {I= r.conclusion\;}
                  \Sinon{I=\textsc{mostFrequentClassExtraction(r)}\;
                  }
                  }
                  \Sinon{I=\textsc{DBpediaInterestExtraction(Q,$S_{l}$)}\;
                  $LB= LB \bigcup \{(S,I)\}$\;
                  }
                  \Retour $Q_{e}= Q\bigcup I$\;
    }
 \caption{\textsc{\emph{QueryEnrichment}}}
  \label{algoSAIRI}}
\end{algorithm}

Le processus d'enrichissement des requêtes est décrit dans l'algorithme \textit{QueryEnrichment}, il prend en input la requête initiale de l'utilisateur Q, \emph{i.e.}, un mot et retourne en output une requête enrichie $Q_{e}$= ($Q \bigcup I$). Les principales étapes de cet algorithme sont les suivantes:
il commence par construire la situation de l'utilisateur décrite par le vecteur de trois dimensions ($S_{l}$,$S_{s}$,$S_{t}$) (\emph{c.f.} ligne 2). Ensuite, dans la ligne 3, un processus de classification est appliqué afin de générer un ensemble de règles d'association à travers la fonction \textsc{situationClassification}. L'étape suivante consiste à analyser cet ensemble pour extraire la conclusion de la règle ayant la situation la plus similaire à celle précédemment construite. Cette conclusion est utilisée pour enrichir la requête initiale (\emph{c.f.} ligne 8). Dans le cas contraire, si aucune requête n'est extraite, un nouvel intérêt est prédit avec la fonction \textsc{DBpediaInterestExtraction} et la base de transaction est enrichie avec une nouvelle transaction T=(S,I) (\emph{c.f.} ligne 11). Une fois identifié, l'intérêt de l'utilisateur est utilisé dans l'enrichissement de sa requête Q (\emph{c.f.} ligne 12).

\begin{table*}[!h]
\centering
\begin{tabular}{|l|}
\hline

  Q : la requête initiale de l'utilisateur\\
  S : la situation sémantique de l'utilisateur extraite de \textsc{DBPEDIA} de la forme ($S_{l}$,$S_{s}$,$S_{t}$)\\
  r : l'ensemble de règles de classification générées\\
  I : l'intérêt de l'utilisateur\\
  LB : la base d'apprentissage contenant des transactions de la forme (S,I)\\
  $Q_{e}$ : la requête enrichie\\
\hline
\end{tabular}
\caption{Notations utilisées dans l'algorithme \textit{QueryEnrichment}}
\label{tab:notations1}
\end{table*}

\subsection{Exemple illustratif du processus d'enrichissement de requêtes}\label{exemple1}
Nous présentons dans cette section un exemple illustraif de notre approche d'enrichissement de requêtes \textsc{SA-IRI}.

Supposons qu'un utilisateur u a lancé la requête \emph{Mona Lisa}. Avec une telle requête ambig\"{u}e, ce n'est pas clair si u est intéréssé par le chef-d'{\oe}uvre de \emph{Leonardo Da Vinci} ou par le film de \emph{Julia Roberts}.
Ainsi, notre approche commence par la construction de la situation de cet utilisateur à partir de sa localisation physique (48.8606349, 2.3375548) et du moment du lancement de la requête Q (dimanche 18 Mars 18:05:00 2012), pour donner:
S= ($S_{l}$,$S_{s}$,$S_{t}$)= (mus\'{e}e,printemps,soir)

Ensuite, en se basant sur le contexte formel décrit par la table \ref{formalContext}, où:
 \begin{itemize}
 \item $\mathcal{O}$ est l'ensemble de transactions: $\mathcal{O}$=\{1,\ldots, 5\}
 \item $\mathcal{I}$ est l'ensemble des attributs \{attribut1, attribut2, attribut3, Classe\}, où:
  \begin{itemize}
  \item attribut1 $\in$ \{"automne", "hiver", "printemps", "\'{e}t\'{e}"\}
  \item attribut2 $\in$ \{"matin", "midi", "soir"\}
  \item attribut3 est le type de la localisation de l'utilisateur
  \item Classe présente l'intérêt de l'utilisateur.
  \end{itemize}
  \item $\mathcal{R}$ est la relation qui lie l'intérêt de l'utilisateur à sa situation actuelle.
 \end{itemize}

\begin{table*}[!htbp]
\begin{center}\scalebox{0.66}{
\begin{tabular} {|c|c|c|c|c|c|c|c|c|c|c|c|c|c|c|c|} \hline
 &\multicolumn{4}{c|}{\textbf{Attribut1}}&\multicolumn{3}{c|}{\textbf{Attribut2}}&\multicolumn{5}{c|}{\textbf{Attribut3}}&\multicolumn{1}{c|}{\textbf{Classe}}\\
\cline{2-14}
& \textsl{$automne$}& \textsl{$hiver$}& \textsl{$printemps$}& \textsl{$\acute{e}t\acute{e}$}& \textsl{$matin$} & \textsl{$midi$}& \textsl{$soir$}& \textsl{$mus\acute{e}e$}& \textsl{$th\acute{e}atre$}& \textsl{$plage$} & \textsl{$montagne$}& \textsl{$centre\_commercial$}& \textsl{$Int\acute{e}r\hat{e}t$}\\
\hline
\textsl{$\textbf{1}$} & $\times$ & & & & & $\times$ & & $\times$ & & & & & art\\
\hline
\textsl{$\textbf{2}$} & & & $\times$ & & & & $\times$ & & $\times$& & & & art \\
\hline
\textsl{$\textbf{3}$} & & & &$\times$ & $\times$ & & &  & &$\times$ & & & surf\\
\hline
\textsl{$\textbf{4}$} & & $\times$& & & $\times$ & & & & & & $\times$ & & camping\\
\hline
\textsl{$\textbf{5}$} & & & &$\times$ & & $\times$ & & & & & & $\times$ & shopping\\
\hline
\end{tabular}}
\end{center}
\caption{Un exemple d'un contexte d'extraction $\mathcal{K}$}
\label{formalContext}
\end{table*}

Nous appliquons la technique de la classification associative générique dérivant les règles de classification suivantes:
\begin{itemize}
\item $R_1$: montagne, \'{e}t\'{e} $\Rightarrow$ camping
\item $R_2$: plage, \'{e}t\'{e} $\Rightarrow$ surf
\item $R_3$: soir, \'{e}t\'{e} $\Rightarrow$ shopping
\item $R_4$: printemps, soir $\Rightarrow$ art
\item $R_5$: midi $\Rightarrow$ art
\end{itemize}

La situation sémantique précédemment construite est comparée à chacune des prémisses de ces règles. La règle $R_4$ est extraite en appliquant la définition \ref{sim}.
Finalement, le concept extrait de la règle $R_4$ est utilisé pour enrichir la requête Q de u et donné $Q_{e}$=Mona Lisa art.

\subsection{Recommandation dynamique d'amis basée sur la découverte des communautés}\label{part2}
La prolifération de la taille des données disponibles dans les réseaux sociaux a motivé plusieurs chercheurs pour multiplier leurs travaux dans ce domaine,  \emph{i.e.}, la découverte des communautés, la recommandation des ressources, l'extension des communautés, etc. \cite{Yan2008}.
Plus particulièrement, dans l'environnement mobile, différentes contraintes s'ajoutent dans la réalisation de ces travaux: (\textit{i}) La dynamique des relations sociales entre les utilisateurs: les systèmes de recommandation sont plus concernés par l'instabilité des cercles sociaux, de la localisation et des intérêts des utilisateurs; et (\textit{ii}) La limitation des relations sociales des utilisateurs: dans les cas réels, les réseaux sociaux sont incomplets. Ainsi, le rôle des systèmes de recommandation est d'étendre les relations entre les utilisateurs afin d'enrichir leurs cercles sociaux.

Par conséquent, nous introduisons dans ce mémoire une nouvelle approche de recommandation dynamique d'amis basée sur la découverte des communautés selon les intérêts des utilisateurs, leurs relations sociales et leurs localisations \cite{BenSassi2013}. Cette approche combine la technique de la marche aléatoire "Random Walk" avec les données du web social, et opère en deux étapes: (\textit{i}) Foaf-A-Walk pour la découverte des communautés: cette étape combine la technique de la marche aléatoire et la modélisation de l'ontologie \textsc{Foaf} (amis, intérêts et localisation) dans la division du réseau social en parties connectées définissant les communautés d'utilisateurs; et (\textit{ii}) Recommandation d'amis: Les membres partageant avec l'utilisateur la même localisation et les mêmes intérêts sont proposés à ce dernier comme nouveaux amis.
Ces étapes sont décrites dans ce qui suit.

\subsubsection{Foaf-A-Walk pour la découverte des communautés}\label{step2_1}
Afin d'améliorer dynamiquement les communautés sociales, nous avons sélectionné les données intéressantes des utilisateurs, \emph{i.e.}, les données pouvant jouer un rôle dans le processus de recommandation d'amis. Ainsi, nous exploitons l'ontologie \textsc{foaf} permettant de décrire les personnes, leurs activités, leurs relations d'amitié et leurs objets. Par conséquent, pour décrire un utilisateur donné nous utilisons trois types de propriétés \textsc{foaf}:
\begin{itemize}
  \item L'ensemble d'amis de l'utilisateur: chaque entrée de cet ensemble est une relation d'amitié exprimée dans le réseau social à l'aide de la propriété \emph{foaf:knows}. Cette propriété est celle la plus utilisée dans l'ontologie \textsc{foaf}. En effet, elle permet de donner explicitement l'ensemble de personnes en relations, \emph{i.e.}, collègues, amis de classe, amis proches, etc.
  \item L'ensemble d'intérêts de l'utilisateur: les intérêts des utilisateurs sont décrits avec la propriété \emph{foaf:interest}. Cette propriété est utilisée dans la découverte des communautés rassemblant les personnes ayant les mêmes intérêts.
  \item La localisation de l'utilisateur: la propriété \emph{foaf:based\_near} est utilisée dans la description de la localisation courante de l'utilisateur et pour le clustering du réseau social en communautés. Dans ce mémoire, chaque communauté regroupe les personnes localisées dans la même ville.
\end{itemize}

Nous nous sommes appuyés sur les travaux de Ding et al. \cite{Ding2005} dans la sélection des propriétés \textsc{foaf} que nous avons utilisées dans le processus de recommandation d'amis. En particulier, en se basant sur les relations qui relient les personnes et leurs intérêts, le calcul des communautés devient plus évident, \emph{i.e.}, la découverte de nouvelles communautés ou l'intégration des communautés anciennes. Par ailleurs, en s'appuyant sur l'hypothèse de l'évolution du web social et de la localisation des utilisateurs, nous pouvons étudier leurs affiliations dynamiques aux communautés.
%Ainsi, dans l'étape suivante nous allons étudier ces deux aspects.

Dans le reste du processus de recommandation, détaillé dans le reste de ce mémoire, nous considérons que les réseaux sociaux sont décrits par une présentation 3-D, où chaque dimension est explorée séparamment, \emph{i.e.}, relation d'amitié, relation géographique et relation par domaine d'intérêt.

La structure en communautés est proposée afin de grouper les personnes ayant la même description, \emph{e.g.}, intérêts, localisations, etc. En effet, les membres d'un réseau social; amis ou non; sont encouragés à partager plus de ressources et à accepter de nouveaux amis dans le but d'élargir leurs cercles sociaux. En se basant sur ce processus, la chance de ces membres de devenir amis augmente et la dynamique de la structure du web social est garantie.

Ainsi, notre idée est de catégoriser les utilisateurs, selon leurs intérêts et leurs localisations courantes, dans un ensemble de communautés étiquetées chacune par un couple <intérêt, localisation>. En effet, Foaf-A-Walk exploite la technique de la marche aléatoire \cite{Lovasz1996} \cite{Randall2006} ayant comme avantages: (\textit{i}) Peut capturer, rapidement, la structure des réseaux sociaux; (\textit{ii}) Peut calculer les communautés d'une manière efficace; et (\textit{iii}) Peut être appliquée dans les algorithmes de classification ascendante pour détecter la structure des communautés.

Dans chaque étape du processus de la marche aléatoire le "promeneur" (ou walker) est un utilisateur qui se déplace d'un noeud à un autre d'une façon aléatoire et uniforme, parmi l'ensemble des utilisateurs auquels il est connecté. Plus précisement, afin d'appliquer cette technique dans notre travail, nous avons utilisé l'algorithme \emph{Walktrap} \cite{Pons2004} basé sur le calcul des distances entre les noeuds du réseau. La formule de calcul de distance entre deux personnes est donnée par l'équation \ref{eq:distance}.

\begin{equation}\label{eq:distance}
    r_{ij}=\sqrt[2]{\sum_{k=1}^n \frac{(P^{t}_{ik}-P^{t}_{jk})^{2}}{d(k)}}
\end{equation}

Où: $P^{t}_{ij}$ (resp. $P^{t}_{jk}$) est la probabilité d'aller d'un utilisateur $i$ à un utilisateur $j$ (resp. aller d'un utilisateur $j$ à un utilisateur $k$) en $t$ étapes, et $d(k)$ le degré de l'utilisateur $k$ (le nombre de personnes connectées à $k$).

Dans la pratique; les réseaux complexes réels; cet algorithme est de complexité $O(n^{2}\log n)$, où: $n$ est le nombre des utilisateurs. Plus particulièrement, il donne les meilleurs résultats, en termes de qualité et de performance, par rapport aux autres algorithmes de découverte des communautés, \emph{e.g.}, l'algorithme utilisé dans \cite{Girvan2002}. Cette comparaison est détaillée dans le chapitre \ref{ch4}, et montre les avantages de notre approche Foaf-A-Walk vu son efficacité côté qualité des résultats et rapidité d'exécution, contrairement à l'approche GN.

Ainsi, nous appliquons cet algorithme deux fois successives. Nous commen\c{c}ons par considérer la propriété \emph{foaf:based\_near} pour définir le type de relations entre les utilisateurs et nous appliquons l'algorithme \emph{Walktrap} pour découvrir les communautés basées sur leurs localisations. Ensuite, en se basant sur ces communautés et en s'appuyant sur la propriété \emph{foaf:interest}, nous calculons la structure des sous-communautés où chacune est étiquetée par un couple <intérêt, localisation>.

\subsubsection{Recommandation d'amis}\label{step2_2}
Vu la saturation et la multiplication des relations dans les réseaux sociaux existants, plusieurs relations d'amitié de la vie réelle ont été négligées.
Toutefois, l'intérêt derrière la recommandation d'amis est d'enrichir ces réseaux et d'augmenter l'intéraction entre leurs membres, \emph{i.e.}, partage de ressources, échange de messages, etc.

Ainsi, en se basant sur les données extraites durant l'étape 1, les communautés calculées dans la même étape sont filtrées pour garder seulement celles étiquetées par la localisation de l'utilisateur et par l'un de ses intérêts. Une fois identifiées, les relations d'amitié entre les membres de ces communautés sont analysées. Dans le cas où aucune relation de type \emph{foaf:knows} n'existe entre l'utilisateur en question et un membre de ces communautés, ce dernier est proposé comme ami à l'utilisateur, en s'appuyant sur notre hypothèse qui dit que \textit{"La probabilité que deux personnes ayant les mêmes localisation et intérêt pour devenir amis est largement élevée"}.

Dans la suite, nous présentons l'algorithme \textit{FriendRecommendation} décrit dans le pseudo-code de l'algorithme \ref{algoRecommendation}. Cet algorithme montre les différentes phases du processus de la recommandation d'amis, où les notations utilisées sont résumées dans le tableau \ref{tab:notations2}.

\begin{algorithm}
  {
    \Donnees{$\mathcal{S}$: un réseau social.}
    \Res{$\mathcal{F}$: un ensemble d'amis recommandés.}
 \Deb{
          \PourCh {$U_{i}\in \mathcal{S}$}
            {
            ($F_{Ui},I_{Ui},L_{Ui}$):= \textsc{userDataExtraction $(U_{i})$}\;
            }
            \{$C_{L1},C_{L2},..,C_{Lp}$\}:= \textsc{Walktrap($\mathcal{S}$)}\;
             \PourCh {$L_{j}; 1 \leq j \leq p$}
            {
            $\mathcal{C}$:= \textsc{Walktrap($C_{Lj}$)}\; //\textit{$\mathcal{C}$:= \{$C^{I1}_{Lj},C^{I2}_{Lj},..,C^{Iq}_{Lj}$\}}
            }
            $\mathcal{C}_{Li}$:=\textsc{userAppartenance($U_{i}$,$\mathcal{C}$)}\;
            //\textit{$\mathcal{C}_{LUi}$:= $\{C^{IU1}_{LUi},C^{IU2}_{LUi},..,C^{IUk}_{LUi}\}$}\\
             \PourCh {$I_{h}; 1 \leq h \leq k$}
            {
            $\mathcal{F}$:=$\mathcal{F}$ $\bigcup$ \textsc{friends($U_{i}$,$C^{Ih}_{Li}$)}\;
            }
            \Retour $\mathcal{F}$\;
             }
 \caption{\textsc{\emph{FriendRecommendation}}}
  \label{algoRecommendation}
  }
\end{algorithm}

L'algorithme \textit{FriendRecommendation} prend en entrée le réseau social $\mathcal{S}$ et donne en sortie l'ensemble d'amis suggéré à l'utilisateur.
Ainsi, il commence par l'extraction des données \textsc{foaf} des utilisateurs décrivant le réseau social $\mathcal{S}$ (\emph{c.f.} ligne 3). Ensuite, dans la ligne 4, l'algorithme \textsc{Walktrap} est appliqué afin de diviser le réseau en communautés basées sur les différentes localisations des utilisateurs, \emph{i.e.}, en utilisant la propriété \emph{foaf:based\_near}. Le même algorithme est appliqué, pour chaque communauté extraite, pour dégager des sous-communautés basées sur les différents intérêts des utilisateurs, \emph{i.e.}, en utilisant la propriété \emph{foaf:interest} (\emph{c.f.} ligne 6). La fonction \textsc{userAppartenance} est utilisée pour déterminer les communautés aux quelles apartient l'utilisateur. Finalement, les membres appartenant à ces communautés sont extraits à l'aide de la fonction \textsc{friends} et l'ensemble $\mathcal{F}$ est suggéré à l'utilisateur contenant un ensemble de nouveaux amis (\emph{c.f.} ligne 12).

\begin{table*}[!h]
\centering
\begin{tabular}{|l|}
\hline
$\mathcal{S}$: un réseau social\\
$\mathcal{F}$: l'ensemble  d'amis recommandés\\
$U_{i}$: utilisateur i\\
$F_{Ui}$: l'ensemble d'amis d'un utilisateur i\\
$I_{Ui}$: l'ensemble d'intérêts d'un utilisateur i\\
$L_{Ui}$: la localisation d'un utilisateur i\\
$C_{Lj}$: une communauté d'utilisateurs basée sur la localisation Lj\\
$C^{Ik}_{Lj}$: une communauté d'utilisateurs basée sur la localisation Lj et l'intérêt Ik\\
\hline
\end{tabular}
\caption{Notations utilisées dans l'algorithme \textit{FriendRecommendation}}
\label{tab:notations2}
\end{table*}

\subsection{Exemple illustratif du processus de recommandation d'amis}\label{exemple2}
\begin{figure*}
\centering
\epsfig{file=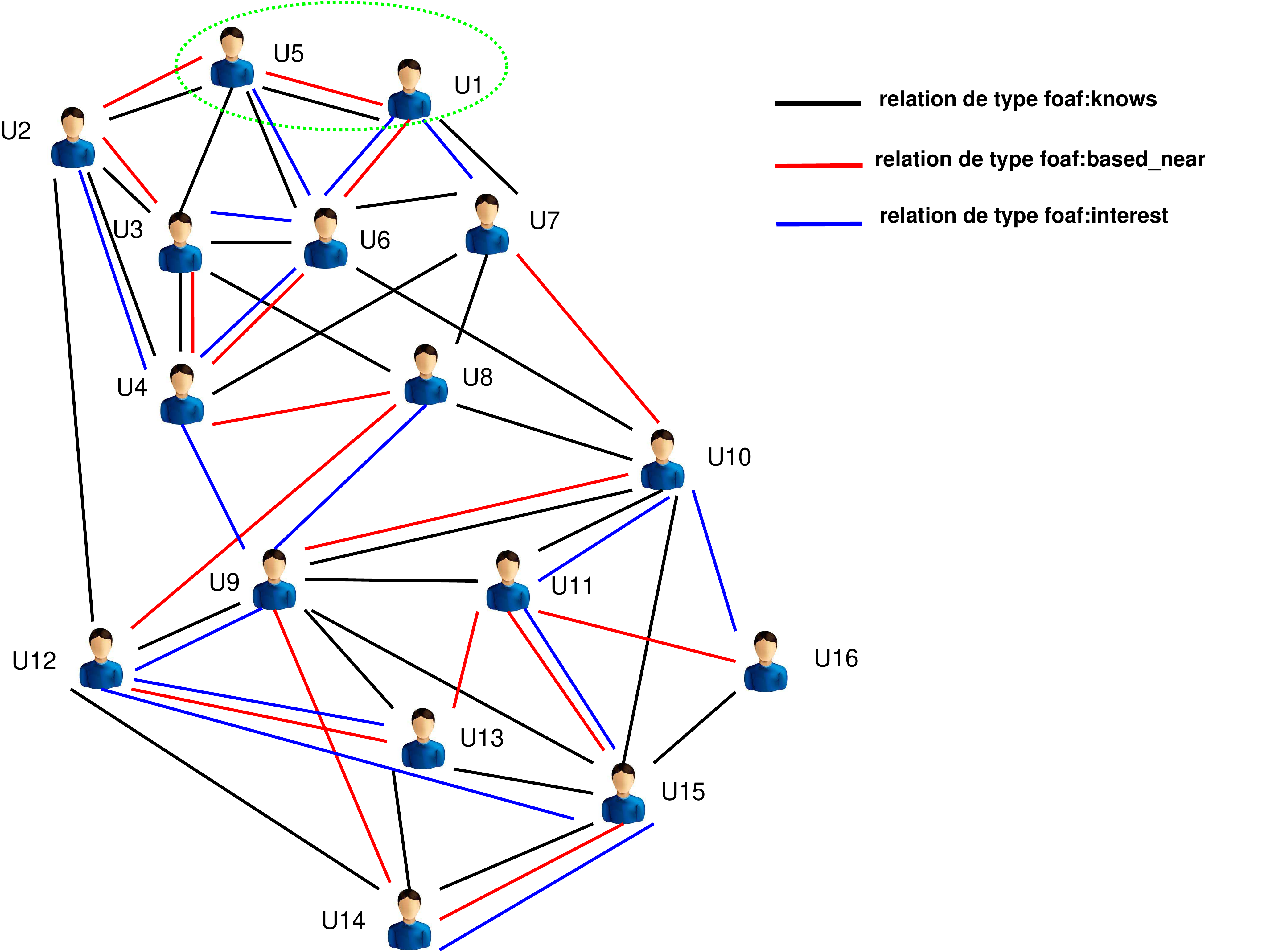, height=5in, width=6in}
\caption{Réseau social S: les liens de type \emph{foaf:knows}, \emph{foaf:interest} et \emph{foaf:based\_near}}
\label{network}
\end{figure*}

Considérons le réseau social S donné par la figure \ref{network} où chaque noeud est décrit par un ensemble de propriétés \textsc{foaf}, \emph{e.g.}, nom, prénom, profession, préférences, etc. Les liens dans ce graphe présentent des relations directes entre les utilisateurs. Ainsi, un trait noir (resp. rouge et bleu) entre deux noeuds montre que les deux utilisateurs sont amis (resp. partagent la même localisation et le même intérêt). Dans cet exemple, nous visons à extraire l'ensemble de personnes pouvant être des futurs amis pour U8.

Notre approche consiste, en un premier lieu, à extraire pour chaque utilisateur trois types d'informations: ses amis, ses intérêts et sa localisation.
Ensuite, nous appliquons l'algorithme \emph{Walktrap} sur le réseau social S en ne considérant que les liens de types \emph{foaf:based\_near}. Par conséquent, nous obtenons quatre communautés:
\begin{enumerate}
  \item $C_{L1}$=$\{U1,U2,U3,U4,U5,U6\}$
  \item $C_{L2}$=$\{U7,U9,U10\}$
  \item $C_{L3}$=$\{U8,U12,U13\}$
  \item $C_{L4}$=$\{U11,U14,U15,U16\}$
\end{enumerate}

En se basant sur l'ensemble des communautés déjà calculé, nous appliquons le même algorithme par rapport à la propriété \emph{foaf:interest}. Nous obtenons comme résultat:
\begin{enumerate}
  \item $C^{I1}_{L1}$=$\{U1,U5\}$. Cette communauté est montrée dans la Figure \ref{network} avec le cercle hachuré en vert.
  \item $C^{I2}_{L1}$=$\{U2,U3\}$
  \item $C^{I3}_{L1}$=$\{U4,U6\}$
  \item $C^{I4}_{L2}$=$\{U7,U9,U10\}$
  \item $C^{I5}_{L3}$=$\{U8,U12,U13\}$
  \item $C^{I6}_{L4}$=$\{U11,U16\}$
  \item $C^{I7}_{L4}$=$\{U14,U15\}$
\end{enumerate}

La sous-communauté $C^{I5}_{L3}$ est la seule communauté qui contient l'utilisateur U8. Ainsi, U12 et U13 sont proposés à ce dernier comme amis condidats puisque aucune relation de type \emph{foaf:knows} n'existe entre eux et U8.

\subsection{De la RI contextuelle à la recommandation d'amis dans un environnement mobile}\label{RI&SR}
\begin{landscape}
\begin{figure*}
\centering
\epsfig{file=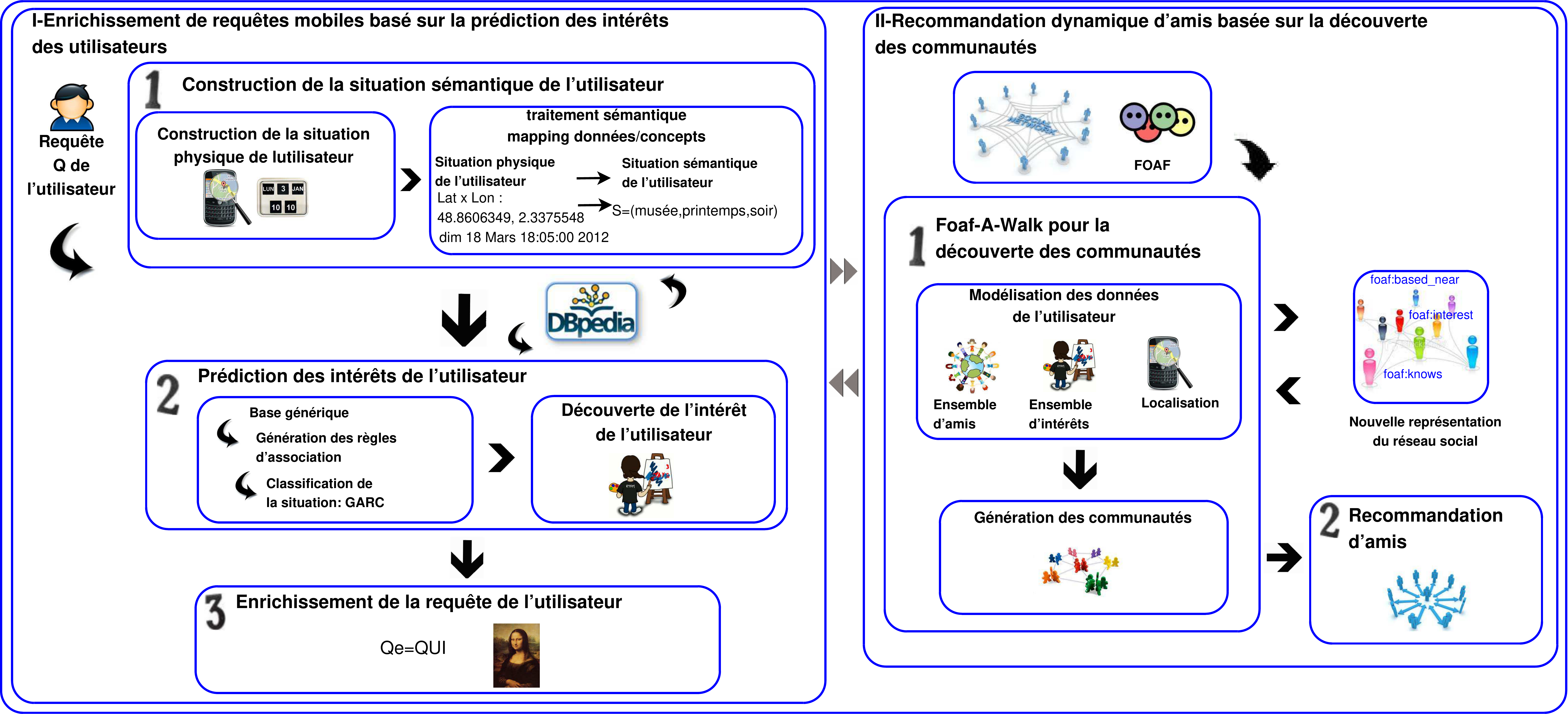, height=4.5in, width=9in}
\caption{Architecture générale de l'approche proposée}
\label{approachfig}
\end{figure*}
\end{landscape}

Dans cette section, nous présentons l'intéraction entre les deux parties de notre approche: l'enrichissement de requêtes et la recommandation d'amis.
En effet, lorsqu'un nouveau intérêt est prédit pour enrichir la requête d'un utilisateur donné, une nouvelle transaction est ajoutée à la base d'apprentissage contenant les corrélations entre les situations de l'utilisateur et ses intérêts. De plus, la description de l'utilisateur dans le réseau social est enrichie avec ce nouveau intérêt.
Par conséquent, à chaque fois o\`{u} un intérêt est prédit à cet utilisateur à partir de \textsc{Dbpedia}, son appartenance communautaire change et un nouveau ensemble d'amis possibles est sélectionné. Ainsi, la prédiction des intérêts des utilisateurs joue un double rôle, \emph{i.e.}, dans l'enrichissement de leurs requêtes et l'extension de leurs cercles sociaux. 
Cette intéraction est décrite dans la figure \ref{approachfig} représentant les différentes étapes de notre approche.

\section{Conclusion}

Dans ce chapitre nous avons proposé une nouvelle approche de RI contextuelle dans le cadre d'un environnement mobile. Ce travail est basé d'une part, sur la prédiction des intérêts de l'utilisateur dans le but de l'enrichissement de sa requête, et d'autre part, sur la découverte des communautés d'utilisateurs, rassemblés par intérêts et localisations, afin de recommander à cet utilisateur l'ensemble de personnes pouvant être ses futurs amis.

Dans le chapitre suivant, nous présentons une étude expérimentale qui va nous permettre d'évaluer les performances de l'approche que nous avons proposée. 
\chapter{Expérimentations \& \'{E}valuations} \label{ch4} \vspace*{3cm}
\section{Introduction}

Vu les contraintes et les spécificités techniques des appareils mobiles, \emph{i.e.}, les difficultés de saisie des requêtes, la zone d'affichage limitée, etc., l'effort émis par l'utilisateur pour saisir un mot avec le clavier d'un téléphone mobile est largement supérieur par rapport à la saisie standard avec un clavier AZERTY. De ce fait, nous nous intéressons à un type de recherche différent de celui des requêtes traditionnelles. En effet, des études sur les logs des requêtes des mobinautes \cite{Kamvar2007} ont montré que les requêtes des utilisateurs mobiles sont plus courtes et plus ambig\"{u}es avec une moyenne de 1.7 mots et 16.8 caractères par requête. De plus, un grand nombre d'utilisateurs ne consultent que la première page parmi l'ensemble des pages qu'ils ont re\c{c}ues comme résultat.
Ainsi, nous avons choisi de se limiter aux 10 premières ressources retournées par le navigateur Google pour tester la performance de notre approche d'enrichissement de requêtes.

De plus, en se basant sur le changement successif de la situation des utilisateurs dans l'environnement mobile, le contenu des réseaux sociaux devient très dynamique. En effet, les intérêts des utilisateurs évoluent et changent au cours du temps avec le changement de leur environnement physique, ce qui entra\^{\i}ne obligatoirement le changement de leurs préférences en terme d'amis. Ainsi, nous avons visé à vérifier l'impact de la situation des utilisateurs dans le changement de leurs cercles sociaux.

Dans le chapitre précédent, nous avons introduit une nouvelle approche de personnalisation de la RI basée sur la situation mobile des utilisateurs. En effet, notre approche est divisée en deux parties liées:
\begin{itemize}
  \item Enrichissement des requêtes mobiles basé sur la prédiction des intérêts des utilisateurs. En effet, notre approche SAI-RI combine entre une technique de fouille de données (la classification associative) et un traitement sémantique basé sur \textsc{Dbpedia} dans la prédiction des intérêts des utilisateurs utilisés dans l'enrichissement de leurs requêtes
  \item Recommandation dynamique d'amis basée sur la découverte des communautés. Cette partie se base sur l'approche Foaf-A-Walk qui permet de faire la fusion d'un algorithme de découverte de communauté avec la description \textsc{FOAF} dans le but de recommander de nouveaux amis aux utilisateurs.
\end{itemize}
Dans ce chapitre, nous allons mener une étude expérimentale réalisée à l'aide de plusieurs types de bases de test: une collectée avec une étude journalière et d'autres collectées du web. Ces expérimentations ont le but de prouver que notre approche donne les meilleurs résultats de point de vue prédiction des intérêts, précision des résultats retournés et qualité des amis recommandés aux utilisateurs.

\section{Description des bases de test}
Afin de prouver la performance de notre approche de personnalisation de la RI, nous avons réalisé une étude expérimentale basée sur trois bases de test.
\subsection{\'{E}tude journalière}
La plus grande limite que nous avons rencontrée dans l'évaluation de notre approche d'enrichissement de requêtes, est l'absence de benchmark permettant d'évaluer les approches de RI basées sur le contexte des utilisateurs.
Ainsi, en se basant sur les études de Sohn \textit{et al.,} \cite{Sohn2008}, nous avons mené une étude journalière (diary study) où chaque utilisateur est invité à enrigistrer le temps, la date, sa localisation et sa requête. 6 participants (3 hommes et 3 femmes) dont l'âge varie entre 24 et 34 ans, ont participé à cette étude (principalement des personnes de notre laboratoire de recherche) ayant tous une expérience dans les recherches web. Cette étude a duré 2 mois et a généré 60 entrées, avec une moyenne de 10 requêtes par personne (minimum=2 et maximum=25).

Dans ce contexte, une entrée utilisateur prend la forme d'un quadruplet (requête, temps, localisation, intérêt). Par conséquent, chaque requête est dépendante de la situation de l'utilisateur, ainsi que de ses intérêts.

Le tableau \ref{dataexample} montre quelques exemples des requêtes collectées, chacun est décrit par l'identifiant de l'utilisateur, le temps, la date et le lieu du lancement de la requête.

\begin{table*}[!h]
\centering\scalebox{0.85}{
\begin{tabular}{l l l l l}
\hline\hline
\textbf{Utilisateur} & \textbf{Temps} & \textbf{Localisation} & \textbf{Intérêt} & \textbf{Requête} \\
  \hline
  1 & Sam Dec 31 13:04:00 2011 & Centre\_commercial & Shopping & Puma\\
  2 & Dim Fev 5 16:30:00 2012 & Le musée du Louvre & Art & MonaLisa\\
  3 & Lun Jan 2 11:00:00 2012 & La plage de la Marsa & Beach-Sports & Sport\\
  \hline
\end{tabular}}
\caption{Un exemple d'entrées extraites à partir de l'étude journalière}
\label{dataexample}
\end{table*}

Ensuite, nous avons fait le mapping des données liées à la localisation et au temps en concepts manuellement à partir de \textsc{DBPEDIA}.
Pour conclure, cette étude journalière nous a permis de définir une base de test réelle contenant des requêtes utilisateurs dépendantes de leurs situations, pour vérifier l'efficacité de notre approche. La pertinence des ressources retournées par notre approche est jugée par l'utilisateur selon son besoin derrière la requête qu'il a soumise.

\subsection{Le challenge Quaero}
Pour s'assurer des résultats obtenus avec notre approche, nous avons visé à utiliser une collection de test appartenant à un challenge.

Dans la suite, nous décrivons le dataset de \emph{Quaero Evaluation for Task 2.6 on Contextual Retieval - Version 3.1}
\footnote{~\url{http://quaero.profileo.com/modules/movie/scenes/home}} qu'on a utilisé comme deuxième support d'évaluation.
Cette base contient 25 sujets représentant les besoins réels des utilisateurs, où chaque sujet inclut: (\textit{i}) le titre présentant le besoin de l'utilisateur ou sa requête; (\textit{ii}) la géolocalisation de l'utilisateur au moment où il a soumis sa requête; et (\textit{iii}) l'historique de recherche présentant les requêtes précédentes de l'utilisateur et les ressources qu'il a cliquées, \emph{i.e.}, les ressources jugées pertinentes.

Par conséquent, pour chaque sujet, nous regroupons les requêtes de l'utilisateur en deux ensembles: un ensemble d'apprentissage contenant l'historique de ses recherches et un ensemble de test contenant sa requête courante.

\subsection{Le dataset \textsc{FOAF}}
Dans le but de mener une étude expérimentale sur notre approche de recommandation d'amis, nous avons collecté un ensemble de documents \textsc{foaf} du web. Nous avons appliqué un processus de filtrage sur ces documents pour supprimer les informations inutiles et enrichir celles qui sont utiles (les relations de type \emph{foaf:knows}, \emph{foaf:based\_near } et \emph{foaf:interest}). Cette base contient 225 instances de type \emph{foaf:person} connectées initialement à l'aide de la propriété \emph{foaf:knows} (1372 instances de cette propriété), et décrites chacune par les propriétés \emph{foaf:based\_near } (225 instances de cette propriété) et \emph{foaf:interest} (438 instances de cette propriété).

Un exemple de description en \textsc{FOAF} d'une partie de notre base est donné par le tableau \ref{tab:foaf}.

\begin{table*}[!h]
\centering\scalebox{0.85}{
\begin{tabular}{|l|}
\hline
<?xml version="1.0"?>\\
<rdf:RDF>\\
<rdf:Description rdf:about="http://www.ivan-herman.net/foaf\#me">\\
<foaf:name>Ivan Herman</foaf:name>\\
<foaf:mbox\_sha1sum>eccd01ba8ce2391a439e9b052a9fbf37eae9f732</foaf:mbox\_sha1sum>\\
<foaf:givenname>Ivan</foaf:givenname>\\
<foaf:surname>Herman</foaf:surname>\\
<foaf:homepage rdf:resource="http://www.ivan-herman.net"/>\\
<foaf:interest rdf:resource="http://dbpedia.org/resource/Semantic\_Web"/>\\
<foaf:knows rdf:resource="http://www.ivan-herman.net/foafExtras.rdf\#SimonKaplan"/>\\
<foaf:knows rdf:resource="http://www.ivan-herman.net/foafExtras.rdf\#Tonya"/>\\
<foaf:based\_near rdf:resource="http://dbpedia.org/resource/Amsterdam"/>\\
</rdf:Description>\\
</rdf:RDF>\\
 \hline
\end{tabular}}
\caption{Un exemple d'une description \textsc{foaf}}
\label{tab:foaf}
\end{table*}

\section{Résultats et discussions}
L'objectif de cette section est de prouver l'efficacité de notre approche de prédiction d'intérêts des utilisateurs pour la RI et la recommandation d'amis dans le cadre d'un environnement mobile. Pour ce faire, nous avons opté pour différents types d'évaluation que nous allons détailler dans la suite de cette section.

\subsection{\'{E}valuation expérimentale de l'approche d'enrichissement de requêtes}
Dans ce cadre d'évaluation, nous considérons que la mesure de la précison est donnée par la formule suivante:
\begin{equation}
Pr\acute{e}cision = \frac{\{ressources~pertinentes\}\bigcap\{ressources~retourn\acute{e}es\}}{\{ressources~retourn\acute{e}es\}}
\end{equation}
\subsubsection{\'{E}valuation expérimentale basée sur l'étude journalière}
Comme première évaluation, nous appliquons notre approche d'enrichissement de requêtes sur la base collectée avec l'étude journalière. En particulier, nous comparons les résultats que nous avons obtenus en appliquant le processus de personnalisation avec ceux retournés par Google.

Nous commençons par une analyse des ressources retournées en terme de quantité. Ainsi, le nombre des ressources retournées sont comparées avec celles qui sont retournées par Google (\emph{c.f.}, Tableau \ref{numberresources}).

Ensuite, nous testons la performance de notre approche en terme de pertinence des résultats retournés (\emph{c.f.}, Tableau \ref{relevantresources}). En effet, nous utilisons 10 mots-clés pour calculer la précision des deux approches, \emph{i.e.}, notre approche d'enrichissement de requêtes vs celle du moteur de recherche Google. Plus précisement, chaque utilisateur est invité à analyser les deux résultats pour leur attribuer une valeur parmi deux valeurs possibles: pertinentes ou non pertinentes.

\begin{figure}[!h]
\centering
\epsfig{file=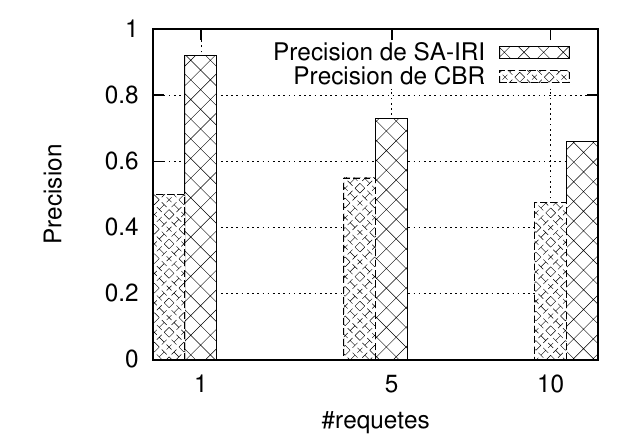, height=2in, width=3.5in}
\caption{Variation des valeurs de précision de notre approche vs celles du CBR}
\label{precision:int}
\end{figure}

Une autre expérimentation consiste à évaluer les intérêts prédits par notre approche par rapport aux intentions de recherche des utilisateurs (\emph{c.f.}, Tableau \ref{predictioninterests}) et par rapport aux intérêts obtenus avec la technique CBR (\emph{c.f.}, Figure \ref{precision:int}).

Ainsi, comme détaillé dans les statistiques du tableau \ref{numberresources}, nous remarquons que le nombre total des ressources retournées par notre approche est largement inférieur par rapport à celles retournées par Google. La requête 5 présente un cas exeptionnel où le nombre des ressources a augmenté. Ce comportement est dû à l'inéxactitude de l'intérêt prédit.

\begin{table*}[!h]
\begin{center}\scalebox{0.8}{
\begin{tabular}{|c|r|r|}
  \hline
  Requêtes & $\#$ Ressources retournées avec Google & $\#$ Ressources retournées avec notre approche\\
  \hline
  Requête1 & 920,000,000 & 21,900,000 \\
  Requête2 & 2,250,000 & 2,250,000 \\
  Requête3 & 13,200,000 & 8,480,000 \\
  Requête4 & 1,860,000,000 & 822,000,000  \\
  Requête5 & 3,640,000,000 & 3,800,000,000 \\
  Requête6 & 5,750,000 & 571,000 \\
  Requête7 & 73,900,000 & 73,100,000 \\
  Requête8 & 177,000,000 & 8,830,000 \\
  Requête9 & 48,400,000 & 1,490,000 \\
  Requête10 & 48,400,000 & 634,000 \\
  \hline
\end{tabular}}
\end{center}
\caption{Comparaison des résultats retournés par notre approche vs ceux de Google}
\label{numberresources}
\end{table*}

D'autre part, le tableau \ref{relevantresources} présente une comparaison, en termes de nombre, des ressources jugées pertinentes retournées par notre approche de personnalisation par rapport à celles retournées par Google. Cette comparaison montre une amélioration au niveau de la précision des ressources obtenues grâce à notre processus d'enrichissement.

\begin{table*}[!h]
\begin{center}\scalebox{0.7}{
\begin{tabular}{|c|r|r|}
  \hline
  Requêtes & Top10 des ressources retournées avec Google & Top10 des ressources retournées avec notre approche \\
  \hline
  Requête1 & 8 parmi 10 ressources pertinentes & 10 parmi 10 ressources pertinentes \\
  Requête2 & 1 parmi 10 ressources pertinentes & 1 parmi 10 ressources pertinentes \\
  Requête3 & 0 parmi 10 ressources pertinentes & 10 parmi 10 ressources pertinentes \\
  Requête4 & 1 parmi 10 ressources pertinentes & 4 parmi 10 ressources pertinentes  \\
  Requête5 & 2 parmi 10 ressources pertinentes & 8 parmi 10 ressources pertinentes \\
  Requête6 & 3 parmi 10 ressources pertinentes & 0 parmi 10 ressources pertinentes \\
  Requête7 & 1 parmi 10 ressources pertinentes & 1 parmi 10 ressources pertinentes \\
  Requête8 & 0 parmi 10 ressources pertinentes & 1 parmi 10 ressources pertinentes \\
  Requête9 & 2 parmi 10 ressources pertinentes & 2 parmi 10 ressources pertinentes \\
  Requête10 & 2 parmi 10 ressources pertinentes & 10 parmi 10 ressources pertinentes \\
  \hline
\end{tabular}}
\end{center}
\caption{Comparaison de la précision de notre approche vs celle de Google}
\label{relevantresources}
\end{table*}

Par ailleurs, nous avons réalisé une autre expérimentation permettant de mesurer l'efficacité de notre processus de prédiction des intérêts. En effet, le tableau \ref{predictioninterests} présente le nombre d'intérêts prédits jugés pertinents, par l'utilisateur, par rapport au nombre total d'intérêts prédits par notre approche.

\begin{table*}[!h]
\begin{center}\scalebox{0.8}{
\begin{tabular}{|c|r|r|}
  \hline
  Requêtes & Les intérêts pertinents prédits  & Le nombre total des intérêts prédits \\
  \hline
  Requête1 & 1 & 1\\
  Requête2 & 1 & 5 \\
  Requête3 & 3 & 4\\
  Requête4 & 1 & 2 \\
  Requête5 & 3 &3 \\
  Requête6 & 1 & 1\\
  Requête7 & 2 & 3 \\
  Requête8 & 2 & 3 \\
  Requête9 & 3 & 5 \\
  Requête10 & 3 & 3\\
  \hline
\end{tabular}}
\end{center}
\caption{Précision de la prédiction des intérêts}
\label{predictioninterests}
\end{table*}

\begin{figure}[!h]
\centering
\epsfig{file=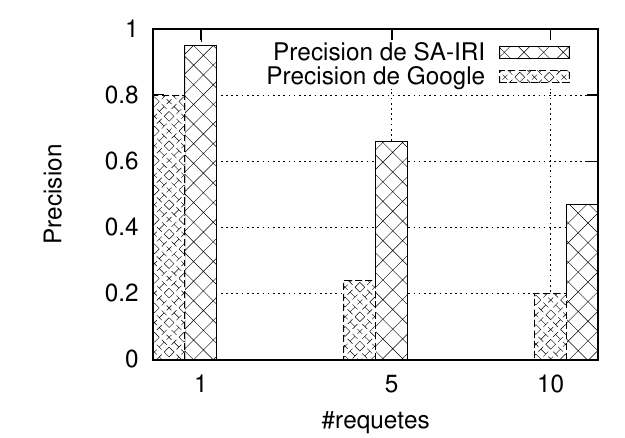, height=2in, width=3.5in}
\caption{Variation des valeurs de précision de notre approche vs celles de Google}
\label{precision:res}
\end{figure}

D'ailleurs, la précision de notre approche \textsc{SA-IRI} est donnée dans la figure \ref{precision:res}. Cette figure donne une comparaison entre la précision de notre approche et celle de Google. Plus précisémment, pour $\#$Requête=1, la moyenne de précision de notre approche est de 0.95, devant celle de Google égale à 0.8, \emph{i.e.}, une amélioration de la moyenne de précision de l'ordre de 15\%.

Par ailleurs, la figure \ref{precision:int} souligne la variation de la précision de la technique de classification que nous avons employée dans le processus de prédiction des intérêts, par rapport à la technique du raisonnement sur la base des cas utilisée dans \cite{Bouidghaghen2010}, selon le nombre de requêtes de test. Ainsi, pour $\#$Requête=1, notre précision est de 0.92, tandis que, pour la technique du CBR elle est de 0.5.
Cependant, pour $\#$Requête=5, nous avons une moyenne de précision égale à 0.73, \emph{i.e.}, une diminution de 19\%.
Les valeurs que nous avons obtenues montrent la différence entre la qualité des règles d'association et le raisonnement par cas. Ces résultats sont justifiés par la compacité et la fléxibilité de la présentation des connaissances sous forme de règles génériques d'association, devant les limites de la présentation sous forme de cas. Plus précisement, la technique de raisonnement à base de cas vérifie l'hypothèse de \emph{"Bon pour moi maintenant, donc bon pour tout le monde toujours"}. Ainsi, en se basant sur les spécificités de l'environnement mobile, un même utilisateur peut avoir deux intérêts différents dans une même situation se répétant deux fois successives. Dans un cas similaire, en employant la technique du CBR, seulement le premier intérêt prédit va être pris en considération. Cette particularité, donne impréssion que les intérêts des utilisateurs sont invariants, ce qui contredit les propriétés du contexte mobile.

\subsubsection{\'{E}valuation expérimentale basée sur le défi Quaero}
\begin{figure}[!h]
\centering
\epsfig{file=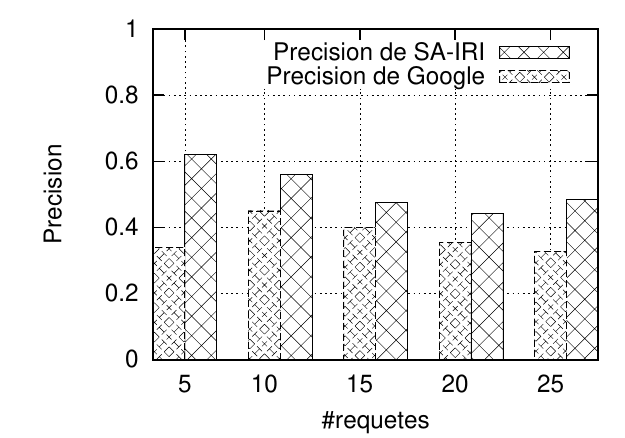, height=2in, width=3.5in}
\caption{Moyenne de précision des ressources retournées par notre approche vs celles de Google}
\label{precision:quaero}
\end{figure}

Dans le but de s'assurer des résultats obtenus avec l'étude journalière, nous avons appliqué notre approche d'enrichissement de requêtes sur la base offerte dans le cadre du défi Quaero.
Ainsi, nous avons fait des expérimentations pour comparer la précision de notre approche et celle de Google (\emph{c.f.}, Tableau \ref{precision:quaero}).
La variation de la précision des resultats est expliquée par le fait que les sujets composant la base sont divisés en deux catégories: 20 sujets sensibles aux situations des utilisateurs et 5 indifférents par rapport à ces situations.
Plus particulièrement, pour $\#$Requête=20, la moyenne de précision de notre approche est égale à 0.393. Cependant, elle s'améliore pour $\#$Requête=25 et atteint 0.404.
Par ailleurs, nous retrouvons les mêmes résultats pour la comparaison avec Google. En effet, pour $\#$Requête=25 la moyenne de précision de Google est de 0.328. Ces résultats montrent l'efficacité de notre approche d'enrichissement de requêtes même dans le cas des requêtes non sensibles aux situations des utilisateurs.

\subsection{\'{E}valuation expérimentale de l'approche de recommandation d'amis} \label{XPRecommandation}
Dans cette section, nous visons à mesurer les performences de l'approche de recommandation d'amis basée sur la découverte des communautés que nous avons proposée.
Ainsi, nous commençons par donner une étude comparative entre notre approche \textbf{Foaf-A-Walk} et \textbf{GN} \cite{Girvan2002} côté précision et temps d'exécution.
En effet, le tableau \ref{quality} montre, en terme de qualité, que l'approche Foaf-A-Walk est meilleure que l'approche GN. Dans le cas des petits réseaux, où le nombre des utilisateurs ne dépasse pas 100, les deux approches donnent pratiquement la même précision, \emph{e.g.}, 0.7 et 0.6. Par ailleurs, avec l'augmentation de la taille du réseau, \emph{i.e.} 200 utilisateurs, notre approche garde la même performance et même une meilleure précision avec une valeur égale à 0.73. Cependant, l'approche GN perd sa performance et donne une précision de l'ordre de 0.5.
De plus, la performance de Foaf-A-Walk est également prouvée avec la croissance du nombre des utilisateurs. Ainsi, nous avons mesuré une précision égale à 0.7 pour 100 utilisateurs et 0.74 pour 225 utilisateurs, \emph{i.e.}, une amélioration de l'ordre de 5.7\%.

\begin{table}[!h]
\centering
\begin{tabular}{|c|c|c|c|c|c|c|}
\hline
Utilisateurs & 100 &  125 &  150  & 175 & 200 & 225\\
\hline
\hline
Foaf-A-Walk & 0.7 &  0.71 &  0.71  & 0.72 & 0.73 & 0.74\\
\hline
GN & 0.6 & 0.58 & 0.55 & 0.53 & 0.5 & 0.48\\
\hline
\end{tabular}
\caption{Précision de l'approche \emph{Foaf-A-Walk} vs celle de GN en fonction de la taille du réseau social}
\label{quality}
\end{table}

Le tableau \ref{time} détaille une comparaison, par rapport au temps de réponse, entre les deux approches Foaf-A-Walk et GN. Nos expérimentations ont prouvé que la première a l'avantage d'être meilleure vu sa vitesse d'exécution.
Plus particulièrement, pour 125 utilisateurs, Foaf-A-Walk prend 0.0003 secondes dans la découverte des communautés. Ainsi, que GN prend 0.1 secondes pour la même phase.

\begin{table}[!h]
\centering
\begin{tabular}{|c|c|c|c|c|c|c|}
\hline
Utilisateurs & 100 &  125 &  150  & 175 & 200 & 225\\
\hline
\hline
Foaf-A-Walk & 0.0001 &  0.0003 &  0.001  & 0.005 & 0.005 & 0.005\\
\hline
GN & 0.01 & 0.1 & 0.37 & 0.65 & 1.39 & 4\\
\hline
\end{tabular}
\caption{Temps d'exécution de l'approche \emph{Foaf-A-Walk} vs celle de GN en fonction de la taille du réseau social (en secondes)}
\label{time}
\end{table}

En se basant sur ces expérimentations, l'approche Foaf-A-Walk a l'avantage d'être meilleure par rapport à sa précision et sa vitesse d'exécution.
En pratique; les réseaux complexes; la compléxité de cette approche est de l'ordre de $O(n^{2}\log n)$ où: $n$ est le nombre des utilisateurs (le nombre de n{\oe}uds du graphe).

\begin{figure}[!h]
\centering
\epsfig{file=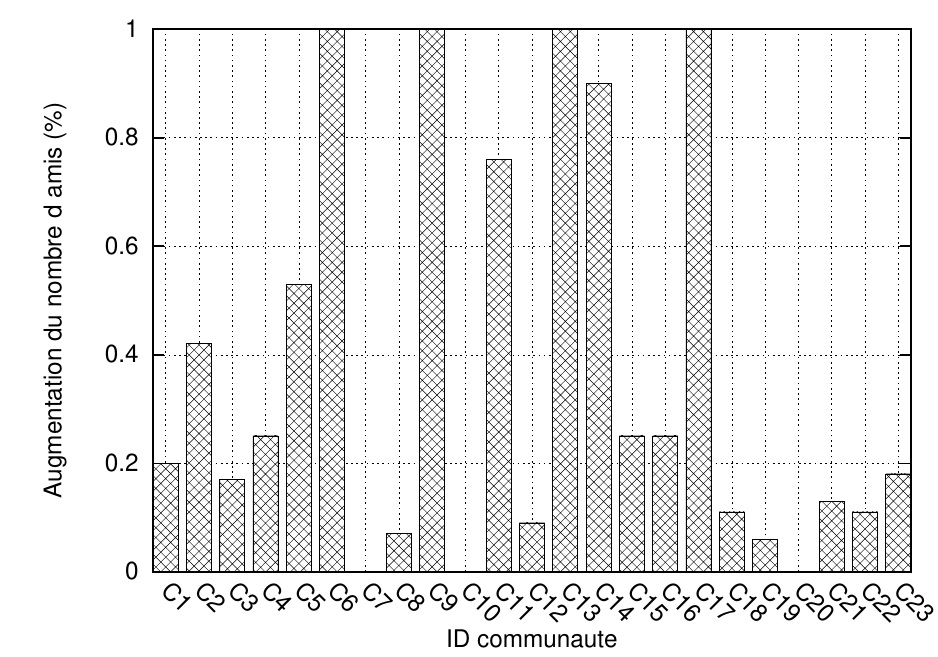, height=2.3in, width=3.8in}
\caption{Le pourcentage d'augmentation de la moyenne des amis dans chaque communauté}
\label{communities}
\end{figure}

Nous avons visé aussi, à tester l'impact du processus de recommandation sur le nombre d'amis dans chacune des communautés découvertes.
Ainsi, la figure \ref{communities} détaille, en terme de poucentage, la moyenne d'augmentation du nombre d'amis dans chaque communauté. Plus précisemment, dans le cas de la communauté 22, la moyenne du nombre d'amis était initialement de 257 personnes. Après l'extenssion du réseau, ce nombre atteint 287 avec un pourcentage d'augmentation égal à 11\%.
Des cas particuliers sont présentés par les communautés C7, C10 et C20, où l'augmentation d'amis est de 0\% vu la structure du graphe utilisé (aucune relation ne lie l'ensemble d'utilisateurs).

\begin{figure}[!h]
\centering
\epsfig{file=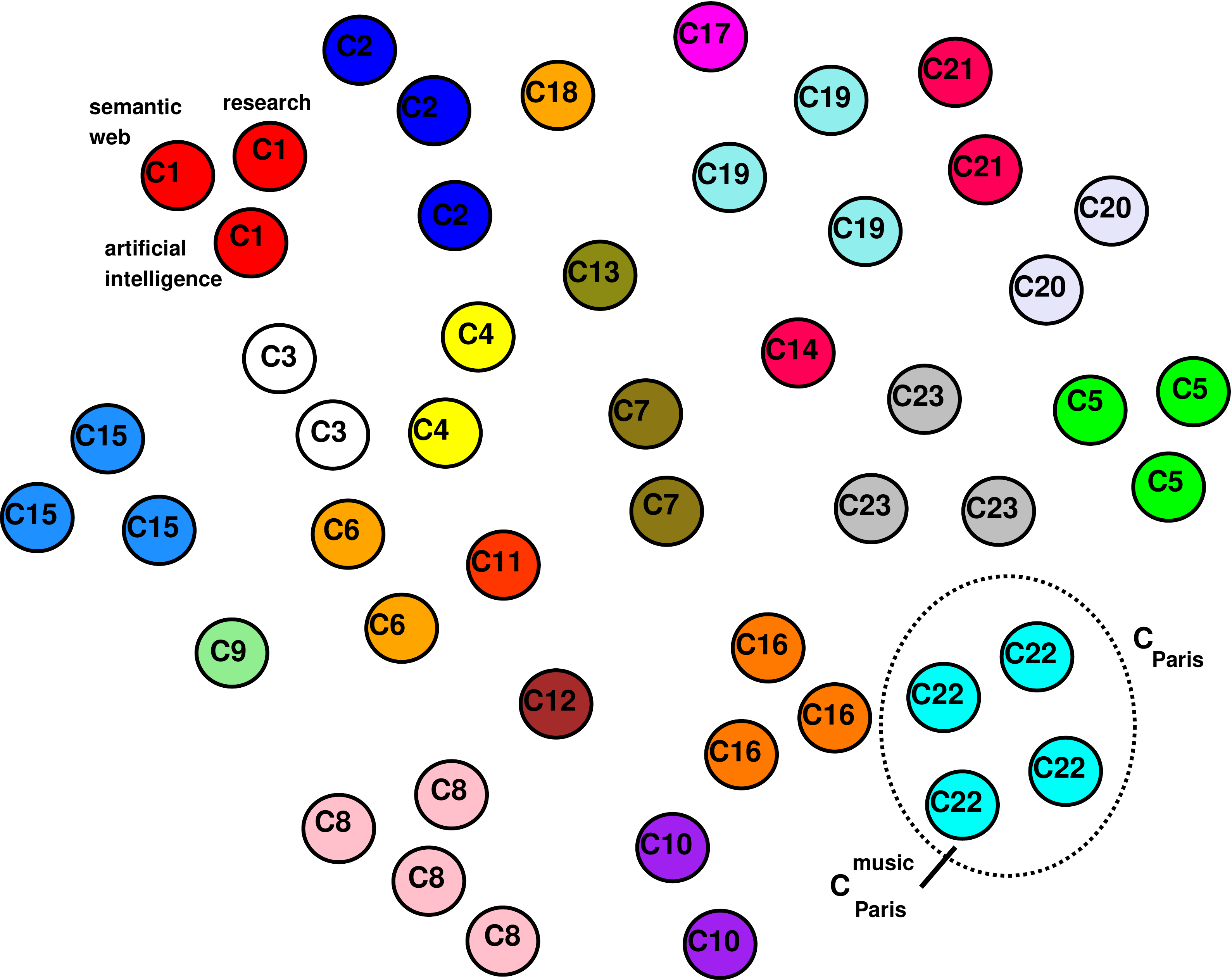, height=3.5in, width=5in}
\caption{La répartition des intérêts dans les communautés basée sur les localisations}
\label{communitiesdiscover}
\end{figure}

Finalement, la répartition des intérêts dans les communautés découvertes par rapport au paramètre de localisation, \emph{e.g.}, l'\'{E}tat de Massachusetts, est décrite dans la figure \ref{communitiesdiscover}. En effet, cette communauté est divisée en trois sous-communautés: Massachusetts/semantic\_web (8.33 \% de la population de cette communauté), Massachusetts/research (54.16 \% de la population de cette communauté) et Massachusetts/artificial\_intelligence (37.5 \% de la population de cette communauté).

\section{Conclusion}

Dans ce chapitre, nous avons réalisé une étude expérimentale de notre approche sur la base de plusieurs datasets. Ainsi, nous avons commencé par prouver la pertinence des résultats retournés par notre approche d'enrichissement de requêtes à l'aide de deux bases de test: une base collectée avec l'étude journalière et une deuxième base obtenue à partir du challenge Quaero. En effet, nous avons mesuré la précision de notre approche \textsc{SA-IRI} et nous avons effectué une comparaison entre la technique de classification utilisée dans notre approche et celle du CBR utilisée dans \cite{Bouidghaghen2010}. A partir de ces expérimentations, nous avons pu prouver que les résultats de notre approche sont largement meilleurs par rapport à ceux de Google.
Par ailleurs, nous avons pu prouver la performence de notre approche de recommandation d'amis en mesurant sa précision et en réalisant une comparaison entre notre approche Foaf-A-Walk et l'approche GN utilisée dans \cite{Girvan2002}. 
%%%%%%%%%%%%%%%%%%%%%%%%%%%%%%%%%%%%%%%%%%%%%%%%%%%%%%%%%%%%%%%%%%%%%%
%   Fichier :   Conclusion gnrale %
%%%%%%%%%%%%%%%%%%%%%%%%%%%%%%%%%%%%%%%%%%%%%%%%%%%%%%%%%%%%%%%%%%%%%%

\Conclusion{Conclusion g\'{e}n\'{e}rale} \label{Conclusion}
\markboth{CONCLUSION G\'{E}N\'{E}RALE}{Conclusion g\'{e}n\'{e}rale}

\begin{quotation}
\begin{flushright}
\textit{``Every Beginning Has an End,
\\ But Every End is the Start of a New Beginning..''\\}
\end{flushright}
\end{quotation}

\bigskip

\bigskip

\bigskip

\PARstart{A}{vec} la popularité des smartphones et la performance des téléphones mobiles, plusieurs contraintes sont imposées et ont obligé les nouvelles technologies à intégrer l'exploitation des contextes des utilisateurs dans leurs processus de selection des informations, susceptibles d'être utiles pour un utilisateur donné dans le but d'améliorer la précision de leurs ressources retournées.

Ainsi, beaucoup d'approches ont tenté d'intégrer le contexte dans leurs travaux de recommandation et de RI.
Plus précisement, ces travaux ont essayé de donner une modélisation unifiée du contexte en utilisant plusieurs sources de données, \emph{e.g.}, les coordonnées GPS, l'horloge système, les logs des requêtes, les réseaux sociaux, etc.

Partant de l'objectif d'exploiter le contexte mobile des utilisateurs dans le cadre de la RI et de la recommandation d'amis, nous avons proposé une nouvelle approche basée sur l'enrichissement de leurs requêtes mobiles et l'extension de leurs cercles sociaux.
Cette approche est constituée de deux grandes parties: (\textit{i}) Enrichissement de requêtes mobiles basé sur la prédiction des intérêts des utilisateurs: La situation de l'utilisateur est construite à travers un mapping entre sa situation physique et les concepts sémantiques extraits à partir de \textsc{Dbpedia}. Ensuite, la similarité entre cette situation et les situations précédentes est calculée afin de sélectionner celle la plus adéquate au profil utilisateur et l'exploiter dans le processus de la RI; et (\textit{ii}) Recommandation dynamique d'amis basée sur la découverte des communautés: Cette recommandation vise à étendre le cercle social de l'utilisateur. En effet, le réseau social décrit à l'aide de l'ontologie \textsc{Foaf} est analysé pour découvrir l'ensemble des communautés permettant de proposer à un utilisateur donné les personnes pouvant lui être intéressantes.
Ainsi, nous avons appliqué le processus de prédiction d'intérêts des utilisateurs à partir de leurs situations dans l'enrichissement de leurs requêtes (la séléction de leurs intentions de recherche) et l'extension de leurs relations sociales (la sélection des personnes qui partagent avec eux les mêmes préférences).

Nous avons validé notre approche contextuelle avec une étude expérimentale utilisant plusieurs datasets. En outre, nous avons commencé par la vérification de la pertinence des résultats retournés par notre approche d'enrichissement de requêtes à l'aide de deux bases de test: une base collectée avec l'étude journalière et une deuxième base obtenue à partir du challenge Quaero.
La précision de notre approche \textsc{SA-IRI} est mesurée grâce à une comparaison entre la technique de classification utilisée dans cette approche et celle du CBR utilisée dans \cite{Bouidghaghen2010}. A partir de ces expérimentations nous avons pu prouver que les résultats de notre approche sont largement meilleurs par rapport à ceux de Google.
Par ailleurs, nous avons visé à prouver la performence de notre approche de recommandation d'amis, avec une base \textsc{Foaf}, en mesurant sa précision et en réalisant une comparaison entre notre approche Foaf-A-Walk et l'approche GN utilisée dans \cite{Girvan2002}.

\subsection*{\textsc{Perspectives}}
Bien que les résultats de nos expérimentations sont assez encourageants, ce mémoire ouvre diverses perspectives. Nous citons dans la suite celles que nous trouvons les plus intéressantes:

\begin{enumerate}
\item Elargir la situation de l'utilisateur, utilisée dans notre approche \textsc{SA-IRI}, en ajoutant d'autres informations contextuelles pour sa construction. Ainsi, nous pouvons ajouter des informations liées à l'état émotionnel de l'utilisateur (triste ou content), des données collectées avec des capteurs physiques, \emph{e.g.}, son (fort ou faible), mouvement (se déplace ou fixe), température (élevée ou basse), etc. Cette modification n'a pas d'effet sur la performance de notre approche, vu les avantages de la technique que nous utilisons dans la manipulation de l'ensemble de situations, \emph{i.e.}, classification associative.

\item Comme notre approche de RI contextuelle \textsc{SA-IRI} a donné des résultats encourageants sur des bases de taille réduite, nous envisageons à l'appliquer sur des bases de test de plus grande taille. Il s'agit d'étaler la durée de l'étude journalière (un an par exemple).

\item Viser à généraliser la recommandation selon le besoin des utilisateurs et les données disponibles pour leur description. A cet effet, nous comptons proposer la recommandation des localisations (sur la base d'amis et d'intérêts) et la recommandation des intérêts (sur la base d'amis et des localisations).

\item Appliquer notre approche de recommandation d'amis sur des bases plus dynamiques et plus volumineuses. En effet, nous comptons appliquer notre approche sur des bases décrites à l'aide de l'ontologie \textsc{OPO} (Online Presence Ontology) vu son aspect dynamique par rapport à \textsc{Foaf}.
    Plus précisément, la représentation \textsc{OPO} de l'utilisateur via son contexte social (amis, coordonnées personnelles, etc.) et son contexte spatio-temporel (localisation et temps exact de la description), permet de définir une représentation socio-contextuelle de ce dernier.

\end{enumerate} 

%%%%%%%%%%%%%%%%%%%%%%%%%%%%%%%%%%%%%%%%%%%%%%%%%%%%%%%%%%%%%%%%%%%%%%%%
% Liste des figures et des tableaux - Liste des abrviations
%\cleardoublepage \addcontentsline{toc}{chapter}{\listtablename}  \listoftables  %
%%%%%%%%%%%%%%%%%%%%%%%%%%%%%%%%%%%%%%%%%%%%%%%%%%%%%%%%%%%%%%%%%%%%%%%%

\bibliographystyle{apalike-fr}

\bibliographystyle{plain}

% Annexes
\appendix

\end{document}